\documentclass[12pt]{article}
\usepackage{amsmath,amssymb,amsfonts}
\usepackage[dvips]{color}
\usepackage{xcolor}
\usepackage{graphicx}
\usepackage{caption}
\usepackage{natbib, har2nat}
\usepackage{rotating}
\usepackage{afterpage}
\usepackage{cancel}
\usepackage{booktabs}
\usepackage{setspace}
\usepackage{enumerate}
\usepackage{mathabx}
\usepackage{float}
\usepackage{footnote}
\usepackage{multirow}
\usepackage[flushleft]{threeparttable}
\usepackage{caption}
\usepackage{subcaption}

\newcommand{\fra}[2]{\frac{\displaystyle #1}{\displaystyle #2}}
\newcommand{\bc}{\begin{center}}
	\newcommand{\ec}{\end{center}}
\newcommand{\be}{\begin{equation}}
\newcommand{\ee}{\end{equation}}
\newcommand{\bea}{\begin{eqnarray}}
\newcommand{\eea}{\end{eqnarray}}
\newcommand{\bean}{\begin{eqnarray*}}
	\newcommand{\eean}{\end{eqnarray*}}
\newcommand{\bt}{\begin{tabular}}
	\newcommand{\et}{\end{tabular}}
\usepackage{url}       %
\usepackage{xcolor}
\usepackage[margin=1in]{geometry}
\usepackage{graphicx}

\usepackage[pdftex,bookmarks,colorlinks]{hyperref}

\newtheorem{theorem}{Theorem}

\newtheorem{proposition}[theorem]{Proposition}

\newcommand{\argmin}{\operatorname*{argmin}}

\newcounter{saveeqn}

\newcommand\numberthis{\addtocounter{equation}{1}\tag{\theequation}}

\numberwithin{equation}{section}
\numberwithin{theorem}{section}

\hypersetup{
	colorlinks,
	citecolor = blue!70,
	linkcolor = red!90,
	urlcolor = blue}

\makeatletter
\def\@fnsymbol#1{\ensuremath{\ifcase#1\or \dagger\or \ddagger\or
		\mathsection\or \mathparagraph\or \|\or **\or \dagger\dagger
		\or \ddagger\ddagger \else\@ctrerr\fi}}
\makeatother

\begin{document}

\setlength{\abovedisplayskip}{5pt}
\setlength{\belowdisplayskip}{7pt}

\title{ \vspace{-1.5cm} \bf Forecasting with Bayesian Grouped Random Effects in Panel Data}

\author{
	Boyuan Zhang\thanks{Department of Economics, Perelman Center for Political Science and Economics, University of Pennsylvania, 133 S. 36th St.,  Philadelphia, PA 19104-6297. Email: \href{mailto:boyuanz@sas.upenn.edu}{boyuanz@sas.upenn.edu}.  We would like to thank Karun Adusumilli, Francis Diebold, Maximilian Göbel, Philippe Goulet Coulombe, Frank Schorfheide for helpful comments and suggestions.} \\
    }


\date{\vspace{-0.4cm}
	University of Pennsylvania\\[2ex]%
	\small
	First Version: June 30, 2020 \\
	This Version: \today \\
	\vspace{0.4cm}
	\large
}

\maketitle

\begin{abstract}
	In this paper, we estimate and leverage latent constant group structure to generate the point, set, and density forecasts for short dynamic panel data. We implement a nonparametric Bayesian approach to simultaneously identify coefficients and group membership in the random effects which are heterogeneous across groups but fixed within a group. This method allows us to flexibly incorporate subjective prior knowledge on the group structure that potentially improves the predictive accuracy. In Monte Carlo experiments, we demonstrate that our Bayesian grouped random effects (BGRE) estimators produce accurate estimates and score predictive gains over standard panel data estimators. With a data-driven group structure, the BGRE estimators exhibit comparable accuracy of clustering with the \textit{Kmeans} algorithm and outperform a two-step Bayesian grouped estimator whose group structure relies on \textit{Kmeans}. In the empirical analysis, we apply our method to forecast the investment rate across a broad range of firms and illustrate that the estimated latent group structure improves forecasts relative to standard panel data estimators.
\end{abstract}

\noindent JEL CLASSIFICATION: C11, C14, C23, C53, G31


\thispagestyle{empty}
\setcounter{page}{0}
\newpage

\newpage


\section{Introduction} \label{sec:intro}

With the increasing availability of panel data, many works have examined and demonstrated its central role in the empirical research throughout the social and business sciences. Analysis of panel data has various edges over that of pure cross-sectional or time-series data. The most important one is that the panel data provide researchers with a flexible way to model both heterogeneity among individuals, firms, regions, and countries and possible structural changes over time. Apart from the principal role in the model estimation, it is interesting and essential to study their relevance for forecasting. Among novel methods emerged recently, the latent group structure in the heterogeneity attracts wide attention. In this paper, we allow for grouped patterns of unobserved heterogeneity in the dynamics panel data models and evaluate whether this latent structure improves the predictive performance in an extensive collection of short time series.


In the dynamics panel data model, it is common to assume that each cross-sectional unit has unique intercept. This assumption introduces a large number of parameters that become a burden in estimation. In models that have as many parameters as individual units, fixed effects estimators are known to suffer from the ``incidental parameters" problem \citep{neyman1948}, which can bring about significant biases in estimates of common parameters. This problem becomes severe in short panels even if the number of units goes to infinity \citep{chamberlain1980,nickell1981}, and the fixed-effects themselves are often poorly estimated. An unreliable estimate leads to concerns about the predictive power of panel data models as inaccurate estimates affect forecasts in all aspects.

To address this issue\footnote{Another important strand of literature implements generalized method of moments (GMM) methods to eliminate bias, see \citet{arellano1991}, \citet{arellano1995} and \citet{blundell1998}. Though successfully solved the ``incidental parameters" problem, this set of methods doesn't allow for any latent group structure.}, econometricians attempt to reduce the number of unknown parameters by dividing units into a finite number of groups. The premise of this idea is that units in the same group share the unit-specific parameters. Previous works include \citet{bonhomme2015}, \citet{ando2016}, \citet{su2016}, \citet{bester2016}, \citet{su2019}, \citet{bonhomme2019}, and \citet{cheng2019}. Moreover, finite mixture model provide a well-known probabilistic approach to model-based clustering \citep{mcnicholas2010,fruhwirth2011a}. With a finite number of groups, econometricians could avoid ``incidental parameter” problem under several particular assumptions and derive consistent estimators for the common parameters.

However, the convenience of the group structure does not come without any cost. The number of groups is an unknown but fixed quantity, and the need to specify the number in advance is deemed one of the significant drawbacks of applying these methods in a clustering context. Many methods have been suggested to estimate the optimal number a posteriori from the data such as BIC \citep{keribin2000,bonhomme2015}, marginal likelihoods \citep{fruhwirth2004}, or the integrated classification likelihood \citep{biernacki2000}. Bayesian approaches sometimes pursue a similar strategy, often adding the DIC to the list of model choice criteria, e.g., \citet{celeux2006} and \citet{kim2019}. If both $N$ and $T$ are large enough, the information criterion could select the true group structure.  However, with a short time span, these criteria might fail to achieve their goal. As noted in \citet[hereafter BM]{bonhomme2015},  the choice of the number of groups is crucial to estimation and inference for model parameters. Misspecification in group number forces the algorithm to consider incorrect group membership. We will later show that it is the information criteria that substantially affects the performance of Grouped Fixed Effects (GFE) estimator proposed by BM.

The contributions of this paper are fourfold. First, closely following \citet[hereafter KW]{kim2019} and \citet{liu2020}, we develop a posterior sampling algorithm that addresses the nonparametric estimation of latent grouped effects and proposes Bayesian Grouped Random Effects (BGRE) estimator. The number of groups is treated as an unknown parameter that is estimated jointly with the component-specific parameters under the assumption that group membership remains constant over time. Instead of using the Finite Mixture model, which needs to preset the number of groups, we use Dirichlet Process (DP) prior, in particular the stick-breaking prior, that allows for infinite potential groups. The entire posterior sampler builds upon the blocked Gibbs sampling\footnote{Unlike the Pólya urn Gibbs sampler \citep{escobar1995}, blocked Gibbs sampler approach avoids marginalizing over the prior and thus allows for direct sampling of the nonparametric posterior, leading to computational and inferential advantages.} proposed by \citet{ishwaran2001}.

Second, we leverage the researcher's prior knowledge of the latent group structure to improve the estimation and forecasting. In particular, we summarize and incorporate the information of subjective group structure in the prior distribution of the membership probabilities. If the subjective prior on the group structure is more precise than the random guess, even with incorrect presumed number of groups, we show that including it in the prior improves the performance of the BGRE estimators as it guides the group membership estimates.

Third, we explore the potential link between the proposed BGRE estimators and unsupervised machine learning method. Theoretically, we show that our block Gibbs sampler for the BGRE estimator is closely related to the \textit{Kmeans} algorithm \citep{macqueen1967} under certain assumptions. In particular, both algorithms assign units to the closest centroid when forming the clusters and recalculate the means of the new cluster afterward. To compare the performance of clustering, we modify our algorithm to incorporate \textit{Kmeans} and construct a two-step BGRE estimator where individuals are clustered in the first step using \textit{Kmeans}, and the group-specific heterogeneity is estimated in the second step. In the simulation section, we document that our BGRE estimators dominate the two-step GRE estimator in terms of the performance of both clustering and forecasting. We also find that the two-step GRE estimators with \textit{Kmeans} algorithm severely underestimate the number of groups under all data generating processes, whereas BGRE estimators deliver accurate estimates.

Last but not least, we examine the performance of BGRE estimators using various sets of simulated data and real data. The Monte Carlo study presents that grouped heterogeneity brings gains in estimating group structure and one-step ahead point, set, and density forecasting relative to commonly used predictors with different parametric priors on individual effects. In particular, our estimators outperform BM's Grouped Fixed Effects (GFE) estimator in various settings of the data generating process. The better performance is primarily due to the accurate estimate of the group structure. Regarding other predictors, we show that failing to model group structure and to pool information across units severely deteriorates the results for both estimation and forecasting. Finally, we use our method to forecast the investment rate across a broad range of firms. The BGRE estimators offer better performance than the standard panel data models in forecasting. This reveals that incorporating the latent group structure provides a great amount of flexibility and improves the predictive power of the underlying panel data model.

Our paper relates to several branches of the literature. Our work is closely related to BM, KW, and \citet[hereafter BLM]{bonhomme2019}.  All of these three papers aim to estimate the unobserved heterogeneity in a linear dynamic panel data model and develop statistical inference methods. BM estimates the parameters of the model using the GFE estimator that minimizes the least-squares criterion for all possible groupings of the cross-sectional units. They jointly estimate the individual types and the model's parameter given the number of groups and perform model selection afterward. One the other hand, BLM modify this method and split the procedure into two steps with \textit{Kmeans} clustering algorithm is used in the first step. From Bayesian's point of view, KW proposes a full Bayesian estimator that simultaneously estimates the group structure and parameters. Unfortunately, none of these works examine the potential forecasting gain when considering the group structure. Our work will fill in this gap and examine the performance of the BGRE estimator in various scenarios.

The paper that most related to ours is \citet{liu2020}. She considers a linear dynamic panel data model and implements the finite mixture model to estimate the underlying distribution of unit-specific intercepts. However, the assumptions of the underlying model in \citet{liu2020} are different from ours. She assumes fully heterogeneous intercepts in a panel data model, which amounts to the standard random-effects model. The finite mixture model in her setting serves as a method to pool information across units. In our work, we specify a group-specific intercept in the population level, and the semiparametric method is the critical ingredient to deliver not only the estimates but also the group structure. Moreover, her main object of interest is to construct individual-specific density forecasts for a panel, while our work includes point, set, and density forecasts and, most importantly, evaluates the performance of group clustering.

This paper also relates to the literature on nonparametric Bayesian approach in the group structure estimation problem. We model the unknown distribution of the heterogeneous coefficient (including grouped intercept and innovation variance) as the Dirichlet Process of Normals with potential infinite groups. The idea of sampling from the Dirichlet Process Model has been widely used by a number of authors including \citet{escobar1995}, \citet{neal2000}, \citet{ishwaran2001}, \citet{molitor2010}, \citet{yau2011}, \citet{hastie2015}, \citet{liverani2015}, \citet{liu2019}, and \citet{liu2020}. To make the infinite-dimensional problem operable, our blocked Gibbs sampler relies on the slice sampling described by \citet{walker2007}, a more efficient version was later proposed by \citet{kalli2011}.

We proceed as follows. In section \ref{sec:model}, we present the specification of a linear dynamic panel data model and discuss the construction and evaluation of point, set, and density forecasts. Section \ref{sec:bayesianEstimation} provides details on nonparametric Bayesian priors and subjective group structure priors. It also documents the connection to the \textit{Kmeans} algorithm. We conduct various Monte Carlo experiments in section \ref{sec:mcmc} to examine the performance of the proposed estimator in a controlled environment in the light of point, density, and set forecasts. We also examine the performance of a few variants of the BGRE estimator. In section \ref{sec:emp_result}, we conduct empirical analysis in which we forecast the investment rate across firms. Finally, we conclude in section \ref{sec:conclusion}. A description of the data sets, additional empirical results, and derivations are relegated to the appendix.


\section{The modeling framework} \label{sec:model}

\subsection{Model}

We consider a panel with observations for cross-sectional units $i=1, \ldots, N$ in periods $t=1, \ldots, T$. Given a panel data set $\{ (y_{it}, x_{it})\}$, a simple linear dynamic panel data model with grouped patterns of heterogeneity takes the following form:
\begin{align} \label{bm_model}
	y_{it} &= \alpha_{g_{i} t} + \rho y_{it-1} + \beta_i^{\prime} x_{it} + \varepsilon_{i t}, \quad \varepsilon_{i t} \stackrel{iid}{\sim} N \left(0, \sigma_{g_{i}}^{2}\right).
\end{align}
where $x_{i t}$ are a $p \times 1$ vector of exogenous variables, they are uncorrelated with $\varepsilon_{i t}$ but is allowed to be arbitrarily correlated with $\alpha_{g_{i} t}$. $\alpha_{g_{i} t}$ denote the time-varying group-specific heterogeneity. The subscript $g_{i} \in \{1,..., K \}$ is the group membership variable with unknown and unconstrained $K$. $y_{it-1}$ is the lagged outcome variable. $\rho$ is the homogeneous AR(1) parameters that are common for all cross-sectional units, and $\beta_{i}$ is a $p \times 1$ vector of unit-specific slope
coefficients. $\varepsilon_{i t}$ is the idiosyncratic error term featured by zero mean and grouped heteroskedasticity $\sigma_{g_{i}}^{2}$, with cross-sectional homoskedasticity being a special case where $\sigma_{g_{i} }^{2}=\sigma^{2}$. This setting leads to a heterogeneous panel with group pattern modeled through $\alpha_{g_{i} t}$ and $\sigma_{g_{i}}^{2}$.

By stacking all observations for unit $i$, we get an aggregated model:
\begin{align}
	y_{i} = \alpha_{g_{i}} + \rho y_{i,-1} + x_{i} \beta_i + \varepsilon_{i}, \quad \varepsilon_{i} \stackrel{iid}{\sim}  N\left(\mathbf{0}, \boldsymbol{\Sigma}_{g_{i}}\right).
\end{align}
where $y_{i} = \left[y_{i 1}, y_{i 2}, \ldots, y_{i T}\right]^{\prime}$, $y_{i,-1} = \left[y_{i 0}, y_{i 1}, \ldots, y_{i T-1}\right]^{\prime}$, $T \times 1$, $x_{i} = \left[x_{i 1}, x_{i 2}, \ldots, x_{i T}\right]^{\prime}$,  $\alpha_{g_{i}} = \left[\alpha_{g_{i} 1}, \alpha_{g_{i} 2}, \ldots, \alpha_{g_{i} T}\right]^{\prime}$, $\varepsilon_{i} = \left[\varepsilon_{i 1}, \varepsilon_{i 2}, \ldots, \varepsilon_{i T}\right]^{\prime}$, $\boldsymbol{\Sigma}_{g_{i}} = \sigma_{g_{i}}^{2} \mathbf{I}_T$.  To indicate the component from which each observation stems, we introduce a group membership variable $G = \left[ g_{1}, \ldots, g_{N}\right]$ taking values in $\{1, \ldots, K\}^{N}$. Define a set of unit that belongs to group $k$: $C_k = \left \{i \in \{1,2,...,N\} | g_i = k \right \}$. Let $|C_k|$ denote the cardinality of the set $C_k$.

Following \citet{sun2005}, \citet{lin2012} and BM, we assume that individual group membership does not vary over time. In addition, for any group $i \neq j$, we assume that they have different paths of random effects, e.g., $\alpha_{i} \neq \alpha_{j}$, and no single unit can simultaneously belong to these two groups: $C_{i} \bigcap C_{j} = \emptyset$.

The main goal of this paper is to estimate the grouped random effects $\alpha_{g_i}$, common parameter $\rho$, hetergenous coeffecients $\beta_{i}$ and group membership $G$ using full sample and provide the point, set, and density forecasts of $y_{i t+h}$ for each unit $i$. Throughout this paper, we focus on the one-step ahead forecast where $h=1$. For the multiple-step forecast, the procedure can be extended by iterating $y_{i T+h}$ in accordance with (\ref{bm_model}) given the estimate of parameters and realizations of data.


\subsection{Estimation and Forecast Evaluation}

\subsubsection{Posterior Predictive Densities}
Our goal is to generate one-step ahead forecasts of $y_{i, T+1}$ for $i = 1,...,N$ conditional on the history of observations,
\begin{align*}
	Y &= [y_1, y_2, ..., y_N ], y_i = [y_{i1},y_{i2},...,y_{iT}]', \\
	X &= [x_1, x_2, ..., x_N ], x_i = [x_{i1},x_{i2},...,x_{iT}]'.
\end{align*}
and newly available exogenous variables $x_{i T+1}$ at $T+1$. For illustration purpose, we drop $X$ and $x_{i T+1}$ from notations but we always condition on these exogenous variables.

The posterior predictive distribution for unit $i$ is given by
\begin{align}
	p(y_{i T+1} | Y) = \int p(y_{i T+1} | Y, \Theta) p(\Theta | Y) d \Theta,
\end{align}
where $\Theta$ is a vector of parameters $\Theta = \{\rho, \beta_i, \alpha_{g_i}, \Sigma_{g_i}, g_i \}$. This density is the posterior expectation of the following function,
\begin{align}
	p(y_{i T+1} | Y, \Theta) = \sum_{k=1}^{K} \mathbf{1}(g_i = k) p\left(y_{i T+1}| Y,  \rho, \beta_i, \alpha_{k}, \Sigma_{k}\right),
\end{align}
which is invariant to relabeling the components of the mixture. Therefore, given $M^*$ posterior draws, the density estimated from the MCMC draws is
\begin{align}
	\hat{p}(y_{i T+1} | Y) = \frac{1}{M^*} \sum_{j=1}^{M^*} \left( \sum_{k=1}^{K^{(j)}} \mathbf{1}(g_i = k) p\left(y_{i T+1}| Y, \rho^{(j)}, \beta^{(j)}_i, \alpha_{k}^{(j)}, \Sigma_{k}^{(j)}\right) \right).
\end{align}
Therefore, we can draw samples from $\hat{p}(y_{i T+1} | Y)$ by simulating (\ref{bm_model}) forward conditional on the posterior draws of $\Theta $ and observations.

\subsubsection{Point Forecasts}
We evaluate the point forecasts via the Root Mean Square Forecast Error (RMSFE) under the quadratic loss function averaged across units. Let $\hat{y}_{i T+1} $ represent the predicted value conditional on the observed data up to period $T$, the loss function is written as
\begin{align}
	L\left(\widehat{y}_{1:N, T+1}, y_{1: N, T+1}\right) = \frac{1}{N} \sum_{i=1}^{N} \left(\hat{y}_{i T+1} - y_{i T+1}\right)^{2} =  \frac{1}{N} \sum_{i=1}^{N} \hat{\varepsilon}_{iT+1}^2,
\end{align}
where $y_{i, T+1}$ is the realization at $T+1$ and $\hat{\varepsilon}_{iT+1}$ denote the forecast error.

The optimal posterior forecast under quadratic loss function is obtain by minimizing the posterior risk,
\begin{align*}
	\hat{y}_{1:N, T+1} & = \argmin_{\hat{y}  \in \mathbb{R}^N} \int_{-\infty}^{\infty} L\left(\hat{y} , y_{1: N, T+1}\right) p(y_{1: N, T+1} | Y) d y_{1: N, T+1} \\
	& = \argmin_{\hat{y} \in \mathbb{R}^N} \frac{1}{N} \sum_{i=1}^{N} \mathbb{E} \left[\left(\hat{y} - y_{i T+h}\right)^{2} | Y \right]. \numberthis
\end{align*}
This implies optimal posterior forecast is the posterior mean,
\begin{align}
	\hat{y}_{i, T+1} = \mathbb{E} \left(y_{i T+1} | Y \right), \text{ for } i=1, \ldots, N.
\end{align}


\subsubsection{Set Forecasts}
We construct set forecasts $CS_{i T+1}$ from the posterior predictive distribution of each unit. In particular, we adopt a Bayesian approach and report the highest posterior density interval (HPDI), which is the narrowest connected interval with coverage probability of $1-\alpha$. Put differently, it requires that the probability of $y_{i T+h} \in CS_{iT+1}$ conditional on having observed the history $Y$ is at least $1-\alpha$, i.e.,
\begin{align}
	P ( y_{i T+1} \in CS_{i T+1} ) \geq 1-\alpha, \quad \text {for all } i,
\end{align}
and this interval is the shortest among all possible single connected candidate sets. Let $\delta^{l}$ be the lower bound and $\delta_{u}$ be the upper bound, then $CS_{i T+1} = \left[\delta_{i}^{l}, \delta_{i}^{u}\right]$.

The assessment of set forecasts in simulation studies and empirical applications is based on two metrics: (1) the cross-sectional coverage frequency,
\begin{align}
	Cov_{T+1} = \frac{1}{N} \sum_{i=1}^{N} \mathbb{I} \left\{y_{i T+1} \in CS_{i T+1} \right\},
\end{align}
and (2) the average length of the sets $C_{i T+1}$,
\begin{align}
	AvgL_{T+1} = \frac{1}{N} \sum_{i=1}^{N} (\delta_{i}^{u} - \delta_{i}^{l}).
\end{align}

\subsubsection{Density Forecasts}
To compare the performance of density forecast for various estimators, we examine the continuous ranked probability score (CRPS) across units. The CRPS is frequently used to assess the respective accuracy of two probabilistic forecasting models. It is a quadratic measure of the difference between the forecast cumulative distribution function (CDF), $F^{T+1}_i(y)$, and the empirical CDF of the observation with the formula as follows,
\begin{align*}
	CRPS_{T+1} &= \frac{1}{N} \sum_{i=1}^{N} CRPS(F^{T+1}_i, y_{i T+h}) \\
	&= \frac{1}{N} \sum_{i=1}^{N} \int_{0}^{\infty}\left(F^{T+1}_{i}(y)-\mathbb{I}\left\{y_{i T+h} \leq y\right\}\right)^{2} dy, \numberthis
\end{align*}
where $y_{i T+h}$ is the realization at $T+1$.


Moveover, we report another metric called the average log predictive scores (LPS) to assess the performance of the density forecast from the view of the probability distribution function (PDF). As suggested in \citet{geweke2010}, the LPS for a panel reads as,
\begin{align}
	LPS_{T+1} = & \frac{1}{N} \sum_{i=1}^{N} \ln p\left(y_{i T+1} | Y \right),
\end{align}

\subsubsection{Estimation}

To evaluate the statistical superiority of pooling within $K$ clusters, we report the bias, standard deviation, average length of 95\% credible set, and frequentist coverage of the posterior mean estimate of $\rho$ across Monte Carlo repetitions. For the random effects $\alpha$, we only present the average bias as it may not be of interest for most empirical analysis.

To estimate the number of groups, we derive a point estimator from its posterior distribution, typically, the posterior mean, which is consistent with a quadratic loss function. In the empirical analysis, we also consider the posterior mode suggested by \citet{malsiner2016}, which is equal to the most frequent number of non-empty components visited during MCMC sampling. These approaches constitute an automatic and straightforward strategy to estimate the unknown number of groups without using model selection criteria or marginal likelihoods.

\subsection{Extension}
Until now, our main focus is the group structure for the intercepts $\alpha_i$ while $\rho$ and $\beta_i$ are left unchanged as in a standard panel data model. Here, we can easily extend our model and allow for joint group-specific heterogeneity in $\alpha$, $\beta$ and $\rho$. Then, the extended model is written as,
\begin{align} \label{bm_model_ex1}
	y_{it} &= \theta_{g_{i}}^{\prime} \widecheck{x}_{it} + \varepsilon_{i t} , \quad \varepsilon_{i t} \stackrel{iid}{\sim} N \left(0, \sigma_{g_{i}}^{2}\right),
\end{align}
where $\widecheck{x}_{it} = [1 \; y_{it-1} \; x'_{it}]' $, and $\theta_{g_{i}} = [\alpha_{g_{i}} \; \rho_{g_{i}} \; \beta_{g_{i}}^{\prime}]'$.

From a group $k$, with a joint conjugate prior for parameters $\theta_{k}$, we modify our block Gibbs Sampler to draw $\alpha_k$, $\rho_k$ and $\beta_k$ simultaneously from their joint posterior distribution. The detailed derivation of the posterior distributions are presented in Appendix \ref{appendix:post_fullGroupStructure}.

\section{Bayesian Estimation}  \label{sec:bayesianEstimation}

In this section, we provide details in Bayesian analysis. In Section \ref{subsec:prior}, we document the specification of the prior distribution for all parameters, including the auxiliary variable in the random coefficient model, and the subjective group prior if econometricians have prior knowledge on group structure. Section \ref{subsec:post} outlines the posterior sampler and the proposed algorithm is shown in the Appendix \ref{subsec:algo}. Finally, we provide preliminary thoughts on the connection between our Bayesian method and unsupervised machine learning methods in Section \ref{subset:linktoML}.

%


\subsection{Nonparametric Bayesian Prior} \label{subsec:prior}

Two sets of specifications of prior distributions are considered in this section. In the first prior specification, we concentrate on random effects models and implement a full Bayesian analysis. In addition, we specify a hyperprior for the distribution of unobserved heterogeneity and then construct a joint posterior for the coefficients of this hyperprior as well as the actual unit-specific and common coefficients. In the second specification, econometricians could provide useful information on the latent group structure and incorporate it in the prior.


\subsubsection{Random Coefficients Model} \label{subsec:RE_model}
In this paper, we focus on the random coefficients model where heterogeneous parameters $\alpha_{g_{i} t}$ and $\sigma_{g_i}$ are independent and are assumed to be independent of the initial value of each unit $y_{i0}$. The specification can be extended to correlated random coefficients model by modeling the joint distribution of heterogeneous parameters and initial values $y_{i0}$.

A typical choice in the nonparametric Bayesian literature is the Dirichlet Process (DP) prior or stick-breaking prior. With group probabilities $\pi_k$ and parameter in prior: (mean, variance) = $(\mu_{\alpha}, \Sigma_{\alpha})$, a draw of $\alpha_{g_{i} t}$ from the DP prior could be viewed as a mixture of point mess with the probability mass function,
\begin{align} \label{prior:a}
	\alpha_{g_{i} t} \sim \sum_{k=1}^{K} \pi_k \delta_{\alpha_{k t}}, \text{ with } \alpha_{k t} \sim N(\mu_{\alpha}, \Sigma_{\alpha}),
\end{align}
where $\delta_{x}$ denotes the Dirac-delta function concentrated at $x$, each $\alpha_{k t}$ is drawn from a normal distribution and $K$ is unknown. $\mu_{\alpha}$ are set to the OLS estimate of $\alpha$ assuming $K = 1$ and $ \Sigma_{\alpha}$ equals $200 \times \hat{\Sigma}_{\alpha}$ where $\hat{\Sigma}_{\alpha}$ is the standard deviation of the OLS estimator. In the same fashion, we can define the DP prior for grouped heteroskedasticity $\sigma^2_{g_{i}}$ given identical group probabilities $\pi_k$:
\begin{align}
	\sigma^2_{g_{i}} \sim \sum_{k=1}^{K} \pi_k \delta_{\sigma^2_{k}}, \text{ with } \sigma^2_{k} \sim IG \left(\frac{\nu_{\sigma}}{2}, \frac{\delta_{\sigma}}{2}\right), \nu_{\sigma} = 12,\text{ and } \delta_{\sigma} = 10,
\end{align}
where each component is drawn from inverse-Gamma distribution.

Put together, the posterior draws of grouped related coefficients can be characterized by a grouped triplet $\{\pi_k, \alpha_{k}, \sigma^2_{k}\}$ for $k = 1,2,..$, and $\alpha_k = [\alpha_{k1}, \alpha_{k2},..., \alpha_{kT}]'$.  Importantly, the distributions of both $\alpha_{k}$ and $\sigma^2_{k}$ are discrete, because draws can only take the values in the set $\left\{(\alpha_{k}, \sigma^2_{k}): k \in \mathbb{Z}^{+}\right\}$. This nonparametric nature makes the Dirichlet Process prior an ideal choice for clustering problems especially when the distinct number of clusters is unknown beforehand. The group parameters $(\alpha_{k}, \sigma^2_{k})$ are assumed to follow the base distribution $B_0$ which is an independent (non-conjugate) Multivariate Normal-Inverse-Gamma (IMNIG) distribution.

On the other hand, the group probability is formalized through an infinite-dimensional stick-breaking prior governed by the concentration parameter $a$,
\begin{align}
	\pi_k &= \xi_k \prod_{j<k} (1-\xi_j) \text{ for } k > 1,  \text{ and } \pi_1 = \xi_1,
\end{align}
where $\xi_k$, which are called stick lengths, are independent random variables drawn from the beta distribution $Beta(1, a)$. This construction can be viewed as a stick-breaking procedure, where at each step, we independently and randomly break the leftover of a stick of unit length and assign the length of this break to the current value of $\pi_{k}$. The smaller $a$ is, the less of the stick will be left for subsequent values (on average), yielding more concentrated distributions.

The concentration parameter $a$ specifies how strong this discretization is. As $a \to 0$, the realizations are all concentrated at a single value, while when $a \to \infty$, the realizations become continuous-valued from its based distribution.  \citet{escobar1995} shows that the number of estimated groups under a DP prior is sensitive to $a$, which indicates that a data-driven estimate should be more reasonable. To determine how discrete we want and how many groups are needed given the data, it is convenient to treat $a$ as a parameter under the nonparametric Bayesian framework. Put differently, we can set up a relatively general hyperprior for $a \sim Gamma \left(0.4, 10\right)$, and update it based on the observations. This step generates a posterior estimate of $a$, which implicitly chooses the optimal $K$ without re-estimating the models with different numbers of groups.

Finally, the prior distribution for the common parameter $\rho$ is chosen to be a normal distribution,
\begin{align}
	\rho \sim N (0, \sigma^2_{\rho}) \text{ with }  \sigma^2_{\rho} = 100.
\end{align}
The prior of heterogeneous parameter $\beta_i$ follows,
\begin{align}
	\beta_i \sim N (0, \Sigma_{\beta}) \text{ with }  \Sigma_{\beta} = 100 \times \mathbf{I}_p.
\end{align}

To sum up, in the random coefficients model, we specify the Dirichlet Process priors for group random effects $\alpha_{g_i t}$ and heteroskedasticity $\sigma^2_{g_{i}}$, a stick-breaking process for group probabilities $\pi_k$, a hyperprior for the concentration parameter $a$ and a normal prior for the common parameter $\rho$ and heterogeneous parameter $\beta_i$.

%
%
%
%
%

\subsubsection{Subjective Priors With Knowledge on Groups} \label{subsec:subPrior}

Frequently, researchers could provide a group structure on all or at least part of the units based on personal expertise and the nature of individuals. For example, firms coming from the same industry may share a similar growth pattern with relative high probability; countries having the same level of development form comparable fiscal policies. Though this presumed group structure might be subjective or purely based on theoretical analysis, it is still valuable to integrate this group information as it guides estimation when it enters the algorithm via a prior.


To integrate the prior knowledge with our model, we introduce the prior for the membership probability $\omega_{i} =  [\omega_{i1}, \omega_{i2},..., \omega_{iK^p}]'$ for all unit $i = 1,..,N$, where $K^p$ is the preset number of groups. Namely, before estimating the group membership, the researcher assigns each unit to different groups with a set of subjective group-specific probabilities, and these probabilities will enter the algorithm through a prior distribution for $\omega_{i}$. We name this prior for $\omega_{i}$ as \textit{Subjective Group Probability (SGP) Prior}. In practice, one could provide a table (for example, Table \ref{tab:subPriorExample}) documenting the subjective group probability of a unit falling into a specific group.

\begin{table}[htp]
	\begin{center}
		\caption{An example of prior group probability, $K^p = 4$}
		\label{tab:subPriorExample}
		\begin{tabular}{c | cccc}
			\toprule
			& \multicolumn{4}{c}{Subjective Group} \\
			\cmidrule(lr){2-5}
			Unit & 1 & 2 & 3 & 4 \\
			\midrule
			1 & 0.75 & 0.20 & 0.05 & 0 \\
			2 & 0.30 & 0.30 & 0.20 & 0.20 \\
			3 & 0 & 0 & 0.50 & 0.50 \\
			4 & 0 & 1 & 0 & 0 \\
			\vdots & \vdots & \vdots & \vdots & \vdots \\
			\bottomrule
		\end{tabular}
	\end{center}
\end{table}

It is worth noting that such a prior design is rather general and flexible, which is able to account for different scenarios. A special case of the SGP prior is that, for each $i = 1,2,..., N$, the researcher is fairly confident in her knowledge and sets one of $\omega_{ik}$ to 100\%. This is equivalent to the case where the researcher exactly partitions $N$ units into $K^p$ predetermined groups.

Building on the prior for the random effects model in Section \ref{subsec:RE_model}, we allow for incorporating the researchers' prior knowledge while inheriting the feature of reallocating units and changing the number of groups along the MCMC sampling. These flexible features enable the block Gibbs sampler to automatically correct and update the imprecise subjective prior, especially when $K^p$ doesn't match the true number of groups.

To incorporate these subjective group probabilities, it is important to choose a proper prior for $\omega$. Dirichlet distribution is an applicable candidate among assorted densities since it is the conjugate prior of the multinomial distribution, which facilitates direct sampling, and, most importantly, provides a natural channel to integrate subjective group probability.

To see this, we use the simplest case where the number of potential groups ($K^*$) in an iteration equals the presumed $K^p$. Let $\omega_{i} =  [\omega_{i1}, \omega_{i2},..., \omega_{iK^p}]$ be the vector of group-specific probability for unit $i$, we set the prior density for $\omega_{i}$ as an unsymmetric Dirichlet distribution,
\begin{align}
	\omega_{i} \sim Dir (a_{i1}, a_{i2},..., a_{iK^p}),
\end{align}
where $a_{ik}$ are concentration parameters and strictly positive. Conditional on $\omega_{i}$, the group membership $g_i$ is assumed to be drawn from a multinomial distribution,
\begin{align}
	g_i \sim Multinomial (\omega_i), \text{ i.e.}, P(g_i = k | \omega_i) = \omega_{i k} \text{, for } k=1, \ldots, K^p.
\end{align}
It can be shown posterior probability of $\omega_{i}$ given $g_i$ is also a Dirichlet distribution with modified hyperparameters: $Dir (a_{i1} + \mathbf{1}(g_i = 1), a_{i2}+ \mathbf{1}(g_i = 2),..., a_{iK^p}+ \mathbf{1}(g_i = K^p))$. Hence we can direct sample $\omega_{i}$ from it posterior distribution.

Another important property of Dirichlet distribution that enables itself to be the most suitable prior is that we can tie our prior probability directly with the expected value of $\omega_{ik}$,
\begin{align}
	E(\omega_{ik}) = \frac{a_{ik}}{\sum_{i = k}^{K^p} a_{ik}}.
\end{align}

To integrate the researcher's prior knowledge, one only need to deliberately choose a set of $\{a_{ik}\}$ such that the expected probability matches her subjective probability on groups.

Since the membership probabilities $\omega_{ik}$ are updated based on observations, and we allow for reallocating units and changing the number of groups along the MCMC sampling, a revision of our block Gibbs sampler is needed to adjust for such changes. The details of the new algorithm are presented in Appendix \ref{appendix:post_subPrior2}. In practice, we can restrict $\sum_{i = k}^{K^p} a_{ik}$ to be 1 so that $a_{ik}$ represents both the subjective group probability for unit $i$ belonging to group $k$ and the prior mean of $\omega_{ik}$.

\subsection{Posterior Sampling} \label{subsec:post}
Draws from the joint posterior distribution can be obtained by using blocked Gibbs sampling. The proposed algorithm is based on \citet{ishwaran2001} and \citet{walker2007}. Though the algorithm in \citet{ishwaran2001} has been widely used for sampling stick-breaking priors, it alone can't fulfill our need for estimating the number of groups without any predetermined level or upper bound since it requires a finite-dimensional prior and truncation. To avoid approximation and a predetermined number of groups $K$, we implement slice-sampling proposed by \citet{walker2007} and modify the framework of \citet{ishwaran2001} with additional posterior sampling steps. Using the conjugate priors specified in Section \ref{subsec:prior}, each parameter is directly drawn from its posterior distribution.

In the Appendix \ref{appendix:algo}, we provide detailed derivations for the conditional distributions over which the Gibbs sampler iterates. We focus on the time-varying grouped random effects model with grouped heteroskedasticity, which is the most sophisticated specification. Other specifications can be estimated by merely ignoring time effects in $\alpha$'s or shutting down the heteroskedasticity.

\subsection{Potential Link to Unsupervised Learning} \label{subset:linktoML}
One feature of our proposed block Gibbs sampler is that it partitions $N$ units into $G$ groups, and, at the same time, generates posterior draws for parameters. This Gibbs sampler and our BGRE estimator inevitably remind us of one of the most popular clustering algorithms in the area of unsupervised machine learning: the Kmeans algorithm. Indeed, the Kmeans algorithm plays a crucial role in BM and BLM, who estimate the grouped fixed-effects from the frequentists' point of view. In this subsection, we seek to illustrate the similarity and connection between our block Gibbs sampler and the Kmeans algorithm in the limit.

We start with the Kmeans clustering algorithm. Given a set of observations $\left(\mathbf{z}_{1}, \mathbf{z}_{2}, \ldots, \mathbf{z}_{N}\right),$ where each observation contains the dependent variables and covariates, $(y_i, x_i')$. Kmeans clustering aims to partition the $N$ observations into $K$ sets  so as to minimize the within-cluster sum of squares,
\begin{align}
	\underset{\{C_{k}\}_{k=1}^K}{\min } \sum_{k=1}^{K} \sum_{i \in C_{k}}\left\|\mathbf{z}_i - \boldsymbol{\mu}_{k}\right\|^{2} \text{  where } \boldsymbol{\mu}_{k}=\frac{1}{\left|C_{k}\right|} \sum_{i \in C_{k}} \boldsymbol{z}_i.
\end{align}

The algorithm alternates between reassigning points to clusters and recomputing the means. For the assignment step, one computes the squared Euclidean distance from each point to each cluster mean, and then assign each observation to the cluster with the nearest mean. The update step of the algorithm recalculates centroid for observations assigned to each cluster and updates $ \boldsymbol{\mu}_{k}$ for all $k$.

According to the block Gibbs sampler, we assign unit $i$ to group $k$ conditional on the draws of other parameters (Eq. (\ref{post_G_app})) with probability,
\begin{align*}
 	&	p(g_i = k | \rho, \beta, \alpha, \Sigma,  G^{(i)}, Y,X) \\
	= \; & \frac{p(y_i | \rho, \beta_i, \alpha_k, \sigma^2_k, Y,X) \mathbf{1} (u_i < \pi_{k})}{ \sum_{j=1}^{K^*} p(y_i | \rho, \beta_i, \alpha_j, \sigma^2_j, Y,X) \mathbf{1} (u_i < \pi_j)}.\\\
	= \;  & \frac{ c_{ik} \exp \left[ - \frac{1}{2} (y_i - \rho y_{-1,i} - x_i \beta_i - \alpha_k)' \Sigma_{k}^{-1} (y_i - \rho y_{-1,i} - x_i \beta_i - \alpha_k) \right]}{ \sum_{j=1}^{K^*} c_{ij} \exp \left[ - \frac{1}{2} (y_i - \rho y_{-1,i} - x_i \beta_i - \alpha_j)' \Sigma_{j}^{-1} (y_i - \rho y_{-1,i} - x_i \beta_i  - \alpha_j) \right]} \\
	= \;  & \frac{ c_{ik} \exp \left[ - \frac{1}{2} (\tilde{y}_i - \alpha_k)' \Sigma_{k}^{-1} (\tilde{y}_i  - \alpha_k) \right]}{ \sum_{j=1}^{K^*} c_{ij} \exp \left[ - \frac{1}{2} (\tilde{y}_i - \alpha_j)' \Sigma_{j}^{-1} (\tilde{y}_i - \alpha_j) \right]}, \numberthis
\end{align*}
where $c_{ik} = (2\pi)^{-\frac{T}{2}}  \Sigma_{k}^{-\frac{1}{2}} \mathbf{1} (u_i < \pi_k)$, $\tilde{y}_i = y_i - \rho y_{-1,i} - x_i \beta_i $, and $y_{-1,i}$ are the lagged values of $y_i$. If we assume homoskedasticity, i.e.,  $\Sigma_{k} = \Sigma$ for all $k$, then in the limit as $\Sigma \rightarrow 0,$ the value of $p(g_i = k | \rho, \beta, \alpha, \Sigma,  G^{(i)}, Y,X)$ approaches zero for all $k$ except for the one corresponding to the smallest weighted distance $(\tilde{y}_i - \alpha_k)' \Sigma_{k}^{-1} (\tilde{y}_i  - \alpha_k)$. In this case, this step is akin to the assignment step of Kmeans but using a weighted Euclidean distance. Then, conditional on newly estimated group membership, we update the group random effects $\alpha_k$ through Bayesian linear regression only using the units of group $k$. This step exactly recalculates the means of the new clusters, establishing the equivalence of the update step.

Having this similarity in mind, it is natural to include the Kmeans algorithm in our Monte Carlo experiment and explore its performance relative to our BGRE estimator in terms of accuracy of clustering. Notably, following BLM, we construct a 2-step GRE estimator equipped with K-mean algorithm in the first step. The performance of this 2-step estimator is assessed in Section \ref{sec:2stepGRE}.





\section{Monte Carlo Simulation} \label{sec:mcmc}
In this section, we conducted Monte Carlo simulation experiments to examine the performance of various Grouped Random Effects (GRE) estimators under different data generating processes (DGPs) and prior assumptions. These DGPs differ in whether the random effects are time-invariant or time-varying and whether to introduce heterogeneity in the variance of innovations. Such designs allow us to examine not only how our approach performs under DGPs with particular features, but also the reliability of appropriately estimating the number of clusters.

We consider a setting with sample size $N = 100$, and time span $T = 11$. The last observation of each unit forms the hold-out sample for evaluation as we focus on one-step ahead forecasts. A similar framework can be applied to multiple-step-ahead forecasts by iterating from period to period. We set the true numbers of groups $K^0 = 4$. Given $N$ and $K^0$, we partition the entire sample into $K^0$ balanced blocks with $N/K_0$ units in each block\footnote{If $N/K_0$ is not an integer, use $\lfloor N/K_0 \rfloor$ for group 1,2,..,$K_0-1$, the last group contains the residual units.}. For each DGP, 100 datasets are generated, and we run the block Gibbs samplers for each data set with $M = 10,000$ iterations after a burn-in of 5,000 draws.

\subsection{Data Generating Process} \label{sec:dgp}
The Monte Carlo simulation is based on the dynamic panel data model in (\ref{bm_model}), in which we suppress the exogenous predictors $x_{i t}$ for simplicity. In short, we consider four linear dynamic DGPs in this section. DGP1 and DGP2 involve time-invariant random effects while time-varying random effects are allowed in DGP3. Moreover, DGP1 and DGP3 consider homoskedasticity, but DGP2 has heteroskedastic innovations. DGP4 is the standard panel data model without a group structure. Throughout these DGPs, the random effects $\alpha_{k}$ and idiosyncratic error $\varepsilon_{i t}$ are standard normal distributed, independent across $i$, $k$, and $t$, and mutually independent. $\varepsilon_{i t}$ is independent of all regressors. The data are simulated according to the following DGPs:

\noindent {\bf DGP1:} Time-invariant grouped random effects, homoskedasticity {\bf(Grp Ti-Homo)}. \\
This DGP is the most naive panel data model with group pattern in the random effect.
\begin{align*}
	y_{it} &= \alpha_{g_{i}} + \rho y_{it-1} + \varepsilon_{i t},
\end{align*}
with $\rho = 0.7$, $\varepsilon_{i t} \stackrel{iid}{\sim} N \left(0, 0.8 \right)$ and $\alpha_k \stackrel{iid}{\sim} N \left(k, 0.5^2 \right)$ for $k = 1,2,...,K^0$.

\noindent {\bf DGP2:} Time-invariant grouped random effects, heteroskedasticity {\bf(Grp Ti-Hetero)}. \\
This DGP aims to incorporate heteroskedasticity, which leads to a slightly complicated process that is hard to estimate.
\begin{align*}
	y_{it} &= \alpha_{g_{i}} + \rho y_{it-1} + \varepsilon_{i t},
\end{align*}
with $\rho = 0.7$, $\varepsilon_{i t} \stackrel{iid}{\sim} N \left(0, \sigma_{g_{i}}^{2}\right)$ where $\sigma_k^2 = 1.5\left( 1 - \frac{k-1}{K^0} \right)^2$, and $\alpha_k \stackrel{iid}{\sim} N \left(k, 0.5^2 \right)$ for $k = 1,2,...,K^0$.

\begin{figure}[H]
	\caption{Mean of Random Effects for GDP3, $K^0 = 4$}
	\begin{center}
		\includegraphics[scale= 0.6]{{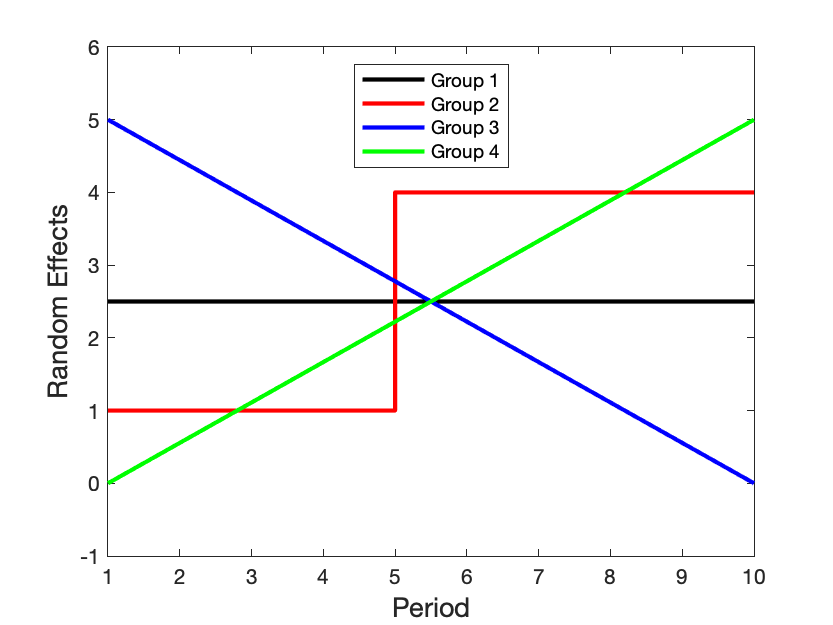}}
	\end{center}
	\label{plot:dgp3_mean}
\end{figure}

\noindent {\bf DGP3:} Time-varying grouped random effects, homoskedasticity {\bf(Grp Tv-Homo)}. \\
So far, we have focused on time-invariant models. But when estimated on real data, it's reasonable to believe the random effect could a have time pattern. Hence, we introduce various time-varying patterns of the random effects while keeping the assumption of homoskedasticity to avoid over-parameterization.
\begin{align*}
	y_{it} &= \alpha_{g_{i}} + \rho y_{it-1} + \varepsilon_{i t},
\end{align*}
with $\rho = 0.7$, $\varepsilon_{i t} \stackrel{iid}{\sim} N \left(0, 1 \right)$ and $\alpha_{it} \stackrel{iid}{\sim} N \left(\underline{\alpha}_{g_i t}, 0.5^2 \right)$ where $\underline{\alpha}_{g_i t}$ varies across periods and groups as depicted in Figure \ref{plot:dgp3_mean}. To enrich the patterns of time-varying random effects, we construct 4 different paths. Group 1 has a constant mean for $\alpha_{i}$. The means for $\alpha_{i}$ in group 2 are also constant but experience a structure change at $T=5$. Group 3 and group 4 are equipped with monotonically increasing/decreasing means. 

\noindent {\bf DGP4:} Time-invariant random effects, homoskedasticity, no group structure {\bf(Std Ti-Homo)}. This is the standard panel data model with unit-specific random effects and identical variance for the innovations.
\begin{align*}
	y_{it} &= \alpha_{i} + \rho y_{it-1} + \varepsilon_{i t},
\end{align*}
with $\rho = 0.7$, $\varepsilon_{i t} \stackrel{iid}{\sim} N \left(0, 0.8 \right)$ and $\alpha_{i} \stackrel{iid}{\sim} N \left(0, 0.5^2 \right)$.



\subsection{Simulation Results}

We consider four types of BGRE estimators that differ on the assumptions made on the random effects (RE) and the variance of errors: (1) time-invariant grouped RE with homoskedasticity (Ti-Homo); (2) time-invariant grouped RE with heteroskedasticity (Ti-Hetero); (3) time-varying grouped RE with homoskedasticity (Tv-Homo); (4) time-varying grouped RE with heteroskedasticity (Tv-Hetero)\footnote{For the Tv-Homo and Tv-Hetero estimator, as we allow for time effects in $\alpha_i$, we use the most recent $\alpha_{iT}$ to make one-step ahead prediciton. This is equivalent to assume the law of motion of $\alpha_{it}$ is a random walk. Modeling the trend of $\alpha_{it}$ would result in a more accurate forecast, but this is beyond the Scope of this paper.}. For instance, Ti-Homo estimator assumes the true model to have time-invariant grouped random effects, and the variance of error terms to be constant across units. Besides the results we will show below, we modify the DGPs and conduct more experiments using (a) larger variance of error terms $\sigma^2$, (b) shorter time span, and (c) different true number of groups $K^0$. These additional results are available in Appendix \ref{appendix:extra_simul}.

Regarding alternative estimators, we consider the following Bayesian estimators that have different prior assumptions on the random effects $\alpha_i$. (1) Bayesian pooled estimator (Pooled): $\alpha_i$ is treated as a common parameter as $\rho$ does, this means all units share the same prior level of $\alpha_i$; (2) flat-prior estimator (Flat): assume $p(\alpha_i) \; \propto \; 1$, this amounts to draw samples from a posterior whose mode is the MLE estimate. Given the estimate of common parameter, there is no pooling across units, $\alpha_i$'s are estimated only using their own history; (3) Parametric-prior estimator (Param): assume $\alpha_i \sim N(\mu, \pi^2)$, where a Normal-Inverse-Gamma hyperprior is further imposed on $(\mu, \pi^{2})$\footnote{The Normal-Inverse-Gamma hyperprior for $(\mu, \pi^{2})$ used in the Monte Carlo simulation is as follow: $\mu | \pi^{2} \sim N(m, v\pi^{2})$ with $m$ equating to the pooled OLS estimator of $\alpha_i$ and $v = 1$; $\pi^{2}$ follows $IG(\nu_{\pi}/2, \delta_{\pi}/2)$ with $\nu_{\pi} = 6$ and $\delta_{\pi} = 4$.}, this prior be thought of as a limit case of the DP prior when the concentration parameter $a \rightarrow 0$, so there is only one cluster, and $\left(\mu, \pi^{2}\right)$ are directly drawn from the base distribution.

\subsubsection{Point Estimates}
Table \ref{tab:MC_est} shows the estimate comparison among alternative predictors. For the DGP1, Ti-Homo and Ti-Hetero estimator are the best in every aspect. This is as expected since they correctly model the time-invariant random effects. Among these two estimators, when we allow for group‐level heteroskedasticity, the optimal number of groups decreases as Ti-Hetero underestimates the number of groups. The coverage probability, however, is not well-controlled, both of which are below the nominal coverage of 0.95. The Flat estimator also has good performance in terms of RMSE of $\rho$. Nonetheless, its coverage probability is relatively low: only 23\% of credible sets successfully contain the true values. This is due to the relatively large biases for $\alpha_{i}$. The rest predictors are considerably worse. This implies that completely ignoring group structure (Pooled, Param) results in a significantly inferior estimate, so does wrongly modeling time-varying random effects (Tv-Homo, Tv-Hetero). Regarding the performance of clustering, Ti-Homo slightly underestimates the number of groups with an average $K$ equals to 3.60 while the truth is 4.

For the case of DGP2, we keep time-invariant random effects while assuming heteroskedasticity. Tv-Homo and Tv-Hetero are still dominating. Tv-Hetero generates the best results with an accurate estimate of the number of groups as it correctly models heteroskedasticity which in turn improves the estimation efficiency. The Flat estimator closely follows them, and the rest are worse. Regarding DGP3, when time-varying random effects are introduced in the model, Tv-Homo and Tv-Hetero estimator yield the best performance and the estimated average $K$s are close to the truth. But the biases are arguably low for these two estimators in sacrifice for small standard deviation and short credible intervals. It is worth noting that, unlike Ti-Homo in DGP1 and Ti-Hetero in DGP2, though correctly specified, the bias for $\alpha_{i}$ is still comparatively high. This is because, for simplicity, we don't model the law of motion for $\alpha_{it}$ and simply assume $\alpha_{iT+1} = \alpha_{iT}$, which results in large bias in $\alpha_i$. As regards the DGP4 that doesn't have a group structure, the Flat estimator is the best since it doesn't pool cross-sectional information but estimate the unit-specific random effects. All of the BGRE estimators have almost identical performances with the estimated number of groups close or equal to 1. Since the Pooled and Param estimators assume no group structure, both have similar estimates as the BGRE estimators.

\begin{table}[htp]
	\begin{center}
		\caption{Monte Carlo Experiment: Point Estimates}
		\label{tab:MC_est}
		\begin{tabular}{ll|rrrrr|r|c}
			\toprule
			& & \multicolumn{5}{c}{$\hat{\rho}$} & \multicolumn{1}{c}{$\hat{\alpha_i}$} & Cluster \\
			\cmidrule(lr){3-7} \cmidrule(lr){8-8} \cmidrule(lr){ 9-9} \noalign{\smallskip}
			& & \multicolumn{1}{c}{RMSE } & \multicolumn{1}{c}{Bias } & \multicolumn{1}{c}{Std } & \multicolumn{1}{c}{AvgL } & \multicolumn{1}{c}{Cov } & \multicolumn{1}{c}{Bias } & \multicolumn{1}{c}{Avg K} \\
			\midrule
			\multirow{7}{*}{\shortstack{DGP 1 \\ (Grp Ti Ho.)}}
													& Ti-Homo & 0.0198 & 0.0113 & 0.0120 & 0.0468 & 0.83 & -0.0744 & 3.60 \\
													& Ti-Hetero & 0.0202 & 0.0118 & 0.0120 & 0.0468 & 0.78 & -0.0776 & 3.55 \\
													& Tv-Homo & 0.2403 & 0.2387 & 0.0187 & 0.0712 & 0.06 & -1.5255 & 1.82 \\
													& Tv-Hetero & 0.2405 & 0.2389 & 0.0108 & 0.0689 & 0.07 & -1.5301 & 1.89 \\
													& Pooled & 0.2449 & 0.2447 & 0.0069 & 0.0268 & 0.00 & -1.5588 & 1 \\
													& Flat & 0.0369 & -0.0344 & 0.0121 & 0.0469 & 0.23 & 0.2166 & 100 \\
													& Param & 0.2711 & 0.2437 & 0.1148 & 0.4545 & 0.38 & -1.5474 & 1 \\

			\midrule
			\multirow{7}{*}{\shortstack{DGP 2 \\ (Grp Ti He.)}}
													& Ti-Homo & 0.0226 & 0.0097 & 0.0150 & 0.0583 & 0.85 & -0.0681 & 3.74 \\
													& Ti-Hetero & 0.0112 & 0.0036 & 0.0082 & 0.0321 & 0.95 & -0.0261 & 3.98 \\
													& Tv-Homo & 0.1924 & 0.1885 & 0.0234 & 0.0893 & 0.14 & -1.2289 & 11.44 \\
													& Tv-Hetero & 0.0965 & 0.0894 & 0.0255 & 0.0979 & 0.30 & -0.5925 & 3.42 \\
													& Pooled & 0.2318 & 0.2316 & 0.0079 & 0.0310 & 0.00 & -1.4905 & 1 \\
													& Flat & 0.0493 & -0.0469 & 0.0146 & 0.0567 & 0.06 & 0.2984 & 100 \\
													& Param & 0.2576 & 0.2303 & 0.1115 & 0.4407 & 0.38 & -1.4792 & 1 \\
			\midrule
			\multirow{7}{*}{\shortstack{DGP 3 \\ (Grp Tv Ho.)}}
													& Ti-Homo & 0.2726 & 0.2724 & 0.0119 & 0.0463 & 0.00 & -2.1937 & 2.00 \\
													& Ti-Hetero & 0.2741 & 0.2738 & 0.0121 & 0.0470 & 0.00 & -2.2031 & 2.32 \\
													& Tv-Homo & 0.0580 & 0.0525 & 0.0221 & 0.0860 & 0.33 & -0.3679 & 3.89 \\
													& Tv-Hetero & 0.0589 & 0.0534 & 0.0222 & 0.0863 & 0.33 & -0.3743 & 3.85 \\
													& Pooled & 0.1925 & 0.1923 & 0.0081 & 0.0314 & 0.00 & -1.6381 & 1 \\
													& Flat & 0.3230 & 0.3227 & 0.0126 & 0.0492 & 0.00 & -2.5462 & 100 \\
													& Param & 0.2172 & 0.1912 & 0.1021 & 0.4033 & 0.54 & -1.6269 & 1 \\

			\midrule
			\multirow{7}{*}{\shortstack{DGP 4 \\ (Std Ti Ho.)}}
												   & Ti-Homo & 0.2177 & 0.2170 & 0.0164 & 0.0635 & 0.01 & 0.0038 & 1.03 \\
												   & Ti-Hetero & 0.2168 & 0.2159 & 0.0165 & 0.0644 & 0.01 & 0.0035 & 1.02 \\
												   & Tv-Homo & 0.2216 & 0.2210 & 0.0161 & 0.0627 & 0.00 & 0.0037 & 1.01 \\
												   & Tv-Hetero & 0.2204 & 0.2198 & 0.0164 & 0.0638 & 0.00 & 0.0038 & 1.16 \\
												   & Pooled & 0.2204 & 0.2198 & 0.0161 & 0.0628 & 0.00 & 0.0037 & 1 \\
												   & Flat & 0.1838 & -0.1817 & 0.0277 & 0.1076 & 0.00 & -0.0032 & 100 \\
												   & Param & 0.2321 & 0.2201 & 0.0714 & 0.2856 & 0.06 & 0.0154 & 1 \\
			\bottomrule
		\end{tabular}
	\end{center}
\end{table}

\subsubsection{Point, Set and Density Forecast}

Table \ref{tab:MC_fcst} reports the predictive performance of a range of parametric forecasts. For the DGP1, the best forecasts are generated by the Ti-Homo estimator, as it is correctly specified in this environment. It has the smallest RMSFE, the shortest average length of the credible set, correct coverage probability, the largest LPS, and the smallest CRPS. Although allowing for heteroskedasticity along with the time-invariant random effects, Ti-Hetero generates a perfect point forecast as well as Ti-Homo. But Ti-Hetero introduces uncertainty revealed by a slightly wider credible set and worse density forecast. Moreover, estimators involving time-varying random effects (Tv-Homo and Tv-Hetero) worsen the forecast. Finally, incorrectly imposing no latent group pattern substantially deteriorates the predictive performance in all aspects.

For the DGP2, Ti-Hetero is expected to have an edge, and indeed, it dominates the remaining alternatives. Ti-Homo performs as great as Ti-Hetero in point forecast. This is because, from the previous section, we know that Ti-Homo generates accurate point estimates for both $\rho$ and $\alpha_{i}$. But Ti-Homo fails in the set forecast and density forecast, which illustrates the importance of modeling heteroskedasticity. Again, the rest estimators suffer apart from the Flat estimator.

In the DGP3, Ti-Homo and Ti-Hetero are doing badly by not capturing the time effects in $\alpha_{g_{i}}$. Tv-Homo and Tv-Hetero are the best, beating the rest by a large margin, and equally accurate in this setup. The coverage probability for these two estimators is slightly lower than that of Ti-Homo and Ti-Hetero in part due to uncertainty introduced by more parameters of interest. Pooled, Flat, and Param estimator neglect both group structure and time-varying random effects, hence generating poor forecasts.

Regarding DGP4, all estimators beside Param have comparable forecasts as all of them deem no group structure in this environment (all estimated $K$s are close to 1). However, Param suffers from high variance, and it always generates the widest credible interval and worse density forecast as in other DGP's.

\begin{table}[htp]
	\begin{center}
		\caption{Monte Carlo Experiment: Forecast}
		\label{tab:MC_fcst}
		\begin{tabular}{ll|rrr|rr|rr}
			\toprule
			& & \multicolumn{3}{c}{Point Forecast}  & \multicolumn{2}{c}{Set Forecast} & \multicolumn{2}{c}{Density Forecast} \\
			\cmidrule(lr){3-5} \cmidrule(lr){6-7} \cmidrule(lr){8-9} \noalign{\smallskip}
			& & \multicolumn{1}{c}{RMSFE } & \multicolumn{1}{c}{Error } & \multicolumn{1}{c}{Std } & \multicolumn{1}{c}{AvgL } & \multicolumn{1}{c}{Cov } & \multicolumn{1}{c}{LPS } & \multicolumn{1}{c}{CRPS } \\
			\midrule
			\multirow{7}{*}{\shortstack{DGP 1 \\ (Grp Ti Ho.)}}
													& Ti-Homo & 0.8117 & -0.0073 & 0.8073 & 3.2036 & 0.95 & -1.2134 & 0.4592 \\
													& Ti-Hetero & 0.8122 & -0.0064 & 0.8079 & 3.2268 & 0.95 & -1.2158 & 0.4586 \\
													& Tv-Homo & 0.8637 & 0.0171 & 0.8544 & 3.4461 & 0.95 & -1.2754 & 0.4875 \\
													& Tv-Hetero & 0.8638 & 0.0179 & 0.8544 & 3.4550 & 0.95 & -1.2762 & 0.4877 \\
													& Pooled & 0.9619 & 0.3974 & 0.8685 & 4.1093 & 0.97 & -1.3905 & 0.5453 \\
													& Flat & 0.8406 & -0.0810 & 0.8326 & 3.2719 & 0.95 & -1.2485 & 0.4753 \\
													& Param & 0.9660 & 0.3994 & 0.8722 & 7.3886 & 1.00 & -1.6369 & 0.6119 \\

			\midrule
			\multirow{7}{*}{\shortstack{DGP 2 \\ (Grp Ti He.)}}
													& Ti-Homo & 1.0599 & 0.0123 & 1.0537 & 4.1027 & 0.93 & -1.4746 & 0.5798 \\
													& Ti-Hetero & 1.0428 & 0.0024 & 1.0365 & 3.7823 & 0.95 & -1.2650 & 0.5416 \\
													& Tv-Homo & 1.2662 & 0.0076 & 1.2551 & 3.7389 & 0.88 & -1.6472 & 0.6700 \\
													& Tv-Hetero & 1.0782 & 0.0025 & 1.0664 & 3.9051 & 0.95 & -1.3074 & 0.5610 \\
													& Pooled & 1.1839 & 0.3975 & 1.1072 & 4.8924 & 0.94 & -1.5952 & 0.6516 \\
													& Flat & 1.0768 & -0.0814 & 1.0678 & 4.1956 & 0.93 & -1.4962 & 0.5907 \\
													& Param & 1.1883 & 0.3978 & 1.1119 & 7.8562 & 0.99 & -1.7568 & 0.7096 \\

			\midrule
      		\multirow{7}{*}{\shortstack{DGP 3 \\ (Grp Tv Ho.)}}
							            			& Ti-Homo & 1.3740 & 0.3027 & 1.3360 & 5.2078 & 0.94 & -1.7468 & 0.7850 \\
							            			& Ti-Hetero & 1.3554 & 0.3067 & 1.3158 & 5.1512 & 0.95 & -1.7230 & 0.7714 \\
							            			& Tv-Homo & 1.0991 & 0.0127 & 1.0913 & 3.9240 & 0.93 & -1.5167 & 0.6222 \\
							            			& Tv-Hetero & 1.1062 & 0.0127 & 1.0985 & 3.9361 & 0.93 & -1.5239 & 0.6262 \\
							            			& Pooled & 1.8892 & 0.1200 & 1.8825 & 5.5759 & 0.85 & -2.1509 & 1.1049 \\
							            			& Flat & 1.2375 & 0.4145 & 1.1612 & 5.3199 & 0.97 & -1.6427 & 0.7025 \\
							            			& Param & 1.8927 & 0.1205 & 1.8860 & 8.2385 & 0.98 & -2.0679 & 1.0809 \\

			\midrule
			\multirow{7}{*}{\shortstack{DGP 4 \\ (Std Ti Ho.)}}
												     & Ti-Homo & 0.8476 & -0.0217 & 0.8429 & 3.3834 & 0.95 & -1.2571 & 0.4785 \\
												     & Ti-Hetero & 0.8476 & -0.0217 & 0.8429 & 3.3833 & 0.95 & -1.2570 & 0.4784 \\
												     & Tv-Homo & 0.8517 & 0.0018 & 0.8426 & 3.3992 & 0.95 & -1.2620 & 0.4811 \\
												     & Tv-Hetero & 0.8523 & 0.0016 & 0.8433 & 3.3987 & 0.95 & -1.2627 & 0.4812 \\
												     & Pooled & 0.8472 & -0.0218 & 0.8425 & 3.3850 & 0.95 & -1.2567 & 0.4783 \\
												     & Flat & 0.8517 & -0.0251 & 0.8470 & 3.2130 & 0.94 & -1.2626 & 0.4821 \\
												     & Param & 0.8684 & -0.0104 & 0.8426 & 8.5561 & 1.00 & -1.6526 & 0.5961 \\
			\bottomrule
		\end{tabular}
	\end{center}
\end{table}

\subsection{Comparison with Variants of BGRE Estimator}
This section conducts four sets of Monte Carlo simulation experiments aiming to examine the variants of BGRE estimator: (1) the GFE estimator proposed by BM, (2) a two-step GRE estimator with Kmeans, (3) the BGRE estimator with the subjective group prior, and (4) the BGRE estimator imposed with true $K^0$. The main text ignores part of the estimators we consider in the previous section and focuses on the correctly specified estimator for each DGP.

In addition to four DGPs specified in Section \ref{sec:dgp}, we design three new DGPs that inherit the main features from DGP1, DGP2, and DGP 3, including balanced group structure and unit variance structure. However, instead of drawing $\alpha_{g_i}$ from the normal, we choose to use a constant $\alpha_k$ for each group. In this way, we could focus on the clustering result rather than repetitions to average out the randomness brought by random effects. We impose $K^0 = 4$ and use this number throughout the Gibbs sampling.

\noindent {\bf DGP5:} Time-invariant grouped fixed effects, homoskedasticity.
\begin{align*}
	y_{it} &= \alpha_{g_{i}} + \rho y_{it-1} + \varepsilon_{i t},
\end{align*}
with $\rho = 0.7$, $\alpha_k = k$ for $k = 1,2,...,K^0$ and $\varepsilon_{i t} \stackrel{iid}{\sim} N \left(0, 1 \right)$.

\noindent {\bf DGP6:} Time-invariant grouped fixed effects, heteroskedasticity.
\begin{align*}
	y_{it} &= \alpha_{g_{i}} + \rho y_{it-1} + \varepsilon_{i t},
\end{align*}
with $\rho = 0.7$, $\alpha_k = k$ for $k = 1,2,...,K^0$ and $\varepsilon_{i t} \stackrel{iid}{\sim} N \left(0, \sigma_{g_{i}}^{2}\right)$ with $\sigma_k^2 = 1.5\left( 1 - \frac{k-1}{K^0} \right)^2$.

\noindent {\bf DGP7:} Time-varying grouped fixed effects, homoskedasticity.
\begin{align*}
	y_{it} &=  \alpha_{g_{i} t} + \rho y_{it-1} + \varepsilon_{i t},
\end{align*}
with $\rho = 0.7$, $\varepsilon_{i t} \stackrel{iid}{\sim} N \left(0, 1 \right)$ and $\alpha_{g_i t} = \underline{\alpha}_{g_i t}$ where $\underline{\alpha}_{g_i t}$ varies across periods and groups as depicted in Figure \ref{plot:dgp3_mean}.

\subsubsection{GFE Estimator}
In this experiment, we compare our BGRE estimator with BM's GFE estimator . In particular, we assess the performance of point forecast in $y$ and the accuracy of coefficient estimates (group random effects $\alpha$ and common parameter $\rho$).

To compare the performance of estimators, we use the default numerical setting\footnote{The default settings are as follow: (1) Number of groups = 4; Number of covariates = 1; Standard errors: 0 (no standard errors). (2) For algorithm 0,  Number of simulations = 100; (3) For algorithm 1, Number of simulations = 10, Number of neighbors = 10 ,	Number of steps = 10.} in BM. It is worth noticing that BM relies on information criteria to ex post select the optimal number of groups. Hence we consider at most 10 groups and estimate the number of groups $K$ in accordance with the following Akaike information criterion (AIC)\footnote{We also tried the alternative choice $\hat{\sigma}^{2} \frac{k T + N + p + 1}{N T} \ln (N T)$ for the penalty. This corresponds to the default BIC used in \citet{bonhomme2015}. We found that, in this case, BIC selected the smallest possible number of groups for all DGPs, i.e., no group structure, whereas the truth is $K^{0}=4$. Moreover, other forms of BIC could always select the largest $K$ as well. Due to the inaccurate estimate of group structure and substantially poor performance, we don't show the result with the default BIC.}:
\begin{align*}
	AIC(k)=\frac{1}{N T} \sum_{i=1}^{N} \sum_{t=1}^{T}\left(y_{i t} - \hat{\rho}^{(k)} y_{i t-1} - \hat{\beta_i}'^{(k)} x_{i t-1} -\widehat{\alpha}_{\widehat{g}_{i} t}^{(k)}\right)^{2} + 2\widehat{\sigma}^{2(k)} \frac{k(T +N - k)}{N T},
\end{align*}
where $\widehat{\sigma}^{2}$ is a consistent estimate of the variance of $\varepsilon_{i t}$:
\begin{align*}
	\widehat{\sigma}^{2(k)}=\frac{1}{N T-K_{\max } T-N-(p+1)} \sum_{i=1}^{N} \sum_{t=1}^{T}\left(y_{i t} - \hat{\rho}^{(k)} y_{i t-1} - \hat{\beta_i}'^{(k)} x_{i t-1} - \widehat{\alpha}^{(k)}_{\widehat{g}_{i} t}\right)^{2}.
\end{align*}

The results are shown in the Table \ref{tab:MC_vsBM}. As BM proposes two algorithms\footnote{These two algorithms could generate different estimates as shown in the case of DGP6 below.}, we present the results for four versions of the GFE estimator. The first two estimators ($\text{GFE}_{a0}$ and $\text{GFE}_{a1}$) equip with the Iterative and Variable Neighborhood Search algorithm, respectively. We impose the true number of groups $K^0$ for the other two estimators ($\text{GFE}^0_{a0}$ and $\text{GFE}^0_{a1}$), i.e., we don't perform model selection but choose $\hat{K} = K^0 = 4$ directly.

For the DGP5 and DGP6, Ti-Homo and Ti-Hetero estimator outperform the GFE estimators in all aspects. The GFE estimator's poor performance is mainly due to the incorrect estimate of the number of groups. This also emphasizes that an inaccurate estimate of group structure would deteriorate both estimates and point forecasts. It is worthy noting that, even imposing the true number of groups, GFE estimators ($\text{GFE}^0_{a0}$ and $\text{GFE}^0_{a1}$) are still straggling and worse than their counterparts with $\hat{K}$ selected by the information criterion. $\text{GFE}^0_{a0}$ and $\text{GFE}^0_{a1}$ generate relatively high bias for both $\alpha_{i}$ and $\rho$. In the case of DGP7, Tv-Homo and Tv-Hetero still have better performance than the GFE estimators whose models are chosen by the information criterion. Furthermore, Tv-Homo and Tv-Hetero perform only marginally worse than $\text{GFE}^0_{a0}$ and $\text{GFE}^0_{a1}$ estimators that are imposed with the true $K^0$. This is because they overestimate the number of groups in some posterior draws, and hence, on average, the posterior mean forecast and estimate are slightly off.


Moreover, the accuracy of the GFE estimator is profoundly affected by the choice of the information criteria. We implement several information criteria proposed by \citet{bai2002} in this Monte Carlo experiment and find that there is no single criterion that consistently selects the correct number of groups nor close to the truth. As the GFE estimator is designed for the time-varying model, we finally select the AIC mentioned above, which chooses a model that is closest to the true model in time-varying DGPs. But this deliberately selected AIC fails to improve the performance of the GFE estimator in DGP7. Once we switch to other forms of AIC or BIC, these results will not hold anymore. These facts also emphasize the importance of not relying on ex-post model selection and the superiority of our Bayesian estimators.

\begin{table}[htp]
	\begin{center}
		\caption{Monte Carlo Experiment: BGRE vs GFE}
		\label{tab:MC_vsBM}
		\begin{tabular}{llrrrrrr}
			\toprule
			& & \multicolumn{3}{c}{Point Forecast}  & \multicolumn{1}{c}{$\hat{\rho}$} & \multicolumn{1}{c}{$\hat{\alpha}$} & \multicolumn{1}{c}{Group} \\
			\cmidrule(lr){3-5} \cmidrule(lr){6-6} \cmidrule(lr){7-7} \noalign{\smallskip}
			& & \multicolumn{1}{c}{RMSFE } & \multicolumn{1}{c}{Error } & \multicolumn{1}{c}{Std } & \multicolumn{1}{c}{Bias } & \multicolumn{1}{c}{Error } & \multicolumn{1}{c}{Avg K}\\
			\midrule
			\multirow{9}{*}{\shortstack{DGP 5 \\ (Grp Ti Ho.)}}
		    & Ti-Homo & 0.7806 & 0.0261 & 0.7801 & 0.0214 & -0.1359 & 4.63 \\
		    & Ti-Hetero & 0.7829 & 0.0254 & 0.7825 & 0.0212 & -0.1353 & 4.81 \\
		    & Tv-Homo & 0.8062 & 0.0904 & 0.8011 & 0.3261 & -2.1184 & 1.00 \\
		    & Tv-Hetero & 0.7995 & 0.0883 & 0.7946 & 0.3262 & -2.1183 & 1.52 \\
		    & $\text{GFE}_{a1}$ & 0.7882 & 0.0845 & 0.7837 & 0.2891 & -1.8767 & 2 \\
		    & $\text{GFE}^0_{a1}$ & 0.8421 & 0.0496 & 0.8406 & 0.0463 & -0.2935 & 4 \\
		    & $\text{GFE}_{a0}$& 0.7882 & 0.0845 & 0.7837 & 0.2891 & -1.8767 & 2 \\
		    & $\text{GFE}^0_{a0}$ & 0.8421 & 0.0496 & 0.8406 & 0.0463 & -0.2935 & 4 \\
		    & Flat & 0.7881 & -0.0487 & 0.7866 & -0.0258 & 0.1752 & 1 \\

			\midrule
			\multirow{9}{*}{\shortstack{DGP 6 \\ (Grp Ti He.)}}
			& Ti-Homo & 1.1399 & 0.2578 & 1.1104 & 0.0100 & -0.0475 & 4.59 \\
			& Ti-Hetero & 1.0971 & 0.2525 & 1.0677 & 0.0060 & -0.0189 & 4.38 \\
			& Tv-Homo & 1.2850 & 0.3136 & 1.2462 & 0.2692 & -1.7265 & 9.62 \\
			& Tv-Hetero & 1.1356 & 0.2989 & 1.0955 & 0.0970 & -0.6102 & 3.53 \\
			& $\text{GFE}_{a1}$  & 1.1786 & 0.3098 & 1.1371 & 0.1580 & -1.0065 & 2 \\
			& $\text{GFE}^0_{a1}$ & 1.3201 & 0.3119 & 1.2827 & 0.1792 & -1.1441 & 4 \\
			& $\text{GFE}_{a0}$ & 1.1786 & 0.3098 & 1.1371 & 0.1580 & -1.0065 & 2 \\
			& $\text{GFE}^0_{a0}$ & 1.2408 & 0.3117 & 1.2010 & 0.1774 & -1.1324 & 4 \\
			& Flat & 1.1209 & 0.1540 & 1.1103 & -0.0489 & 0.3402 & 1 \\

			\midrule
			\multirow{9}{*}{\shortstack{DGP 7 \\ (Grp Tv Ho.)}}
			& Ti-Homo & 1.3620 & 0.4483 & 1.2862 & 0.2668 & -2.0886 & 2.00 \\
			& Ti-Hetero & 1.3625 & 0.4477 & 1.2869 & 0.2668 & -2.0884 & 2.00 \\
			& Tv-Homo & 0.9908 & 0.1201 & 0.9835 & 0.0453 & -0.2998 & 4.02 \\
			& Tv-Hetero & 0.9953 & 0.1155 & 0.9886 & 0.0519 & -0.3478 & 4.02 \\
			& $\text{GFE}_{a1}$ & 1.1174 & 0.1214 & 1.1108 & 0.0510 & -0.3387 & 3 \\
			& $\text{GFE}^0_{a1}$ & 0.9728 & 0.1146 & 0.9661 & -0.0071 & 0.0673 & 4 \\
			& $\text{GFE}_{a0}$ & 1.1174 & 0.1214 & 1.1108 & 0.0510 & -0.3387 & 3 \\
			& $\text{GFE}^0_{a0}$ & 0.9728 & 0.1146 & 0.9661 & -0.0071 & 0.0673 & 4 \\
			& Flat & 1.2750 & 0.5588 & 1.1460 & 0.3183 & -2.4505 & 1 \\
			\bottomrule
		\end{tabular}
	\end{center}
	Note: The meanings of notation in the second column are as follow: GFE = BM's estimator; $\text{GFE}^0$ = GFE estimator with the true $K_0$; a1 = algorithm 1: Variable Neighborhood Search; a0 = algorithm 0: Iterative Search.
\end{table}

 \subsubsection{GRE Estimator with Subjective Group Prior}

In this section, we explore whether subjective group structure improves the accuracy of forecast and group clustering. We conduct two Monte Carlo experiments corresponding to SGP Prior defined in Section \ref{subsec:subPrior}.


We consider five scenarios, each of which differs in the structure of subjective prior probability and hence the prior group probability $\pi$. The exact specification is characterized in Table \ref{tab:subPrior_specification}. The first three scenarios set the preset number of groups as the truth $K^0$, whereas subjective group probabilities are different in levels. In the scenario 1, the researcher is pretty confident about her decision and assigns 100\% to the right group for each unit, which amounts to knowing the true membership. However, she is entirely uninformed and cluster units with even probability for each group (i.e., $\omega_{ik} = 1/K^0$ for $\forall i,k$) in the scenario 3. Scenario 2 is an intermediate case where one is less confident in her knowledge and correctly assigns a unit to its group with the prior probability of 70\% (other groups equally split the remaining 30\%). For the scenario 4 and 5, the number of groups is different from the truth. We assume the researcher divides all units into $K \ne K^0$ even groups with the prior probability of 100\%\footnote{The last two scenarios aim to evaluate the performance of SGP prior when the number of groups is wrong. Instead of randomly assigning a unit to each groups with a set of probability, we assume the econometrician is confident on her prior and set 100\% for a particular group. In particular, we assign the first $N/K$ units into group 1, the next $N/K$ units into group 2, and so on. We also run other designs for scenario 4 and 5 with different prior probabilities. The results show that, as long as a certain amount of units are correctly clustered into groups, the performances of SPG-RE estimators are slightly better than those of the BGRE estimator.}.

\begin{table}[htp]
	\begin{center}
		\caption{Simulation Design: Subjective Group Probability}
		\label{tab:subPrior_specification}
		\begin{tabular}{c | c | c}
			\toprule
			Scenarios & \# of Groups & Descriptions \\
			\midrule
			1 & $ K^0$ & very confident, assign 100 \% to the correct group \\
			2 & $ K^0$ & less confident, assign 70 \% to the correct group \\
			3 & $ K^0$ & uninformed, evenly assign $1/K^0 \times $100 \% to the each group \\
			4 & $ K^0-1$ & very confident, assign 100 \% to a particular group (might be wrong) \\
			5 & $ K^0+1$ & very confident, assign 100 \% to a particular group (might be wrong) \\
			\bottomrule
		\end{tabular}
	\end{center}
\end{table}

We conduct the simulation experiment for the SGP Prior, where we examine the impact of subjective group probability prior on the performance of estimate and forecast under DGP3 (time-varying random effects model). To adopt the SGP prior, we revise the BGRE estimator and construct the new Bayesian estimator under the assumptions: (1) time-varying random effects and (2) heteroskedasticity, which corresponds to the most general case of a panel data model. We name it as Subjective Group Probability Random Effect (SGP-RE) estimator. Given different specifications of SGP prior, we report the performance of five SGP-RE estimators relative to the Tv-Hetero estimator (benchmark model).

\begin{figure}[tbp]
	\caption{Monte Carlo Experiment: Estimates, SGP Prior}
	\label{fig:MC_SGP_est}
	\centering
	\begin{minipage}{1\textwidth} 
		\includegraphics[width=\linewidth]{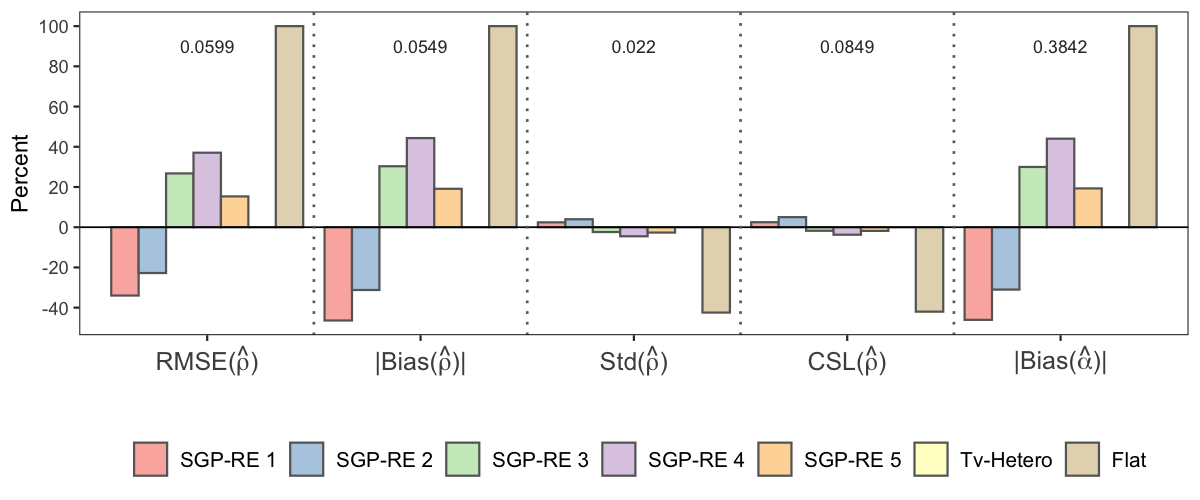}
		{\footnotesize \textit{Note: the value of each relative metric is capped by 100\% to enhance the readability. The number in each subpanel represents the (original) value of metric for the benchmark model (Tv-Hetero).} \par}
	\end{minipage}
\end{figure}

\begin{figure}[tbp]
	\caption{Monte Carlo Experiment: Forecast, SGP Prior}
	\label{fig:MC_SGP_fcst}
	\centering
	\begin{minipage}{1\textwidth} 
		\includegraphics[width=\linewidth]{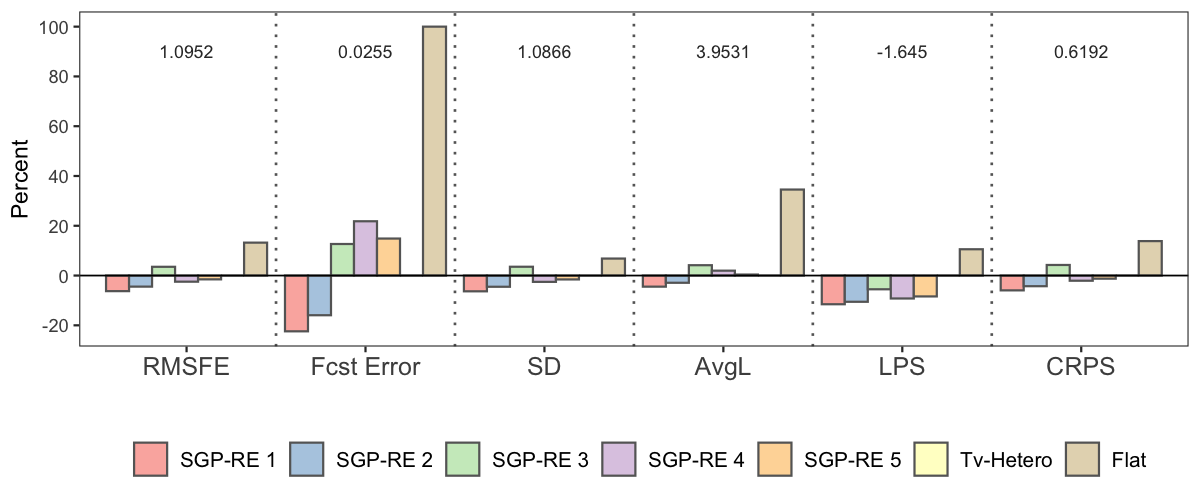}
		{\footnotesize \textit{Note: the value of each relative metric is capped by 100\% to enhance the readability. The number in each subpanel represents the (original) value of metric for the benchmark model (Tv-Hetero).} \par}
	\end{minipage}
\end{figure}

 Figure \ref{fig:MC_SGP_est} and \ref{fig:MC_SGP_fcst} depict the relative performance of various SGP-RE estimators against the Tv-Hetero estimator\footnote{The full results are presented in the Appendix \ref{appendix:subPrior}.}. Remember that the prior knowledge is the most accurate in scenario 1, where the researcher is equivalent to know the true group structure. In this regard, a clear gain in estimate emerges as the RMSE for $\rho$, bias for $\rho$ and $\alpha_i$ generated by SGP-RE1 (100\% confidence) decreases by more than 35\% relative to the Tv-Hetero estimator. As we move from scenario 1 to the other scenarios, the prior information becomes less accurate. SGP-RE2 (70\% confidence) beats the benchmark with moderate improvement on the RMSE ($\sim$22\%) and the bias ($\sim$32\%) while SGP-RE3 (uninformed) underperforms the benchmark by a huge margin. The poor performance of SGP-RE3 is not surprising. Though it correctly specifies $K = K^0$, the uninformative prior forces the algorithm to consider other incorrect groups with a large chance (= $1 - \frac{1}{K^0}$), and hence deteriorates the performance of both estimates and forecasts. In terms of the one-step ahead forecast, SGP-RE1 leads the rest by scoring the lowest values for each metrics (and the highest LPS), closely followed by SGP-RE2. SGP-RE3 is suffering in terms of point and set forecast. As for scenario 4 and 5, despite the flexibility of allowing for changing $a$ and $K^p$ along MCMC sampling, the incorrect specifications of the group number deteriorate the estimation, both of which fail to deliver reliable estimates for $\rho$ and $\alpha$. Nonetheless, such prior structures help point and density forecasts. Both estimators beat the benchmark and generate comparable LPS to SGP-RE1 and SGP-RE2. This valuable improvement mainly results from the fact the algorithm could exploit the prior knowledge on group structure that successfully partitions merely a fraction of units and adapt the number of groups accordingly.

In short, regarding the overall performance of the SGP prior, the best case is that the researcher has a relative confident prior and knows the true number of groups. In this case, the SGP-RE estimator dominates the Tv-Hetero estimator from every angle. However, in practice, we rarely come up with such a precise prior due to the incomplete understanding of the population of the data. Instead, we might be less confident on our knowledge or even specify more/fewer groups than the truth. Under this circumstance, the SGP-RE estimator could still deliver a better density forecast because of the great exploration of the prior information and the adaptive scheme featured by our Bayesian method.

%

\subsubsection{Two-Step GRE Estimator} \label{sec:2stepGRE}

In this section, we compare our BGRE estimator with a two-step GRE estimator, where units are clustered into groups in the first step using \textit{Kmeans} algorithm, and the model is then estimated in the second step with group-specific heterogeneity. Unlike BLM, we implement the Bayesian framework in the second step to echo other Bayesian estimators presented in the previous section. This two-step procedure allows us to examine the clustering accuracy of Kmeans relative to our full Bayesian estimate as two-step GRE estimators can be viewed as a special case of the GRE estimator with group membership determined by Kmeans. The optimal number of clusters in Kmeans is selected by the average silhouette method.



\begin{figure}[htp]
	\caption{Monte Carlo Experiment: Point Estimates, Two-Step GRE with Kmeans}
	\label{fig:MC_2step_est}
	\centering
	\begin{minipage}{0.95\textwidth} 
		\includegraphics[width=\linewidth]{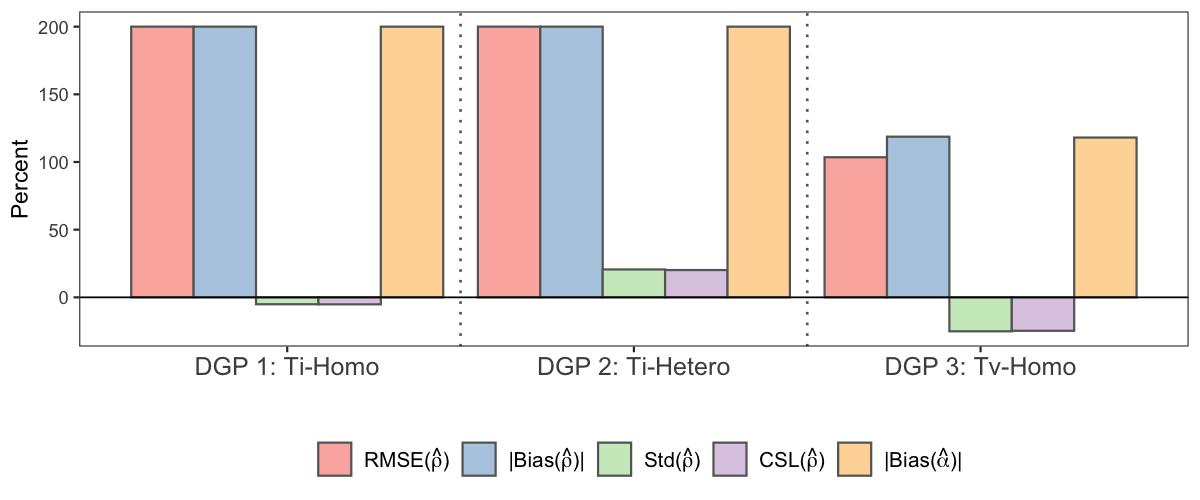}
		{\footnotesize \textit{Note: the value for each relative metric is capped by 200\% to enhance the readability.} \par}
	\end{minipage}
\end{figure}

\begin{figure}[htp]
	\caption{Monte Carlo Experiment: Forecast, Two-Step GRE with Kmeans}
	\label{fig:MC_2step_forecast}
	\centering
	\begin{minipage}{0.95\textwidth} 
		\includegraphics[width=\linewidth]{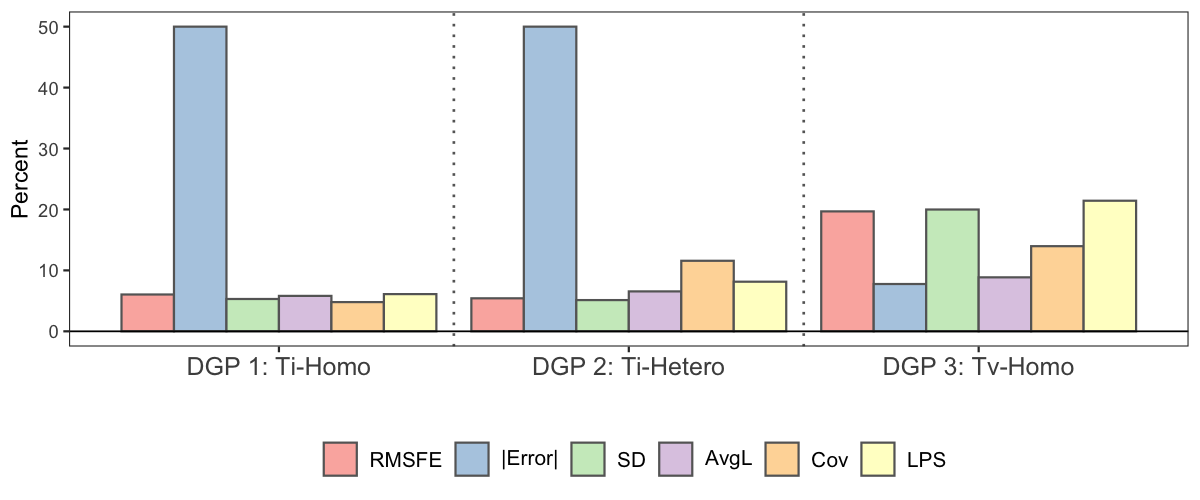}
		{\footnotesize \textit{Note: the value for each relative metric is capped by 50\% to enhance the readability.} \par}
	\end{minipage}
\end{figure}

To avoid cluttering the tables in the main text, we depict the selected results\footnote{The full results are presented in the Appendix \ref{appendix:2step_full_table}.} for estimates and forecast in Figure \ref{fig:MC_2step_est} and \ref{fig:MC_2step_forecast}. Each bar represents the performance of the two-Step GRE estimators against the performance of the original GRE estimator. Above zero indicates the 2-step estimator underperforms the benchmark while a 2-step estimator is better when its bars show negative values. The main models are correctly specified for each DGP, i.e., Ti-Homo for DGP 1, Ti-Hetero for DGP 2, and Tv-Homo for DGP 3.

Figure \ref{fig:MC_2step_est} presents the point estimates for each DGP. We document the root mean squared forecast error, absolute bias, standard deviation, and the average length of the 95\% credible set for the common parameter $\rho$, while the metric for the random effects is the absolute bias. According to these measures, the two-Step GRE estimators perform worse than the Bayesian GRE estimator as they introduce much higher bias in the estimate of $\rho$ (hence larger RMSE for $\rho$) and $\alpha_i$. It is worth noting that the estimator equipped with Kmeans doesn't affect the standard deviation and the average length of 95\% credible set of $\rho$.

The inferior performance of the 2-step GRE estimator is due to the inaccurate estimate of group structure. Table \ref{tab:MC_grp_kmean} reports the estimated number of groups from the two-step GRE estimators with Kmeans and the BGRE estimator. Regarding the performance of clustering, the Kmeans algorithm severely underestimates the number of groups as it prefers much less groups, while the true number is 4. Meanwhile, our BGRE approach accurately estimates the number of groups, though slightly underestimated in the DGP 1.

\begin{table}[htp]
	\begin{center}
		\caption{Number of groups, Kmeans vs. BGRE}
		\label{tab:MC_grp_kmean}
		\begin{tabular}{lccc}
			\toprule
			& \multicolumn{1}{c}{DGP 1} & \multicolumn{1}{c}{DGP 2}& \multicolumn{1}{c}{DGP 3} \\
			\midrule
			Kmeans & 2.20 & 2.20 & 2.26 \\
			BGRE & 3.60 & 3.98 & 3.85 \\
			\bottomrule
		\end{tabular}
	\end{center}
\end{table}


Figure \ref{fig:MC_2step_forecast} shows the point, set, and density forecast for each DGP. As Kmeans fails to estimate the group structure, none of the 2-step GRE estimators outperform the GRE estimators. Namely, Kmeans clustering doesn't help make a more accurate forecast, and instead, it generates a much higher forecast bias and standard deviation.

\section{Empirical Application} \label{sec:emp_result}
In this section, we illustrate the use of Bayesian Grouped Random Effects estimators in a cross-firm study. We revisit the investment regression and use a different version of the dynamic grouped panel model to forecast the investment rate\footnote{The investment rate for a firm in a particular year is defined as the fraction of capital expenditures in property, plant, and equipment in terms of the beginning-of-year capital stock.} for a panel of firms in all industries. Instead of using the traditional Tobin's Q-type investment regression, we implement a new scheme proposed by \citet{gala2019}, who directly estimates the corporate investment rate without Tobin's Q. Again, our main focus is the one-step ahead point, set and density forecast. Due to space limitations, we only report forecast results for the most recent year in the main text. Summary statistics, and additional details of implementation are stored in Appendix \ref{appendix:data} and \ref{appendix:extra_empical}.

\subsection{Model Specification}
We consider a general model with grouped latent heterogeneity in $\alpha_{i}$. Following \citet{hsiao1997} and \citet{gala2019}, the investment equation is specified as,
\begin{align}
	\left(\frac{I}{K}\right)_{it} = \alpha_{g_{i}t} + \rho \left(\frac{I}{K}\right)_{it-1} + \beta_{i1} \left(\frac{CF}{K}\right)_{it-1} + \beta_{i2} \ln K_{i t-1} + \beta_{i3} \ln \left(\frac{Y}{K}\right)_{it-1} + \varepsilon_{it},
\end{align}
where capital stock, $K_{it}$ is defined as net property, plant and equipment; $I_{it}$ is capital investment; $CF_{it}$, is a liquidity variable defined as cash flow minus dividends; $Y_{t}$ is the end-of-year sales; $\varepsilon_{it}$ are the normally distributed error terms. The subscript $i$ denotes companies, and $t$ denotes time. Unlike the commonly specified investment equation using Tobin's Q, the additional terms, including the natural logarithm of lagged capital and sales-to-capital ratio, are based on the regression proposed by \citet{gala2019}. The lagged values of the investment rate are included as explanatory variables to avoid endogeneity problems.

As we focus on forecasting, we can relax a few assumptions to achieve better predictive performance. These assumptions include time-invariant random effects $\alpha_{g_i}$, homoskedasticity in $\sigma_{i}$ and homogeneous coefficients for all dependent variables ($\beta_{i \cdot}$ = $\beta_{i}$). Table \ref{tab:emp_models} summarizes the estimators and their properties we consider in this section. The implementation of time-invariant RE and homoskedasticity is similar to the one in the previous section, i.e., construct four versions of BGRE estimator: Ti-Homo, Ti-Hetero, Tv-Homo, and Tv-Hetero. Despite the fact that the homogeneous slopes have been frequently rejected in empirical studies of estimates and inference, such an assumption could provide potential improvement in forecasts.

 \begin{table}[htp]
 	\begin{center}
 		\caption{Summary of Estimators in the Empirical Analysis}
 		\label{tab:emp_models}
 		\begin{tabular}{ll|c|c|c}
 			\toprule
 			& &Time-invariant $\alpha_{i}$ & Homogeneity & Group Structure \\\hline
 			\multirow{7}{*}{\shortstack{Homogenous \\ Coef.}}
 			& Ti-Homo & $\mathbf{X}$ &  $\mathbf{X}$& $\mathbf{X}$  \\
 			& Ti-Hetero & $\mathbf{X}$ &  & $\mathbf{X}$  \\
 			& Tv-Homo &  &  $\mathbf{X}$ & $\mathbf{X}$  \\
 			& Tv-Hetero &  & & $\mathbf{X}$  \\
 			& Flat &  $\mathbf{X}$ & $\mathbf{X}$ &  \\
 			& Pooled &  $\mathbf{X}$& $\mathbf{X}$ &   \\
 			& Param &  $\mathbf{X}$ & $\mathbf{X}$ &  \\

 			\midrule
 			\multirow{5}{*}{\shortstack{Heterogenous \\ Coef.}}
 			& Ti-Homo & $\mathbf{X}$ &  $\mathbf{X}$& $\mathbf{X}$  \\
 			& Ti-Hetero & $\mathbf{X}$ &  & $\mathbf{X}$  \\
 			& Tv-Homo &  &  $\mathbf{X}$ & $\mathbf{X}$  \\
 			& Tv-Hetero &  & & $\mathbf{X}$  \\
 			& Flat &  $\mathbf{X}$ & $\mathbf{X}$ &  \\
 			\bottomrule
 		\end{tabular}
 	\end{center}
 \end{table}

\subsection{Data}
The individual company data are obtain from COMPUSTAT Annual database. To account for potential structural breaks and the advanced speed of capital accumulation in the recent decades, our sample is composed of a balanced panel of firms for the years 2000 to 2019, that includes firms from all industries with no missing value in accounting data.

We keep only firm-years that have non-missing information required to construct the primary variables of interest, namely: capital stock $K$, investment $I$, liquidity $CF$, and sales revenues $Y$. The further details of constructing the sample can be found in the Appendix \ref{appendix:data}. The final sample comprises 337 firms and the observations for each firm is 20.

To examine the performance of various estimators with limited observations, we choose to use a rolling window of 15 years. In this sense,  we create five balanced panels which end in years 2014, ..., 2018 ($t=T$), respectively. The observations in the next year ($t=T+1$) are reserved for pseudo-out-of-sample forecast evaluation. For illustrative purposes, we will present the results for the year 2019 in the remainder of this section. The full results are presents in Appendix \ref{appendix:extra_empical}.

\subsection{Results}

We begin the empirical analysis by comparing the performance of point, set, and density forecast for the last panel (in-sample periord: 2005 - 2018). We aim to forecast the investment rate in 2019. We consider all the model specifications depicted in the Table \ref{tab:emp_models} and their performance is presented in the Table \ref{table:emp_fcst}. Throughout the analysis, the Flat estimator serves as the benchmark as it essentially assumes individual effects. In Table \ref{table:emp_fcst}, the third column shows the RMSFE for the one-step ahead forecast. For the panel considered in the table, we first notice that the best model is the Tv-Hetero (time-varying random effects, heteroskedasticity) in homogeneous coefficients specification. It outperforms the benchmark -- Flat estimator by 25\%. Ti-Hetero also delivers accurate point forecast, which suggests time effects provide merely marginal improvement. Under heterogeneous coefficients specification, for the BGRE estimators, though all of them beat the Flat estimator, their RMSFEs are relatively larger. This may arise from the fact that heteroskedasticity alone can capture a great amount of individual effects, thus imposing heterogeneous coefficients in $\beta_{i}$ may overfit the model and lead to poor forecasts.


\begin{table}[tbp]
	\begin{center}
		\caption{Empirical Application: Forecast Performance}
		\label{table:emp_fcst}
		\begin{tabular}{llcccccc}
			\toprule
			(1) & (2) & (3) & (4) & (5) & (6) & (7) & (8) \\
			\midrule
			& & \multicolumn{2}{c}{Point} & \multicolumn{2}{c}{Set} & \multicolumn{2}{c}{Density} \\
			\cmidrule(lr){3-4} \cmidrule(lr){5-6} \cmidrule(lr){7-8} \noalign{\smallskip}
			& & RMSFE & Avg K & Cov & Length & LPS & CRPS \\
			\midrule
			\multirow{7}{*}{\shortstack{Homogenous \\ Coef.}}
			& Ti-Homo & 0.1108 & 2 & 0.9614 & 0.5908 & 0.7039 & 0.0605 \\
			& Ti-Hetero & 0.0822 & 7.86 & 0.9021 & 0.4012 & 1.3724 & 0.0464 \\
			& Tv-Homo & 0.1177 & 1 & 0.9525 & 0.5875 & 0.6671 & 0.0634 \\
			& Tv-Hetero & 0.0812 & 6.75 & 0.8813 & 0.3867 & 1.2981 & 0.0454 \\
			& Pooled & 0.1150 & 1 & 0.9555 & 0.5966 & 0.6746 & 0.0627 \\
			& Flat & 0.1100 & 1 & 0.9703 & 0.6041 & 0.6935 & 0.0604 \\
			& Param & 0.2043 & 1 & 1.0000 & 6.8211 & -1.2722 & 0.3554 \\
			\midrule
			\multirow{5}{*}{\shortstack{Heterogenous \\ Coef.}}
			& Ti-Homo & 0.1144 & 7.48 & 0.8724 & 0.2837 & 1.1904 & 0.0485 \\
			& Ti-Hetero & 0.1152 & 6.68 & 0.8724 & 0.2841 & 1.1883 & 0.0485 \\
			& Tv-Homo & 0.1070 & 1 & 0.9703 & 0.6344 & 0.6879 & 0.0615 \\
			& Tv-Hetero & 0.1101 & 7.46 & 0.8338 & 0.2741 & 1.1013 & 0.0486 \\
			& Flat & 0.1164 & 1 & 0.9733 & 0.6753 & 0.6205 & 0.0650 \\
			\bottomrule
		\end{tabular}
	\end{center}
\end{table}

\begin{figure}[tbp]
	\caption{Empirical Application: Distributions of Group Number}
	\begin{center}
		\includegraphics[scale=.38]{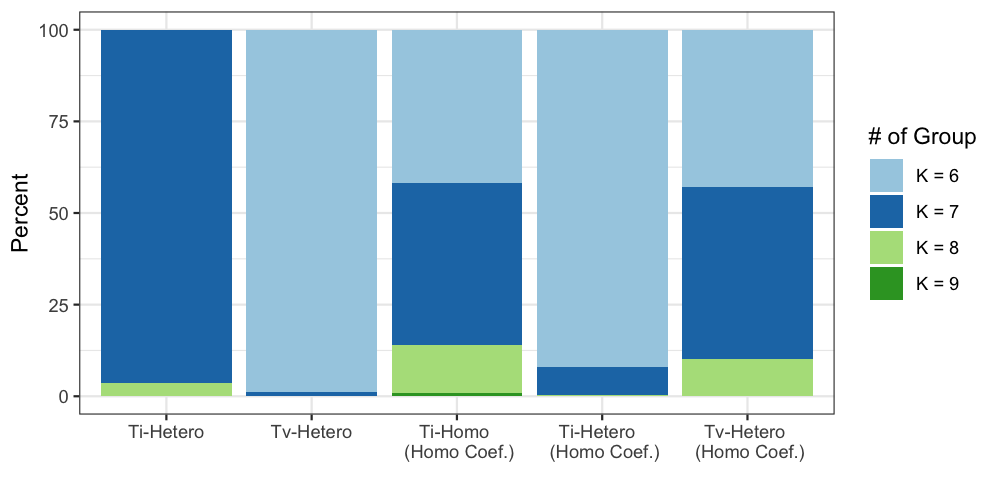}
	\end{center}
	\label{fig:prob}
\end{figure}

The fourth column documents the average number of latent groups in $\alpha_{i}$. Most of our BGRE estimators deem a group structure with more than six underlying components. And as we will discuss later, this rich group structure is the cornerstone of the accurate and flexible density forecast. In Figure \ref{fig:prob}, we present the posterior distribution of the number of groups for those BGRE estimators that have more than two groups. Regardless of the predictive performance, most estimators agree on the number of groups, with the posterior mode ranging from 6 to 7. On the other hand, the SIC code, which categorizes companies into the industries by their business activities, suggests that there are ten different industries in our sample. This indicates that our block Gibbs sampling algorithm reshuffles the default group structure and optimally pools firms from several sectors.

The fifth and sixth columns present the average coverage rate with the nominal coverage probability of 95\% and the average length of the 95\% credible set. In general, the coverage rates generated by the homogeneous coefficients specifications are substantially larger than the sets obtained from the models with heterogeneous coefficients and are closer to the nominal coverage probability of 95\%. However, the homogeneous setting doesn't improve the average length of the credible set. Indeed, the decrease in the average length is evident for the heterogeneous coefficient models, and it becomes even more pronounced once we impose heteroskedasticity. This is because the sizeable cross-sectional variation in the posterior predictive distributions leaves plenty of room for heterogeneous and heteroskedastic models to shorten the credible set while maintaining the coverage rate in a reasonable range.

The last two columns in Table \ref{table:emp_fcst} depict the performance of the density forecast. Consistent with the point forecast, the Ti-Hetero and Tv-Hetero models under homogeneous coefficients specification have comparable performance and dominate the rest with larger LPS and smaller CRPS. The Ti-Hetero has the largest LPS while Tv-Hetero scores the lowest CRPS. These facts emphasize that incorporating homogeneous coefficients and heteroskedasticity is crucial for density forecasting while time effects are not important.

The results for estimation and forecast in other years are presented in Appendix \ref{appendix:extra_empical}. In short, the result for 2019 is representative, as most conclusions discussed above also apply for other years. Although no single estimator consistently dominates the rest across the years, at least one of our BGRE estimators always offers the best performance and beats the standard panel data models.

\begin{figure}[tbp]
	\caption{Empirical Application: Posterior Predictive Density for all industries}
	\begin{center}
		\includegraphics[scale=0.8]{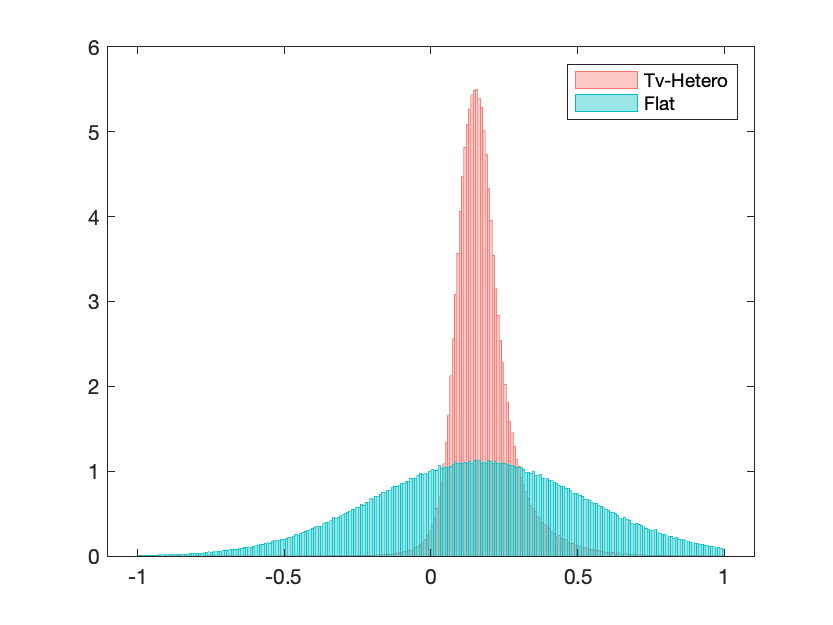}
	\end{center}
	\label{fig:yfcst_all_industry}
\end{figure}

To further investigate the posterior predictive density, we plot the densities of the investment rate generated by the Tv-Hetero and Flat estimator under homogeneous specification in Figure \ref{fig:yfcst_all_industry}. Both posterior predictive densities have similar posterior means while imposing the grouped random effects remarkably sharpens the density around the mean.  The reason is that Tv-Hetero estimator leverages latent group structure and pools the information of firms that share great similarities while the Flat estimator treats individual firm separately and makes a prediction based on limited observations.

\begin{figure}[tbp]
	\caption{Empirical Application: Posterior Predictive Density for selected industries}
	\begin{center}
		\includegraphics[scale=0.75]{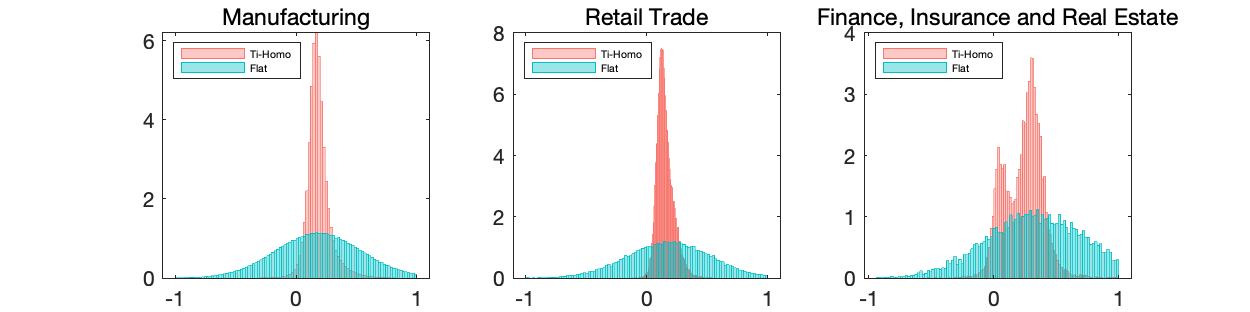}
	\end{center}
	\label{fig:yfcst_selected_industry}
\end{figure}

Figure \ref{fig:yfcst_selected_industry} further aggregates the predictive density over industries. Comparing Tv-Hetero and Flat estimator across industries, several observations stand out. First, Tv-Hetero predictive densities tend to be more concentrated in each industry. This is in line with Figure \ref{fig:yfcst_all_industry}. Second, there is substantial heterogeneity in density forecasts across sectors. While pooling the forecast for all firms yielding a well-behaved bell shape, the posterior predictive densities for the individual industries are in various shapes. This would pose difficulties for the standard estimator to forecast the future and call for a flexible model specification. Third, the Flat estimator is not flexible enough to portray the potential non-normal predictive distribution due to the restrictive normality assumption. In this case, our BGRE estimator, especially the Tv-Hetero estimator, manages to depict the skewed trimodal distribution for the Retail Trade division and bimodality in the Finance, Insurance and Real Estate divison via combining different groups.

%

\section{Conclusion} \label{sec:conclusion}
This paper studies the estimation and prediction of a dynamic panel data model with latent grouped random effects. We adopt a nonparametric Bayesian approach to identify coefficients and group membership in the random effects simultaneously. This approach avoids the severe issue introduced by the ex-post model selection and allows us to incorporate any forms of prior knowledge on group structure. In Monte Carlo experiments, we show that the BGRE estimators have the edge over standard Bayesian estimators. Regarding clustering, the BGRE estimators generate comparable performance with the \textit{Kmeans} algorithm. Our empirical application to investment rates across firms reveals that the estimated latent group structure provides a great amount of flexibility and improves point, set, and density forecasts.

The present work raises interesting issues for further research. First, it may be appealing to consider group structures in the AR(1) parameters and heterogeneous coefficients. This would allow us further to reduce the complexity of a panel data model and may improve predictive performance. Second, more clever attempts could be made to incorporate the subjective prior group structure. Our proposed method summarizes prior information in the prior of the membership probability, which can be further improved to overcome its limitation. Third, our method can be extended to nonstationary panels, where panel units and co-integrating relationships may possess latent group structures. Four, the assumption that an individual cannot change its group identity during the whole sampling period can be relaxed in the next step, leading to an even more flexible specification.

\clearpage
\setstretch{1}
\bibliography{ref}

\clearpage
\setstretch{1.3}
\appendix
\setcounter{saveeqn}{\value{section}}\renewcommand{\theequation}{\mbox{%
                \Alph{saveeqn}.\arabic{equation}}} \setcounter{saveeqn}{1} %
\setcounter{equation}{0}
\renewcommand*\thepage{A-\arabic{page}}
\setcounter{page}{1}
\renewcommand*\thetable{A-\arabic{table}}
\setcounter{table}{0}
\renewcommand*\thefigure{A-\arabic{figure}}
\setcounter{figure}{0}

\bc

{\Large {\bf Supplemental Appendix to \\
		``Forecasting with Bayesian Grouped Random Effects in Panel Data'' }}

{\bf Boyuan Zhang}

\ec



\appendix

\numberwithin{equation}{section}
\numberwithin{table}{section}
\numberwithin{figure}{section}

\section{Posterior Distributions and Algorithms} \label{appendix:algo}

\subsection{Random Effects Model} \label{appendix:algo_RE}
Below, I present the conditional posterior distribution for the time-varying random effects model with heteroskedasticity, which is the most complicated scenarios. For other models, such as its time-invariant counterparts and homoskedastic model, adjustment can be easily made by eliminating time effects and heteroskedasticity.

To facilitate derivation, we stack observations and parameters,
\begin{align*}
	\text{Observations: } Y &= [y_1, y_2, ..., y_N ], y_i = [y_{i1},y_{i2},...,y_{iT}]', \\
	\text{Covariates: } X &= [x_1, x_2, ..., x_N ], x_i = [x_{i1},x_{i2},...,x_{iT}]', \\
	\text{Random effects: }\alpha &= \left[\alpha_{1}, \alpha_{2}, \ldots\right], \\
	\text{Covariance matrices: } \Sigma &= \left[\sigma^2_{1}, \sigma^2_{2}, \ldots\right],\\
	\text{Hetergeneous coefficients: } \beta &= \left[\beta_{1}, \ldots, \beta_{N}\right], \\
	\text{Stick length: } \Xi &= \left[\xi_{1}, \xi_{2}, \ldots\right], \\
	\text{Group membership: } G &= \left[g_{1}, \ldots, g_{N}\right], \\
	\text{Auxiliary varaible: } u &= \left[  u_1, u_2, ..., u_N \right],\\
	\text{Hyper parameters: } \phi &= \left[ \mu_{\alpha}, \Sigma_{\alpha}, \nu_{\sigma}, \delta_{\sigma}\right ].
\end{align*}

The posterior of unknown objects in the random effects model is,
\begin{align*}
	& p(\rho, \beta, \alpha, \Sigma, \Xi, a, G | Y, X) \\
	\propto \; & p(Y | X, \rho, \beta, \alpha, \Sigma, G ) p(\rho, \beta, \alpha, \Sigma, \Xi, a, G) \\
	\propto \; & p(Y | X, \rho, \beta, \alpha, \Sigma, G ) p(\alpha, \Sigma|\phi) p(\Xi | a) p(G | \Xi) p(\rho) p(\beta) p(a) \\
	= \; & \prod_{i=1}^{N} p(y_i | x_i, \rho, \beta_i, \alpha_{g_{i}}, \sigma^2_{g_{i}}) \prod_{j=1}^{\infty} p(\alpha_{j}, \sigma^2_{j} | \phi) \prod_{j=1}^{\infty} p(\xi_j | a) \prod_{i=1}^{N} p(g_i | \Xi) p(\rho) \prod_{i=1}^{N} p(\beta_i) p(a) \\
	= \; & \left[ \prod_{i=1}^{N} p(y_i | x_i, \rho, \beta_i, \alpha_{g_{i}}, \sigma^2_{g_{i}}) p(g_i | \Xi) p(\beta_i) \right] \left[ \prod_{j=1}^{\infty}  p(\alpha_{j}, \sigma^2_{j} | \phi) p(\xi_j | a)\right] p(\rho) p(a). \numberthis
\end{align*}
In the following derivation and algorithm, we adopt the slice sampler \citep{walker2007} that avoids approximation in \citet{ishwaran2001}. \citet{walker2007} augments the posterior distribution with a set of auxiliary variables $u = [u_1, u_2, ..., u_N]$, which are i.i.d. standard uniform random variables, i.e, $u_i \stackrel{iid}{\sim} U(0,1)$. Then the augmented posterior is written as,
\begin{align*}
	& p(\rho, \beta, \alpha, \Sigma, \Xi, a, G, u | Y, X) \\
	\propto \; & \left[ \prod_{i=1}^{N} p(y_i | x_i, \rho, \beta_i, \alpha_{g_{i}}, \sigma^2_{g_{i}}) \mathbf{1} (u_i < \pi_{g_i}) p(\beta_i)\right] \left[ \prod_{j=1}^{\infty}  p(\alpha_{j}, \sigma^2_{j} | \phi) p(\xi_j | a)\right] p(\rho) p(a) \\
	= \; & \left[ \prod_{i=1}^{N} p(y_i | x_i, \rho, \beta_i, \alpha_{g_{i}}, \sigma^2_{g_{i}}) p(u_i | \pi_{g_i}) \pi_{g_i}  p(\beta_i)\right] \left[ \prod_{j=1}^{\infty}  p(\alpha_{j}, \sigma^2_{j} | \phi) p(\xi_j | a)\right] p(\rho) p(a), \numberthis
\end{align*}
where $\pi_{g_i} = p(g_i | \Xi)$, and $\mathbf{1}(\cdot)$ is the indicator function, which is equal to zero unless the specific condition is satisfied. The original posterior can be recovered by integrating out $u_i$ for $i = 1,2,...,N$. As we don't limit the upper bound of the number of groups, it is impossible to sample from an infinite-dimensional posterior density. The merit of slice-sampling is that it reduces the dimensions and allows us to solve a manageable problem with finite dimensions, which we will see below.

With a set of auxiliary variables $u = [u_1, u_2, ..., u_N ]$, we define the largest possible number of potential components as
\begin{align} \label{def:kstar_app}
	K^* = \min_k \left\{ u^* > 1 - \sum_{j=1}^k \pi_j \right\},
\end{align}
where
\begin{align} \label{def:ustar_app}
	u^* = \min_{1 \le i \le N} u_i.
\end{align}
Such specification ensures that for any group $k > K^*$ and any unit $i \in \{1,2,...,N\}$, we have $u_i > \pi_k$\footnote{See proof in proposition \ref{prop:1}}. This crucial property limits the dimension of $\alpha$ and $\Sigma$ to $K^*$ as the densities of $\alpha_k$ and $\sigma_{k}$ equal 0 for $k > K^*$ due to  $\mathbf{1} (u_i < \pi_k) = 0$, which will be clear in the subsequent posterior derivation.

Next, we define the number of active groups
\begin{align} \label{def:k_a_app}
	K^a = \max_{1 \le i \le N} g_i.
\end{align}
It can be shown that $K^a \le K^*$\footnote{See proof in proposition \ref{prop:1}}.

\noindent {\bf Conditional posterior of $\alpha$ (grouped random effects)}.
\begin{align*}
	p(\alpha |\rho, \beta, \Sigma, \Xi, a, G, u,  Y,X)
	\propto \; & \left[ \prod_{i=1}^{N} p(y_i | x_i, \rho, \beta_i, \alpha_{g_{i}}, \Sigma_{g_{i}}) \mathbf{1} (u_i < \pi_{g_i}) \right] \left[ \prod_{j=1}^{\infty}  p(\alpha_{j}, \sigma^2_{j} | \phi) \right],
\end{align*}

\noindent For $k \in \{1,2,...,K^a\}$, define a set of unit that belongs to group $k$,
\begin{align} \label{def:Ck_app}
	C_k = \left \{i \in \{1,2,...,N\} | g_i = k \right \},
\end{align}
then the posterior density for $\alpha_k$ read as
\begin{align*}
	& p(\alpha_k|\rho, \beta, \Sigma, \Xi, a, G, u, Y, X) \\
	\propto \; & \left[ \prod_{i \in C_k} p(y_i | x_i, \rho, \beta_i, \alpha_k, \sigma^2_k) \right] p(\alpha_{k}| \phi) \\
	\propto \; & \exp \left[ - \sum_{i \in C_k} \left( y_i - \rho y_{-1,i} - x_i\beta_i - \alpha_k \right)'\Sigma_k^{-1} \left( y_i - \rho y_{-1,i} - x_i\beta_i  - \alpha_k \right) \right] \exp \left[ - \left( \alpha_k - \mu_\alpha \right)'\Sigma_\alpha^{-1} \left( \alpha_k - \mu_\alpha \right) \right] \\
	\propto \; & \exp \left[ - \left( \alpha_k - \bar{\mu}_{\alpha_k} \right)' \bar{\Sigma}_\alpha^{-1} \left( \alpha_k - \bar{\mu}_{\alpha_k} \right) \right],
\end{align*}
where $y_{-1,i}$ are lagged values for $y_i$. Assuming an independent normal conjugate prior for $\alpha_k$, the posterior for $\alpha_k$ is given by
\begin{align} \label{post_alpha_app}
	\alpha_k |\rho, \beta, \Sigma, \Xi, a, G, u, Y, X \sim N \left( \bar{\mu}_{\alpha_k}, \bar{\Sigma}_{\alpha_k} \right).
\end{align}
where
\begin{align*}
	\bar{\Sigma}_{\alpha_k} &= \left( \Sigma_\alpha^{-1} + \sum_{i \in C_k} \Sigma_i^{-1} \right)^{-1}, \\
	\bar{\mu}_{\alpha_k} &= \bar{\Sigma}_{\alpha_k} \left[\Sigma_\alpha^{-1} \mu_\alpha + \sum_{i \in C_k} \Sigma_i^{-1} \widetilde{y}_i \right], \\
	\widetilde{y}_i &= y_i - \rho y_{-1,i} - x_i\beta_i.
\end{align*}
If group $k$ is empty, we draw $\alpha_k$ from its prior $ N \left( \mu_{\alpha}, \Sigma_{\alpha} \right)$.\\

\noindent {\bf Conditional posterior of $\Sigma$ (grouped variance)}.
Under the cross-sectional independence, for $k = 1,2,...,K^a$,
\begin{align*}
	p(\sigma^2_k|\rho, \beta, \alpha, \Xi, a,G, u, Y, X)
	\propto \; & \left[ \prod_{i \in C_k} p(y_i | x_i, \rho, \beta_i, \alpha_k, \sigma^2_k) \right] p(\sigma_{k}^2 | \phi)
\end{align*}
Assuming a inverse-gamma prior $\sigma_{k}^2 \sim IG \left( \frac{v_\sigma}{2}, \frac{\delta_\sigma}{2} \right)$, the posterior distribution of $\sigma_{k}^2$ is
\begin{align*}
	& p(\sigma_{k}^2 |\rho, \beta, \alpha, G, u, Y, X) \\
	\propto \; & \prod_{i \in C_k} \left[ (\sigma_{k}^2)^{-\frac{T}{2}} \exp \left( - \frac{\sum_{t=1}^T (y_{it} - \rho y_{it-1} - \beta' x_{it} - \alpha_{kt} )^2}{2\sigma_{k}^2} \right) \right] \left( \frac{1}{\sigma_{k}^2} \right)^{\frac{v_\sigma}{2}+1} \exp \left( - \frac{\delta_\sigma}{2\sigma_{k}^2} \right) \\
	= \; & \left( \frac{1}{\sigma_{k}^2} \right)^{\frac{v_\sigma + T|C_k|}{2}+1} \exp \left( - \frac{\delta_\sigma + \sum_{i \in C_k} \sum_{t=1}^T (y_{it} - \rho y_{it-1} - \beta' x_{it} - \alpha_{kt} )^2}{2\sigma_{k}^2} \right).
\end{align*}
This implies
\begin{align} \label{post_Sigma_app}
	\sigma_{k}^2 |\rho, \beta, \alpha, \Xi, a, G, u, Y, X \sim IG \left( \frac{\bar{v}_{\sigma,k}}{2}, \frac{\bar{\delta}_{\sigma,k}}{2} \right),
\end{align}
where
\begin{align*}
	\bar{v}_{\sigma,k} &= v_\sigma + T|C_k|, \\
	\bar{\delta}_{\sigma,kt} &= \delta_\sigma + \sum_{i \in C_k} \sum_{t=1}^T (\widetilde{y}_{it} - \alpha_{kt} )^2, \\
	|C_k| &= \text{occurrence of } g_i = k, \\
	\widetilde{y}_{it} &= y_{it} - \rho y_{it-1} - \beta' x_{it}.
\end{align*}
If group $k$ is empty, we draw $\sigma_{k}^2$ from its prior $ IG \left( \frac{v_\sigma}{2}, \frac{\delta_\sigma}{2} \right)$. \\

\noindent {\bf Conditional posterior of $\rho$ (common coefficient)}.
Using a normal conjugate prior $\rho \sim \textit{N } (\mu_\rho, \Sigma_\rho)$, we could solve standard Bayesian linear regression to get the posterior density of the common coefficient $\rho$,
\begin{align*}
	& p(\rho | \beta, \alpha, \Sigma, \Xi, a, G, u, Y, X) \propto \; \left[ \prod_{i=1}^{N} p(y_i | x_i, \rho, \beta_i, \alpha_{g_{i}}, \Sigma_{g_{i}}) \right] p(\rho) \\
	\propto \; & \exp \left[ - \sum_{i=1}^N (y_i - \alpha_{g_i} - x_i\beta_i - \rho y_{-1,i})' \Sigma_{g_i}^{-1} (y_i - \alpha_{g_i} - x_i\beta_i - \rho y_{-1,i} \right] \exp \left[ -(\rho - \mu_\rho)' \Sigma_\rho^{-1} (\rho - \mu_\rho)\right].
\end{align*}
This implies
\begin{align} \label{post_rho_app}
	\rho |\beta, \alpha, \Sigma, \Xi, a, G, u, Y, X \sim \mathcal{N} \left( \bar{\mu}_{\rho}, \bar{\Sigma}_{\rho} \right),
\end{align}
where
\begin{align*}
	\bar{\Sigma}_{\rho} &= \left( \Sigma_\rho^{-1} + \sum_{i=1}^N  y_{-1,i}' \Sigma_{g_i}^{-1} y_{-1,i} \right)^{-1}, \\
	\bar{\mu}_{\rho} &= \bar{\Sigma}_{\rho} \left[\Sigma_\rho^{-1} \mu_\rho + \sum_{i=1}^N y_{-1,i}' \Sigma_{g_i}^{-1} \hat{y}_i \right], \\
	\hat{y}_i &= y_i - \alpha_{g_{i}} -x_i\beta_i.
\end{align*}

\noindent {\bf Conditional posterior of $\beta$ (heterogeneous coefficients)}.
As $\varepsilon_{i t}$ is independent across units, we solve for $\beta$ for each unit separately. We transform the model into a standard linear model with a known form of heteroskedasticity,
\begin{align*}
	y_{it} - \alpha_{g_i t} -  \rho y_{i t-1}  = \beta'_{i} x_{it} + \varepsilon_{it}, \varepsilon_{it} \sim N(0, \sigma^2_{g_i}).
\end{align*}
Using a normal conjugate prior $\beta_i \sim \textit{N } (\mu_\beta, \sigma^2_\beta)$, for the unit $i$, the posterior distribution is written as,
\begin{align*}
	& p(\beta_i | \rho, \alpha, \Sigma, \Xi, a, G, u, Y, X) \\
	\propto \; &  p(y_i | x_i, \rho, \beta_i, \alpha_{g_{i}}, \sigma^2_{g_{i}}) p(\beta_i) \\
	\propto \; & \exp \left[ -  \frac{ \sum_{t=1}^T (y_{it} - \alpha_{g_i} - \rho y_{it}- x_{it}' \beta_i )^2}{2\sigma^2_{g_i}}\right] \exp \left[ -(\beta_i - \mu_\beta)' \Sigma_\beta^{-1} (\beta_i - \mu_\beta)\right].
\end{align*}
This implies
\begin{align} \label{post_beta_app}
	\beta_i |\rho, \alpha, \Sigma, \Xi, a, G, u, Y, X \sim \mathcal{N} \left( \bar{\mu}_{\beta_i}, \bar{\Sigma}_{\beta_i} \right),
\end{align}
where
\begin{align*}
	\bar{\Sigma}_{\beta_i} &= \left( \Sigma_\rho^{-1} + \sigma^{-2}_{g_i} \sum_{t=1}^T   x_{it} x_{it}' \right)^{-1} \\
	\bar{\mu}_{\beta_i} &= \bar{\Sigma}_{\rho} \left[\Sigma_\rho^{-1} \mu_\rho + \sigma^{-2}_{g_i} \sum_{t=1}^T   x_{it} \hat{y}_{it} \right] \\
	\hat{y}_{it} &= y_{it} - \alpha_{g_{i}} - \rho y_{it-1}
\end{align*}

\noindent {\bf Conditional posterior of $\Xi$ (stick length)}.
\begin{align*}
	p(\Xi | \rho, \beta, \alpha, \Sigma, a, G, u, Y, X)
	\propto \; & \left[ \prod_{i=1}^{N} p(u_i | \pi_{g_i}) \pi_{g_i}\right] \left[ \prod_{j=1}^{\infty} p(\xi_j | a)\right] \\
	\propto \; & \left[ \prod_{i=1}^{N} p(u_i | \pi_{g_i}) \xi_{g_i} \prod_{l < g_i} (1- \xi_l) \right] \left[ \prod_{j=1}^{\infty} p(\xi_j | a)\right]
\end{align*}

For $k = 1, 2, ..., K^a$,
\begin{align*}
	p(\xi_k | \rho, \beta, \alpha, \Sigma, a, G, u, Y, X)
	\propto \; & \left( \prod_{i \in C_k} \xi_k \right) (1-\xi_k)^{\sum \limits_{j=1}^N \mathbf{1}(g_j>k)} (1-\xi_k)^{a-1} \\
	\propto \; & \xi_k^{|C_k|} \left( 1-\xi_k \right)^{a + \sum \limits_{j=1}^N \mathbf{1}(g_j>k) -1}.
\end{align*}

Therefore, posterior distribution of $\xi_k$ is
\begin{align} \label{post_Xi_app}
	\xi_k | \rho, \beta, \alpha, \Sigma, a, G, u, Y, X \sim \textit{Beta} \left( |C_k| + 1, a + \sum_{j=1}^N \mathbf{1}(g_j>k) \right).
\end{align}

Give $\Xi = [\xi_1, \xi_2, ..., \xi_{K^a}]$, update $\pi_1, \pi_2,...,\pi_{K^a}$,

\begin{align} \label{post_p_app}
	\pi_{k} | G,\Xi \sim \left\{
	\begin{array}{ll}
	{\xi_{1},} & {k = 1} \\
	{\xi_{k} \prod_{j<k}\left(1-\xi_{j}\right),} & {k = 2, \ldots, K^a}
	\end{array}
	\right..
\end{align}

\noindent {\bf Conditional posterior of $a$ (concentration parameter)}.
Regarding the DP concentration parameter, the standard posterior derivation doesn't work due to the unrestricted number of components in the current sampler. Instead, we implement the 2-step procedure proposed by \citet{escobar1995} (p.8-9). Following their approach, we first draw a latent variable $\eta$ from
\begin{align} \label{post_eta_app}
	\eta \sim \textit{Beta } (a+1, N).
\end{align}
Then, conditional on $\eta$ and $K^a$, we assume sample $a$ from a mixture of two Gamma distribution:
\begin{align} \label{post_a_app}
	p(a | \eta, K^a) = \pi_a \textit{Gamma } (m+K^a, n-\log(\eta)) + (1 - \pi_a) \textit{Gamma } (m+K^a-1, n-\log(\eta)),
\end{align}
with the weights $\pi_a$ defined by
\begin{align*}
	\fra{\pi_a}{1-\pi_a} = \frac{m+K^a-1}{N[n-\log(\eta)]}.
\end{align*}

\noindent {\bf Conditional posterior of $u$ (auxiliary variable)}.
Conditional on the group ``stick lengths'' $\xi_k$ and group member indices $G$, it is straightforward to show that the posterior density of $u_i$ is a uniform distribution ranging define on $(0, \pi_{g_i})$, that is
\begin{align} \label{post_u_app}
	u_i | \Xi, G \sim \textit{Unif }(0, \pi_{g_i}),
\end{align}
where $\pi_{g_i} = \xi_{g_i} \prod_{j<g_j} (1-\xi_j)$. Moreover, it is worth noting that the values for $K^*$ and $u^*$ need to be updated according to (\ref{def:kstar_app}) and (\ref{def:ustar_app}) after this step.\\

\noindent {\bf Conditional posterior of $G$ (group membership)}.
We derive the posterior distribution of $g_i$ consider on $G^{(i)}$, where $G^{(i)}$ is a set including all member indices except for $g_i$, i.e., $G^{(i)} = G \char`\\ g_i$. Hence, for $k = 1, 2, ..., K^*$,
\begin{align*}
	p(g_i = k | \rho, \beta, \alpha, \Sigma, \Xi, a, G^{(i)}, u, Y,X)
	\propto \; p(y_i | \rho, \beta_i, \alpha_k, \sigma^2_k, Y, X) \mathbf{1} (u_i < \pi_{g_i}).
\end{align*}
As per a discrete distribution, we normalize the point mass to get a valid distribution:
\begin{align} \label{post_G_app}
	p(g_i = k | \rho, \beta, \alpha, \Sigma, \Xi, a, G^{(i)}, u, Y,X)
	= \; \frac{p(y_i | \rho, \beta_i, \alpha_k, \sigma^2_k, Y,X) \mathbf{1} (u_i < \pi_{k})}{ \sum_{j=1}^{K^*} p(y_i | \rho, \beta_i, \alpha_j, \sigma^2_j, Y,X) \mathbf{1} (u_i < \pi_j)}.
\end{align}

\subsubsection{Blocked Gibbs Sampler and Algorithm} \label{subsec:algo}
Initialization:
\begin{enumerate}[(i)]
	\item Preset the initial number of active groups $K_0^a$. As derived by \citet{antoniak1974}, the expected number of unique groups is $E\left[K | a \right] \approx a \log \left(\frac{a + N}{a}\right)$. We set $K_0^a$ to its expected value with concentration parameter $a$ replaced by prior mean.

	\item In ignorance of group heterogeneity ($K = 1$) and heteroskedasticity, run OLS to get $\hat{\alpha}_{OLS}$, $\hat{\rho}_{OLS}$, $\hat{\beta_i}_{OLS}$ and $Cov(\hat{\alpha}_{OLS})$. These OLS estimators serve as the mean and covariance matrix in the related priors.

	\item Generate $K_0^*$ random sample from the distribution $N(\hat{\alpha}_{OLS}, Cov(\hat{\alpha}_{OLS}))$.

	\item Initialize group membership $G$ by sampling from (\ref{post_G_app}) ignoring $\mathbf{1} (u_i < \pi_{g_i})$. Remove empty groups.
	%
\end{enumerate}
For each iteration $s = 1,2,..,N_{sim}$
\begin{enumerate}[(i)]
	\item Number of active groups:
	\begin{align*}
		K^a = \max_{1 \le i \le N} g_i^{(s-1)}.
	\end{align*}
	\item Group ``stick length'': for $k = 1,2,...,K^a$, draw $\xi_k$ from a Beta distribution in (\ref{post_Xi_app}):
	\begin{align*}
		\xi_k | \rho^{(s-1)},\beta^{(s-1)},  \alpha^{(s-1)}, \Sigma^{(s)}, a^{(s-1)}, G^{(s-1)}, u^{(s-1)}, Y, X \sim \textit{Beta} \left( |C_k| + 1, a + \sum_{j=1}^N \mathbf{1}(g_j>k) \right),
	\end{align*}
	and calculate group probability in accordance to (\ref{post_p_app}).

	\item Group heterogeneity: for $k = 1,2,...,K^a$, draw $\alpha_k^{(s)}$ from a normal distribution in (\ref{post_alpha_app}):
	\begin{align*}
		\alpha_k |\rho^{(s-1)},\beta^{(s-1)}, \Sigma^{(s-1)}, a^{(s-1)}, G^{(s-1)}, u^{(s-1)}, Y, X \sim \textit{N} \left( \bar{\mu}_{\alpha_k}, \bar{\Sigma}_{\alpha_k} \right).
	\end{align*}

	\item Group heteroskedasticity: for $k = 1,2,...,K^a$ and $t = 1,2,...,T$, draw $\sigma_{k}^{2^{(s)}}$ from an inverse Gamma distribution in (\ref{post_Sigma_app}):
	\begin{align*}
		\sigma_{k}^2 |\rho^{(s-1)},\beta^{(s-1)}, \alpha^{(s)}, G^{(s-1)}, u^{(s-1)}, Y, X \sim \textit{IG} \left( \frac{\bar{v}_{\sigma,k}}{2}, \frac{\bar{\delta}_{\sigma,k}}{2} \right).
	\end{align*}

	\item Label switching\footnote{Without this step, the one-at-a-time updates of the allocations mean that clusters rarely switch labels, and consequentially the ordering will be largely determined by the (perhaps random) initialization of the sampler.}: after each iteration an additional random permutation step is added to the MCMC scheme which randomly permutes the current labeling of the components. Random permutation ensures that the sampler explores all K! modes of the full posterior distribution and avoids that the sampler is trapped around a single posterior mode. Following \citet{liu2020}\footnote{See Algorithm C.4 in the appendix}, we update $\left \{\alpha_k^{(s)}, \sigma_{k}^{2^{(s)}}, \pi_k^{(s)}, g_i^{(s-1)} \right \}$ by three Metropolis-Hastings label-switching moves developed by \citet{papaspiliopoulos2008} (step (a) and (b)) and \citet{hastie2015} (step (c)). All these label switching moves aim to improve numerical convergence.

	\begin{enumerate} [(a)]
		\item Randomly select two nonempty groups $i$ and $j$, swap group labels $g_i^{(s-1)}$ and $g_j^{(s-1)}$ for all units in these groups, accept new label with probability:
		\begin{align*}
			\min \left(1, \frac{\pi_i^{N_j} \pi_j^{N_i}}{\pi_i^{N_i} \pi_j^{N_j}} \right) = \min \left(1, (\pi_i / \pi_j)^{N_j - N_i} \right),
		\end{align*}
		where $N_i$, $N_j$ are the number of units in the group $i$ and $j$ respectively.

		\item Randomly select two adjacent groups $l$ and $l+1$ such that $\{l,l+1\} \subset \{1,2,...,K^a\}$, swap group label $ g_l^{(s-1)}$ \textbf{and} ``stick length'' $\xi_l^{(s)}$, accept new label and stick length with probability:
		\begin{align*}
			\min \left(1, \frac{\tilde{p}_l^{N_{l+1}} \tilde{p}_{l+1}^{N_l}}{\pi_l^{N_l} \pi_{l+1}^{N_{l+1}}} \right),
		\end{align*}
		where $\tilde{p}_i$ and $\tilde{p}_j$ are new group probabilities derived with new $\xi_{l}^{(s)}$ and $\xi_{l+1}^{(s)}$.

		\item Randomly select two adjacent groups $k$ and $k+1$ such that $\{k,k+1\} \subset \{1,2,...,K^a\}$, swap group label $ g_i^{(s-1)}$, ``stick length'' $\xi_k^{(s)}$ \textbf{and update} group-specific parameter $\{\alpha_k^{(s)},  \sigma_{k}^{2^{(s)}} \}$, accept new new label and stick length with probability
		\begin{align*}
			\min \left\{1, \left( R_1 / \widetilde{R} \right)^{N_{k+1}} \left( R_2 / \widetilde{R} \right)^{N_{k}} \right \},
		\end{align*}
		where
		\begin{align*}
			R_{1} &= \frac{1+a+N_{k+1}+\sum_{l>k+1} N_{l}}{a+N_{k+1}+\sum_{l>k+1} N_{l}}, \\
			R_{2} &= \frac{a+N_{k}+\sum_{l>k+1} N_{l}}{1+a+N_{k}+\sum_{l>k+1} N_{l}}, \\
			\widetilde{R} &= \frac{\pi_{k+1} R_1 + \pi_{k} R_2}{\pi_k + \pi_{k+1}}.
		\end{align*}
		The new group probability is defined as $p'_{k} =  \pi_{k+1} R_1 / \widetilde{R}$ and $p'_{k+1} = \pi_{k} R_2 / \widetilde{R}$. Additionally, we update the ``stick lengths''\footnote{This particular choices of $\xi'_{k}$ and $\xi'_{k+1}$ ensure the group probabilities that are changed are those associated with the the group $k$ and $k+1$, and the rest are unchanged. Moreover, it can be shown that $(1-\xi'_{k})(1-\xi'_{k+1}) = (1-\xi_{k})(1-\xi_{k+1})$. See more details in the appendices of \citet{hastie2015}.} for group $k$ and $k+1$ such that
		\begin{align*}
			\xi'_{k} &= \frac{ p'_{k} }{\prod_{l<c} (1- \xi_l)}, \\
			\xi'_{k+1} &= \frac{ p'_{k+1} }{ (1-\xi'_{k}) \prod_{l<c} (1- \xi_l)}.
		\end{align*}
	\end{enumerate}

	\item Auxiliary variables: for $i = 1,2,...,N$, draw $u_i$ from an uniform distribution in (\ref{post_u_app}):
	\begin{align*}
		u_i | \Xi^{(s)}, G^{(s)} \sim \text{U}(0, p^{(s)}_{g_i}).
	\end{align*}
	Then calculate $u^*$ according to (\ref{def:ustar_app}).

	\item DP concentration parameter:
	\begin{enumerate}[(a)]
		\item Draw latent	variable $\eta$ from a Beta distribution in (\ref{post_eta_app}):
		\begin{align*}
			\eta \sim \textit{Beta} (a+1, K^a)
		\end{align*}
		\item Draw concentration parameter $a$ from a mixture of Gamma distribution in (\ref{post_a_app}):
		\[
		a | \eta, K^a \sim \left\{
		\begin{array}{ll}
			\textit{Gamma} \left (m+K^a, n-\log(\eta) \right) & \mbox{with prob. } \pi_a \\
			\textit{Gamma} \left (m+K^a-1, n-\log(\eta) \right) & \mbox{with prob. } 1-\pi_a
		\end{array}
		\right. ,
		\]
		and $\pi_a$ is defined as
		\begin{align*}
			\fra{\pi_a}{1-\pi_a} = \frac{m+K^a-1}{N(n-\log(\eta))}.
		\end{align*}
	\end{enumerate}

	\item Potential groups: start with $\tilde{K} = K^a$,
	\begin{enumerate}[(a)]
		\item Group probabilities:
		\begin{enumerate} [(1)]
			\item if $ \sum_{j=1}^{\tilde{K}} \pi_{j}^{(s)} > 1-u^*$, set $K^* = \tilde{K}$ and stop
			\item otherwise, let $\tilde{K} = \tilde{K}+1$, draw $\xi_{\tilde{K}} \sim Beta \left(1, \alpha^{(s)}\right),$ update $\pi_{\tilde{K}} = \xi_{\tilde{K}} \prod_{j<\tilde{K}}\left(1- \xi_{j} \right)$ and go to step $(1)$
		\end{enumerate}
		\item Group parameters:  for $k = K+1, \cdots, K^*$, draw $\alpha_{k}^{(s)}$ and $\sigma_{k}^{2(s)}$ from their prior distributions.
	\end{enumerate}

	\item Common AR(1) parameter: draw $\rho$ from a normal distribution in (\ref{post_rho_app}):
	\begin{align*}
		\rho |\beta^{(s-1)}, \alpha^{(s)}, \Sigma^{(s)}, a^{(s)}, G^{(s-1)}, u^{(s)}, Y, X \sim \textit{N} \left( \bar{\mu}_{\rho}, \bar{\Sigma}_{\rho} \right).
	\end{align*}

	\item Heterogeneous parameter: draw $\beta_i$ from a normal distribution in (\ref{post_beta_app}):
	\begin{align*}
		\beta_i |\rho^{(s)}, \alpha^{(s)}, \Sigma^{(s)}, a^{(s)}, G^{(s-1)}, u^{(s)}, Y, X \sim \textit{N} \left( \bar{\mu}_{\beta_i}, \bar{\Sigma}_{\beta_i} \right).
	\end{align*}

	\item Group membership: for $i = 1,2,...,N$ and $k = 1,2,...,K^*$, draw $g_i$ from a multinomial distribution in (\ref{post_G_app}):
	\begin{align*}
		p(g_i = k | \rho^{(s)},\beta^{(s)}, \alpha^{(s)}, \Sigma^{(s)}, \xi^{(s)}, a^{(s)}, G^{(i)}, u^{(s)}, Y, X)
		= \; \frac{p(y_i | \rho^{(s)},\beta_i^{(s)}, \alpha^{(s)}_k, \Sigma^{(s)}_k) \mathbf{1} (u^{(s)}_i < \pi_{k}) }{ \sum_{j=1}^{K^*} p(y_i | \rho^{(s)},\beta_i^{(s)}, \alpha^{(s)}_j, \Sigma^{(s)}_j) \mathbf{1} (u^{(s)}_j < \pi_{g_j})}
	\end{align*}
\end{enumerate}

\subsection{Random Effects Model with Subjective Group Probability Prior} \label{appendix:post_subPrior2}
This algorithm is designed for the random effect model where econometricians have prior knowledge on the group structure and presume the number of groups $K^p$. Building on the algorithm for the random effect model in Section \ref{appendix:algo_RE}, we allow for incorporating the researchers' prior knowledge while inheriting the feature of reallocating units and changing the number of groups along the MCMC sampling.

We use the same notation as in Appendix \ref{appendix:algo_RE},
\begin{align*}
	\text{Observations: } Y &= [y_1, y_2, ..., y_N ], y_i = [y_{i1},y_{i2},...,y_{iT}]', \\
	\text{Covariates: } X &= [x_1, x_2, ..., x_N ], x_i = [x_{i1},x_{i2},...,x_{iT}]', \\
	\text{Random effects: }\alpha &= \left[\alpha_{1}, \alpha_{2}, \ldots, \alpha_{K} \right], \\
	\text{Covariance matrices: } \Sigma &= \left[\Sigma_{1}, \Sigma_{2}, \ldots, \Sigma_{K}\right],\\
	\text{Group membership: } G &= \left[g_{1}, \ldots, g_{N}\right],\\
	\text{Stick length: } \Xi &= \left[\xi_{1}, \xi_{2}, \ldots\right], \\
	\text{Group probability: } \pi &= \left[\pi_{1}, \ldots, \pi_{N}\right],\\
	\text{Membership probability: } \omega &= \left[\omega_{1}, \ldots, \omega_{N}\right], \omega_{i} =  [\omega_{i1},\omega_{i2} ,...,\omega_{iK} ]',\\
	\text{Auxiliary varaible: } u &= \left[  u_1, u_2, ..., u_N \right],\\
	\text{Hyper parameters: } \phi &= \left[ \mu_{\alpha}, \Sigma_{\alpha}, \nu_{\sigma}, \delta_{\sigma}\right ].
\end{align*}

Notice that we define two sets of probabilities, $\pi$ and $\omega$. In practice, they have distinct roles in the algorithm. The group membership $\pi$ captures the groups' probability based on the entire sample, and, most importantly, determines the upper bound of the auxiliary variable $u_{g_i}$ that has direct effect on the potential number of groups $K^*$. On the other hand, the membership probability $\omega_i$ represents the probabilities of a unit $i$ belonging to each of $K$ groups, through which the researcher's prior knowledge enters the algorithm.

As regards the choices of prior, we adopt the independent Multivariate Normal-Inverse-Gamma prior Dirichlet Process priors for group random effects $\alpha_{g_i t}$ and heteroskedasticity $\sigma^2_{g_{i}}$, a normal prior for the common parameter $\rho$, an unsymmetric Dirichlet prior for the membership probability $\omega$ with concentration parameters chosen by the econometrician, a multinomial prior for Group membership $g_i$, a Beta prior for the stick length $\xi$, and a mixture Gamma prior for the concentration parameter $a$

The posterior of unknown objects in this random effects model is:
\begin{align*}
	& p(\rho, \beta, \alpha, \Sigma, G, \omega  | Y, X) \\
	\propto \; & p(Y | X, \rho, \beta, \alpha, \Sigma, G, \omega ) p(\rho, \beta, \alpha, \Sigma, G, \pi) \\
	\propto \; & p(Y | X, \rho, \beta, \alpha, \Sigma, G ) p(\alpha, \Sigma|\phi) p(G | p) p(\rho) p(\beta) p(\omega | a) \\
	= \; & \prod_{i=1}^{N} p(y_i | x_i, \rho, \beta_i, \alpha_{g_{i}}, \sigma^2_{g_{i}}) \prod_{j=1}^{K} p(\alpha_{j}, \sigma^2_{j} | \phi) \prod_{i=1}^N p(g_i| \omega_i)  \prod_{i=1}^N p(\omega_i| a_i)  \prod_{i=1}^N p(\beta_i)  p(\rho) \\
	= \; & \prod_{i=1}^{N} \left[p(y_i | x_i, \rho, \beta_i, \alpha_{g_{i}}, \sigma^2_{g_{i}}) p(g_i| \omega_i) p(\omega_i| a_i)  p(\beta_i)\right] \prod_{j=1}^{K} p(\alpha_{j}, \sigma^2_{j} | \phi) p(\rho).
\end{align*}

To allow for automatically adjustment for the number of groups, we introduce a set of auxiliary variables $u = [u_1, u_2,..., u_N]$ proposed by \citet{walker2007} and rewrite the posterior above as,
\begin{align*}
	& p(\rho, \beta, \alpha, \Sigma, \Xi, a, G, u, \omega, \pi | Y, X) \\
	\propto \; & \prod_{i=1}^{N} \left[p(y_i | x_i, \rho, \beta_i, \alpha_{g_{i}}, \sigma^2_{g_{i}}) p(g_i| \omega_i) p(\omega_i| a_i) \mathbf{1} (u_i < \pi_{g_i}) p(\beta_i)\right] \prod_{j=1}^{K} p(\alpha_{j}, \sigma^2_{j} | \phi) p(\rho).
\end{align*}

The number of potential groups $K^*$ and the number of active groups $K^a$ are defined in the equation (\ref{def:kstar_app}) and (\ref{def:k_a_app}).

\noindent {\bf Conditional posterior of $\alpha$ (grouped random effects)}.  Identical to (\ref{post_alpha_app}).

\noindent {\bf Conditional posterior of $\Sigma$ (grouped variance)}. Identical to (\ref{post_Sigma_app}).

\noindent {\bf Conditional posterior of $\rho$ (common coefficient)}. Identical to (\ref{post_rho_app}).

\noindent {\bf Conditional posterior of $\beta$ (heterogeneous coefficients)}. Identical to (\ref{post_beta_app}).

\noindent {\bf Conditional posterior of $\Xi$ (stick length)}. Identical to (\ref{post_Xi_app}). Then generate $\pi$ in accordance to (\ref{post_p_app}).

\noindent {\bf Conditional posterior of $a$ (concentration parameter)}. Identical to (\ref{post_a_app}).

\noindent {\bf Conditional posterior of $u$ (auxiliary variable)}. Identical to (\ref{post_u_app}).

\noindent {\bf Conditional posterior of $\omega$ (membership probability)}. Sampling from the posterior of $\pi$ can be implemented as follows. As we adopt Dirichlet prior for $\pi$ and Multinomial prior for $g_i $, for $i = 1,..., N$, the posterior is written as,
\begin{align*} \label{post_omega_app}
	p(\omega_i |\rho, \beta, \alpha, \Sigma, G, Y, X)
	\propto \; & p(g_i | \omega_i) p(\omega_i | a_i)\\
	\propto \; & \left(\omega_{i1}^{\mathbf{1}(g_i = 1)} \ldots \omega_{iK^p}^{{\mathbf{1}(g_i = K^p)}}\right) \times \left(\omega_{i1}^{a_{i1}-1} \ldots \omega_{ik}^{a_{iK^p}-1}\right)  \\
	= \; &\omega_{i1}^{a_{i1} + \mathbf{1}(g_i = 1) -1} \ldots \omega_{iK^p}^{a_{i1} + \mathbf{1}(g_i = K^p) -1}. \numberthis
\end{align*}
This implies
\begin{align}
	\omega_i |\rho, \beta, \alpha, \Sigma, G, Y, X \sim Dir \left( a_{i1} + \mathbf{1}(g_i = 1), \ldots, a_{iK^p} + \mathbf{1}(g_i = K^p)\right).
\end{align}

It is worth noting that, during MCMC sampling, we allow for more/fewer groups than the researcher expects, i.e., the potential number of groups $K^{*(s)}$ could be larger or smaller than $K^p$ in some iteration $s$. In such circumstances, we modify the Dirichlet posterior distribution in (\ref{post_omega_app}) to account for such changes. Notably, we present the posterior in three cases.


\textbf{\emph{Case 1}}: $K^{*(s)} = K^p$. The posterior distribution in (\ref{post_omega_app}) is still valid.

%

\textbf{\emph{Case 2}}: $K^{*(s)} > K^p$. We have to address new groups. For the additional groups such that $k^{\dagger} > K^p$, we have, for $\forall i$,
\begin{align*}
	\mathbf{1}(g^{(s-1)}_i = k^{\dagger}) = 0 \text{ if } k^{\dagger} > K^{a(s-1)} \text{ and } a_{ik^{\dagger}} = 0.
\end{align*}
where $K^{a(s-1)} = \max \limits_{1 \le i \le N} g^{(s)}_i$ denotes the number of active groups in the previous iteration.

To ensure a non-negative posterior probability $\omega_{ik^{\dagger}}$ for the new groups, we assume that, for some $\epsilon$\footnote{In the simulation, we set $\epsilon = 0.3$ so that a unit has a chance of 30\% to be assigned to the new groups a priori.},
\begin{align*}
	\sum_{k = k^{\dagger}}^{K^{*(s)}} \tilde{a}_{ik^{\dagger}} = \epsilon \text{ and } \tilde{a}_{i m} = \tilde{a}_{i n}, \forall m, n \ge k^{\dagger}.
\end{align*}
and adjust the prior membership probability $a_{ik}$ if $k<k^{\dagger}$ by multiplying $1-\epsilon$. This step artificially assigns nonzero probabilities to news groups and forms a set of new hyperparameters $\tilde{a}_{ik} $ such that $\sum_{k=1}^{K^{*(s-1)}} \tilde{a}_{ik} = 1$. Then we draw $\omega_i$ from the posterior density,
\begin{align}
	\omega_i |\rho, \beta, \alpha, \Sigma, G, Y, X \sim Dir \left( \tilde{a}_{i1} + \mathbf{1}(g_i = 1), \ldots, \tilde{a}_{iK^{*(s)}} + \mathbf{1}(g_i = K^{*(s)})\right).
\end{align}

\textbf{\emph{Case 3}}: $K^{*(s)} < K^{p}$. We have few groups than the researcher assumes and a unit $i$ might be assigned to a group that is no longer considered in the current iteration (i.e., $K^{*(s)}  < g_i^{(s-1)} $). In this case, we need to select and renormalize a subset of $a_{ik}$ since some groups are dismissed. In this regard, for a unit $i$, we select $K^{*(s)}$ most frequent non-empty groups among the groups visited in the previous iteration $s-1$. If there are not enough candidates, we add back non-selected groups in the first $K^{*(s)}$ out of the $K^{p}$ groups. Then we normalize the selected $a_{ik}$\footnote{In practice, if all selected $a_{ik}$ are zero for some $i$, we simply assume $a_{ik} = 1/K^{*(s)}$.} and get $\hat{a}_{ik}$. Finally we draw $\omega_i$ from the posterior density,
\begin{align}
	\omega_i |\rho, \beta, \alpha, \Sigma, G, Y, X \sim Dir \left( \hat{a}_{i1} + \mathbf{1}(g_i = 1), \ldots, \hat{a}_{iK^{*(s)}} + \mathbf{1}(g_i = K^{*(s)})\right).
\end{align}

\noindent {\bf Conditional posterior of $G$ (group membership)}.
We derive the posterior distribution of $g_i$ consider on $G^{(i)}$, where $G^{(i)}$ is a set including all member indices except for $g_i$, i.e., $G^{(i)} = G \char`\\ g_i$. Hence, for $k = 1, 2, ..., K^*$,
\begin{align*}
	p(g_i = k | \rho, \beta, \alpha, \Sigma, G^{(i)}, \pi, \omega, Y, X)
	\propto \; p(y_i | \rho, \beta_i, \alpha_k, \Sigma_k, Y,X) \omega_{ik} \mathbf{1} (u_i < \pi_{g_i}).
\end{align*}
As per a discrete distribution, we normalize the point mass to get a valid distribution,
\begin{align}
	p(g_i = k | \rho, \beta_i, \alpha, \Sigma, G^{(i)}, \pi, \omega, Y, X)
	= \; \frac{p(y_i | \rho, \beta_i, \alpha_k, \Sigma_k, Y,X) \omega_{ik} \mathbf{1} (u_i < \pi_k) }{ \sum_{j=1}^{K} p(y_i | \rho, \beta_, \alpha_j, \Sigma_j, Y,X)  \omega_{ij} \mathbf{1} (u_i < \pi_j)}.
\end{align}

\subsection{Random Effects Model with Group Structures in $\alpha$, $\rho$ and $\beta$} \label{appendix:post_fullGroupStructure}
In this subsection, I present the conditional posterior distribution for the time-invariant random effects model with group structures in $\alpha$, $\rho$ and $\beta$. The model is,
\begin{align*}
	y_{it} &= \theta_{g_{i}}^{\prime} \widecheck{x}_{it} + \varepsilon_{i t} , \quad \varepsilon_{i t} \stackrel{iid}{\sim} N \left(0, \sigma_{g_{i}}^{2}\right),
\end{align*}
where $\widecheck{x}_{it} = [1 \; y_{it-1} \; x'_{it}]' $, and $\theta_{g_{i}} = [\alpha_{g_{i}} \; \rho_{g_{i}} \; \beta_{g_{i}}^{\prime}]'$.

We use the same notation as in Appendix \ref{appendix:algo_RE} aside from the coefficients $\theta$,
\begin{align*}
	\text{Observations: } Y &= [y_1, y_2, ..., y_N ], y_i = [y_{i1},y_{i2},...,y_{iT}]', \\
	\text{Covariates: } X &= [x_1, x_2, ..., x_N ], x_i = [x_{i1},x_{i2},...,x_{iT}]', \\
	\text{Stacked coefficients: }\theta &= \left[\theta_{1}, \theta_{2}, \ldots\right], \\
	\text{Covariance matrices: } \Sigma &= \left[\sigma^2_{1}, \sigma^2_{2}, \ldots\right],\\
	\text{Stick length: } \Xi &= \left[\xi_{1}, \xi_{2}, \ldots\right], \\
	\text{Group membership: } G &= \left[g_{1}, \ldots, g_{N}\right], \\
	\text{Auxiliary varaible: } u &= \left[  u_1, u_2, ..., u_N \right],\\
	\text{Hyper parameters: } \phi &= \left[ \mu_{\theta}, \Sigma_{\theta}, \nu_{\sigma}, \delta_{\sigma}\right ].
\end{align*}

The posterior of unknown objects in the random effects model is,
\begin{align*}
	& p(\theta, \Sigma, \Xi, a, G, u | Y, X) \\
	\propto \; & \left[ \prod_{i=1}^{N} p(y_i | x_i, \theta_{g_{i}}, \sigma^2_{g_{i}}) \mathbf{1} (u_i < \pi_{g_i})\right] \left[ \prod_{j=1}^{\infty}  p(\theta_{j}, \sigma^2_{j} | \phi) p(\xi_j | a)\right] p(\rho) p(a). \numberthis
\end{align*}

The number of potential groups $K^*$ and the number of active groups $K^a$ are defined in the equation (\ref{def:kstar_app}) and (\ref{def:k_a_app}).

\noindent {\bf Conditional posterior of $\theta$ (grouped random effects)}.
\begin{align*}
	p(\theta |\Sigma, \Xi, a, G, u,  Y,X)
	\propto \; & \left[ \prod_{i=1}^{N} p(y_i | x_i, \theta_{g_{i}}, \Sigma_{g_{i}}) \mathbf{1} (u_i < \pi_{g_i}) \right] \left[ \prod_{j=1}^{\infty}  p(\theta_{j}, \sigma^2_{j} | \phi) \right]
\end{align*}

\noindent For $k \in \{1,2,...,K^a\}$, define a set of unit that belongs to group $k$,
\begin{align} \label{def:Ck_app_fullG}
	C_k = \left \{i \in \{1,2,...,N\} | g_i = k \right \},
\end{align}
then the posterior density for $\theta_k$ read as
\begin{align*}
& p(\theta_k|\Sigma, \Xi, a, G, u, Y, X) \\
	\propto \; & \left[ \prod_{i \in C_k} p(y_i | x_i, \theta_k, \sigma^2_k) \right] p(\theta_{k}| \phi) \\
	\propto \; & \exp \left[ - \sum_{i \in C_k} \left( y_i - \widecheck{x}_{i} \theta_{k} \right)'\Sigma_k^{-1} \left( y_i - \widecheck{x}_{i} \theta_{k}  \right) \right] \exp \left[ - \left( \theta_k - \mu_\theta \right)'\Sigma_\theta^{-1} \left( \theta_k - \mu_\theta \right) \right] \\
	\propto \; & \exp \left[ - \left( \theta_k - \bar{\mu}_{\theta_k} \right)' \bar{\Sigma}_\theta^{-1} \left( \theta_k - \bar{\mu}_{\theta_k} \right) \right].
\end{align*}
Assuming an independent normal conjugate prior for $\theta_k$, the posterior for $\theta_k$ is given by
\begin{align} \label{post_alpha_fullG}
	\theta_k |\Sigma, \Xi, a, G, u, Y, X \sim N \left( \bar{\mu}_{\theta_k}, \bar{\Sigma}_{\theta_k} \right),
\end{align}
where
\begin{align*}
	\bar{\Sigma}_{\theta_k} &= \left( \Sigma_\theta^{-1} + \sigma_{k}^{-2} \sum_{i \in C_k} \widecheck{x}_{i} \widecheck{x}'_{i}\right)^{-1}, \\
	\bar{\mu}_{\theta_k} &= \bar{\Sigma}_{\theta_k} \left[\Sigma_\theta^{-1} \mu_\theta + \sigma_{k}^{-2}\sum_{i \in C_k} \widecheck{x}_{i} y_i \right].
\end{align*}
If group $k$ is empty, we draw $\theta_k$ from its prior $ N \left( \mu_{\theta}, \Sigma_{\theta} \right)$.\\

\noindent {\bf Conditional posterior of $\Sigma$ (grouped variance)}.
Under the cross-sectional independence, for $k = 1,2,...,K^a$,
\begin{align*}
	p(\sigma^2_k|\theta, \Xi, a,G, u, Y, X)
	\propto \; & \left[ \prod_{i \in C_k} p(y_i | x_i, \rho, \beta_i, \alpha_k, \sigma^2_k) \right] p(\sigma_{k}^2 | \phi)
\end{align*}
Assuming a inverse-gamma prior $\sigma_{k}^2 \sim IG \left( \frac{v_\sigma}{2}, \frac{\delta_\sigma}{2} \right)$, the posterior distribution of $\sigma_{k}^2$ is
\begin{align*}
	& p(\sigma_{k}^2 |\rho, \beta, \alpha, G, u, Y, X) \\
	\propto \; & \prod_{i \in C_k} \left[ (\sigma_{k}^2)^{-\frac{T}{2}} \exp \left( - \frac{\sum_{t=1}^T (y_{it} - \theta_{k}^{\prime} \widecheck{x}_{it}  )^2}{2\sigma_{k}^2} \right) \right] \left( \frac{1}{\sigma_{k}^2} \right)^{\frac{v_\sigma}{2}+1} \exp \left( - \frac{\delta_\sigma}{2\sigma_{k}^2} \right) \\
	= \; & \left( \frac{1}{\sigma_{k}^2} \right)^{\frac{v_\sigma + T|C_k|}{2}+1} \exp \left( - \frac{\delta_\sigma + \sum_{i \in C_k} \sum_{t=1}^T (y_{it} - \theta_{k}^{\prime} \widecheck{x}_{it} )^2}{2\sigma_{k}^2} \right).
\end{align*}
This implies
\begin{align} \label{post_Sigma_app_fullG}
	\sigma_{k}^2 |\rho, \beta, \alpha, \Xi, a, G, u, Y, X \sim IG \left( \frac{\bar{v}_{\sigma,k}}{2}, \frac{\bar{\delta}_{\sigma,k}}{2} \right),
\end{align}
where
\begin{align*}
	\bar{v}_{\sigma,k} &= v_\sigma + T|C_k|, \\
	\bar{\delta}_{\sigma,kt} &= \delta_\sigma + \sum_{i \in C_k} \sum_{t=1}^T (y_{it} - \theta_{k}^{\prime} \widecheck{x}_{it}  )^2, \\
	|C_k| &= \text{occurrence of } g_i = k.
\end{align*}
If group $k$ is empty, we draw $\sigma_{k}^2$ from its prior $ IG \left( \frac{v_\sigma}{2}, \frac{\delta_\sigma}{2} \right)$. \\

\noindent {\bf Conditional posterior of $\Xi$ (stick length)}. Identical to (\ref{post_Xi_app}).

\noindent {\bf Conditional posterior of $a$ (concentration parameter)}. Identical to (\ref{post_a_app}).

\noindent {\bf Conditional posterior of $u$ (auxiliary variable)}. Identical to (\ref{post_u_app}).

\noindent {\bf Conditional posterior of $G$ (group membership)}. Identical to (\ref{post_G_app}).

\section{Proofs}
\begin{proposition} \label{prop:1}
	Suppose that we have a model with posterior as given in the section \ref{subsec:post}. Given the definition of the number of potential component $K^*$ (eq.(\ref{def:kstar_app})), the minimum of auxiliary variables $u^*$ (eq.(\ref{def:ustar_app})) and the number of active group $K$ (eq.(\ref{def:k_a_app})), we have
	\begin{enumerate}[(i)]
		\item $u_i > \pi_k$ for $\forall i = 1,2,...,n$ and $\forall k > K^*$;
		\item $K < K^*$.
	\end{enumerate}
\end{proposition}
\textit{Proof:}
\begin{enumerate}[(i)]
	\item By definition, $u^* = \min \limits_{1\le i \le N} u_i$ for $i = 1,2,...,n$, then,
	\begin{align*}
	u_i \ge u^* > 1 - \sum_{j=1}^{K^*} \pi_j =  \sum_{j=K^*}^{\infty} \pi_j \ge \pi_k, \forall k > K^*.
	\end{align*}
	\item Let $i'$ be an unit $i$ such that $g_{i'} = K$. According to the posterior of $G$, the group $K$ exists if $u_{i'} < \pi_{K}$, otherwise $p(g_i = K | \cdot) = 0$. Then by definition,
	\begin{align*}
	u^* \le u_{i'} < \pi_{K} \Rightarrow 1 - u^* > 1 - \pi_{K} = \sum_{j = 1}^{K-1} \pi_j.
	\end{align*}
	Since $K^*$ is the smallest number $s.t.$ $1 - u^* < \sum \limits_{j = 1}^{K^*} \pi_j$, then $K \le K^*$.
\end{enumerate}

\section{Convergence Diagnostic} \label{appendix:conv_check}
To assess convergence,  we assess the trace plot, cumulative mean, and auto-correlation of posterior draws for different coefficients. In particular, the data generating process used here is DGP7, where we assume time-varying grouped random effects and homoskedasticity.  We evaluate the most complicated BGRE estimator: Tv-Hetero (time-varying $\alpha_{i}$, heteroskedasticity), and report the convergence diagnostics for $\alpha_{5,1}$, $\sigma^2_{10}$ and $\rho$\footnote{Due to time effects and heteroskedasticity, we randomly present one of the $\alpha$ for unit $i=5$ and in period $t=1$, and the variance of error term $\sigma^2$ for unit $i=10$.}.

\begin{figure}[H]
	\caption{Convergence Diagnostics, $\alpha_{5,1}$ ($i = 5$, $t=1$)}
	\begin{center}
		\includegraphics[scale= 0.5]{{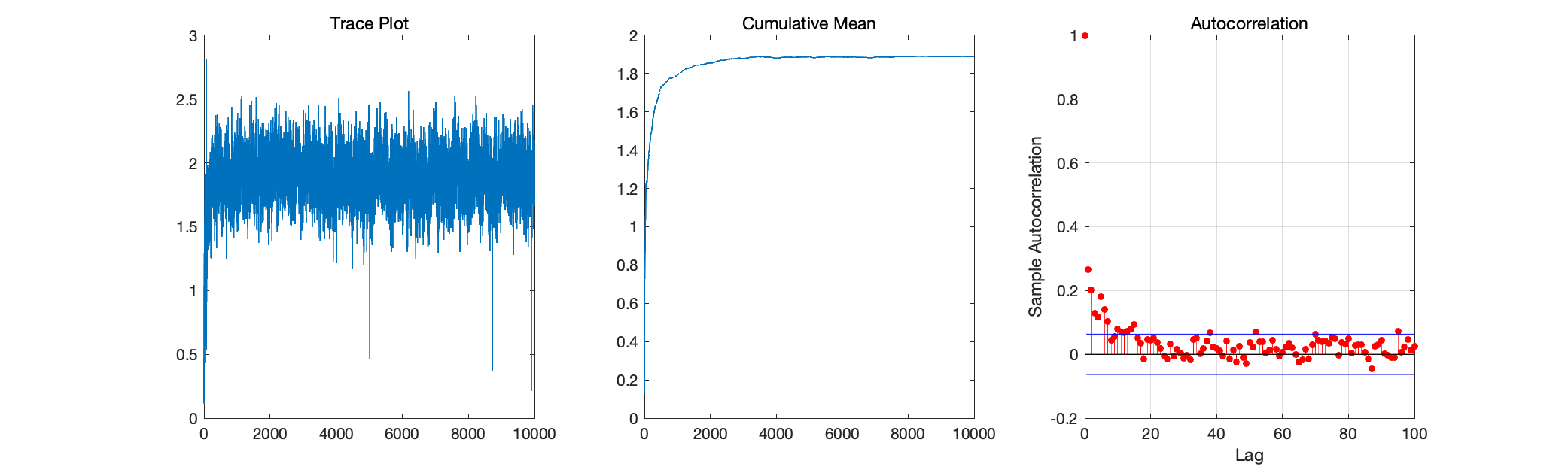}}
	\end{center}
\end{figure}

\begin{figure}[H]
	\caption{Convergence Diagnostics, $\sigma_{10}$, ($i = 10$)}
	\begin{center}
		\includegraphics[scale= 0.5]{{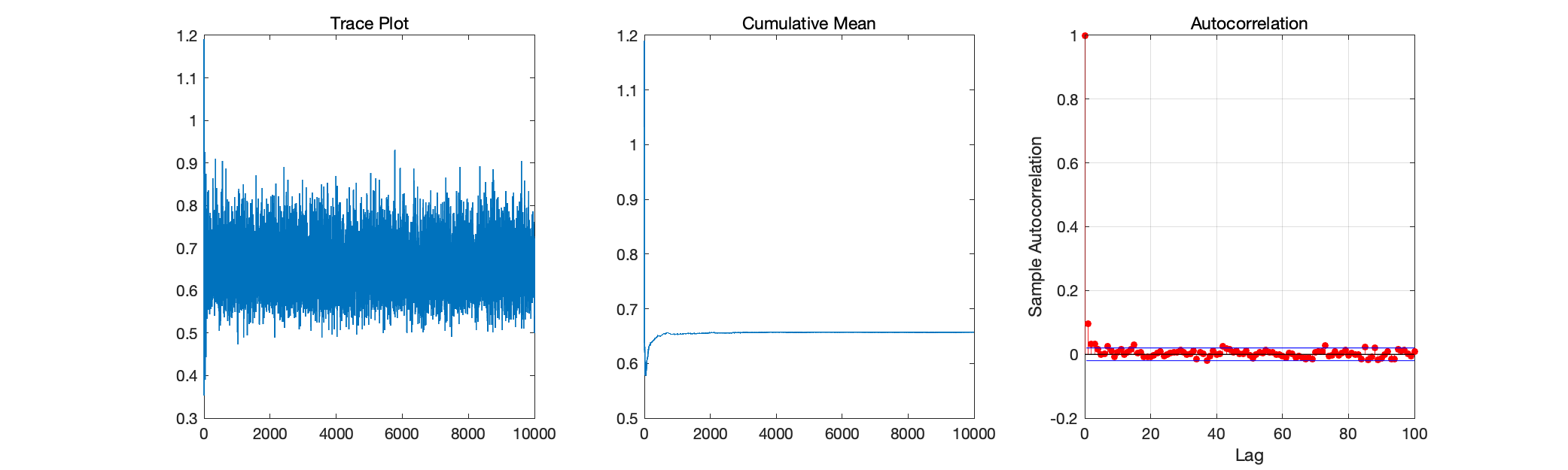}}
	\end{center}
\end{figure}

\begin{figure}[H]
	\caption{Convergence Diagnostics, $\rho$}
	\begin{center}
		\includegraphics[scale= 0.5]{{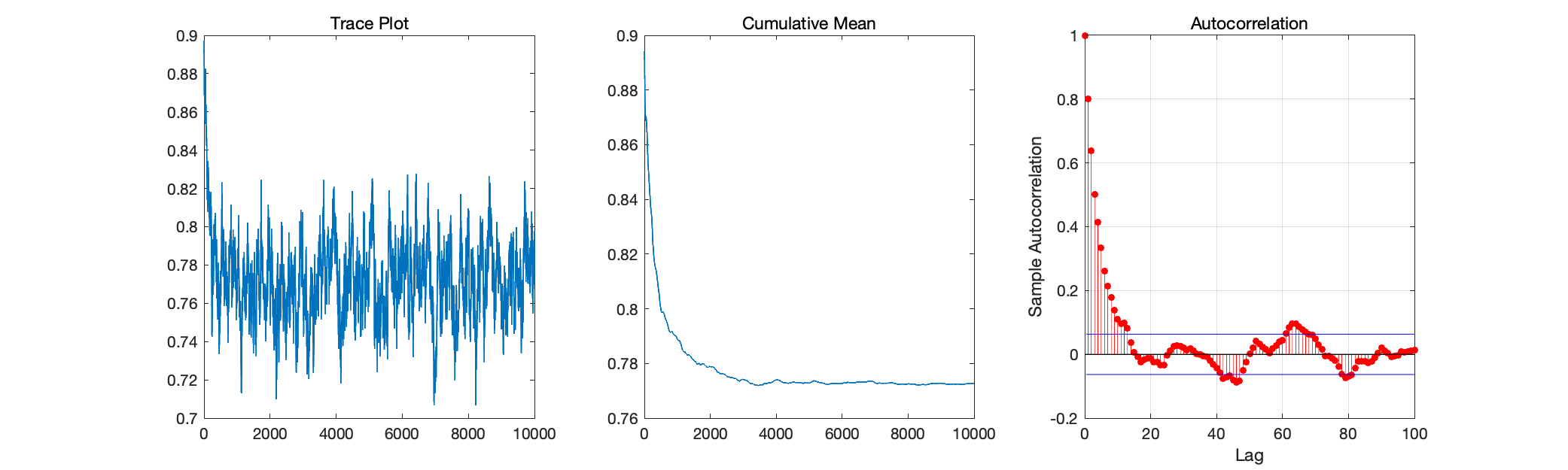}}
	\end{center}
\end{figure}

\section{Computation Details} \label{appendix:com_detail}

\section{Additional Simulation Results} \label{appendix:extra_simul}

\subsection{Main MC Simulation: Larger variance}

In this section, we present the additional simulation results of DGP1, DGP2 DGP3 and DGP4 with larger variance with $\sigma^2 = 1.2^2$. The anther settings remain the same: $N = 100$, $T = 10$ and the true number of groups is $K^0 = 4$.

Table \ref{tab:MC_est_large_sigma} and \ref{tab:MC_fcst_large_sigma} shows the estimate and forecast comparison among alternative predictors. For DGP1 and DGP2, the results are similar to those of smaller variance in the main text: the best models are Ti-Homo and Ti-Hetero, respectively. However, in the DGP3, the Tv-Homo and Tv-Hetero estimators, which are expected to stand out since they correctly model the time effects, don't offer the best performance. The potential reason is that the estimation becomes substantially difficult in the presence of both time-varying random effects and much noisier error terms, making it hard to accurately determine group structure. Regarding the DGP4, Ti-Homo and Ti-Hetero deliver outstanding performance relative to other alternative estimators. As there is no group structure in this DGP, the Flat estimator should be the best, which indeed generate accurate estimates and forecast but Ti-Homo and Ti-Hetero can still stand out. This is mainly because Ti-Homo and Ti-Hetero optimally partition similar units into several groups, which averages out the noisy error terms and, hence, scores great performance. These exciting results also suggest that, though no group structure in the sample, our BGRE estimators have the edge over other estimators who either pool all information (Pooled) or treat each unit separately (Flat).

\begin{table}[H]
	\begin{center}
		\caption{Monte Carlo Experiment: Point Estimates, Larger $\sigma^2$}
		\label{tab:MC_est_large_sigma}
		\begin{tabular}{llrrrrrrc}
			\toprule
			& & \multicolumn{5}{c}{$\hat{\rho}$} & \multicolumn{1}{c}{$\hat{\alpha_i}$} & Cluster \\
			\cmidrule(lr){3-7} \cmidrule(lr){8-8} \cmidrule(lr){ 9-9} \noalign{\smallskip}
			& & \multicolumn{1}{c}{RMSE } & \multicolumn{1}{c}{Bias } & \multicolumn{1}{c}{Std } & \multicolumn{1}{c}{AvgL } & \multicolumn{1}{c}{Cov } & \multicolumn{1}{c}{PBias } & \multicolumn{1}{c}{Avg K} \\
			\midrule
			\multirow{7}{*}{\shortstack{DGP 1 \\ (Grp Ti Ho.)}}
			& Ti-Homo & 0.0336 & 0.0227 & 0.0169 & 0.0658 & 0.70 & -0.1446 & 3.08 \\
			& Ti-Hetero & 0.0342 & 0.0241 & 0.0165 & 0.0643 & 0.63 & -0.1548 & 2.96 \\
			& Tv-Homo & 0.2386 & 0.2363 & 0.0181 & 0.0686 & 0.05 & -1.5126 & 1.30 \\
			& Tv-Hetero & 0.2436 & 0.2419 & 0.0155 & 0.0597 & 0.05 & -1.5488 & 1.41 \\
			& Pooled & 0.2217 & 0.2215 & 0.0088 & 0.0342 & 0 & -1.4113 & 1 \\
			& Flat & 0.0665 & -0.0642 & 0.0164 & 0.0639 & 0.03 & 0.4058 & 100 \\
			& Param & 0.2493 & 0.2205 & 0.1119 & 0.4429 & 0.51 & -1.4000 & 1 \\

			\midrule
			\multirow{7}{*}{\shortstack{DGP 2 \\ (Grp Ti He.)}}
			& Ti-Homo & 0.0146 & 0.0058 & 0.0104 & 0.0404 & 0.94 & -0.0405 & 4.19 \\
			& Ti-Hetero & 0.0084 & 0.0033 & 0.0062 & 0.0242 & 0.91 & -0.0231 & 3.96 \\
			& Tv-Homo & 0.2231 & 0.2215 & 0.0193 & 0.0744 & 0.02 & -1.4385 & 10.59 \\
			& Tv-Hetero & 0.1639 & 0.1611 & 0.0218 & 0.0827 & 0.09 & -1.0595 & 3.08 \\
			& Pooled & 0.2495 & 0.2494 & 0.0063 & 0.0245 & 0 & -1.6040 & 1 \\
			& Flat & 0.0262 & -0.0237 & 0.0105 & 0.0409 & 0.32 & 0.1506 & 100 \\
			& Param & 0.2743 & 0.2480 & 0.1135 & 0.4488 & 0.29 & -1.5926 & 1 \\

			\midrule
			\multirow{7}{*}{\shortstack{DGP 3 \\ (Grp Tv Ho.)}}
			& Ti-Homo & 0.2074 & 0.2064 & 0.0172 & 0.0670 & 0.02 & 0.0052 & 1.03 \\
			& Ti-Hetero & 0.2063 & 0.2054 & 0.0169 & 0.0659 & 0.01 & 0.0050 & 1.03 \\
			& Tv-Homo & 0.2102 & 0.2095 & 0.0168 & 0.0653 & 0 & 0.0052 & 1.03 \\
			& Tv-Hetero & 0.2113 & 0.2106 & 0.0169 & 0.0658 & 0 & 0.0051 & 1.17 \\
			& Pooled & 0.2102 & 0.2096 & 0.0166 & 0.0646 & 0 & 0.0050 & 1 \\
			& Flat & 0.1870 & -0.1849 & 0.0277 & 0.1080 & 0 & -0.0047 & 100 \\
			& Param & 0.2162 & 0.2098 & 0.0505 & 0.2012 & 0.01 & 0.0168 & 1 \\

			\midrule
			\multirow{7}{*}{\shortstack{DGP 4 \\ (Std Ti Ho.)}}
			& Ti-Homo & 0.0145 & 0.0063 & 0.0097 & 0.0376 & 0.88 & -0.0439 & 4.45 \\
			& Ti-Hetero & 0.0148 & 0.0069 & 0.0098 & 0.0384 & 0.89 & -0.0481 & 4.23 \\
			& Tv-Homo & 0.3096 & 0.3093 & 0.0137 & 0.0530 & 0 & -1.9922 & 1.99 \\
			& Tv-Hetero & 0.3108 & 0.3105 & 0.0135 & 0.0522 & 0 & -1.9998 & 2.04 \\
			& Pooled & 0.2529 & 0.2528 & 0.0062 & 0.0240 & 0 & -1.6281 & 1 \\
			& Flat & 0.0256 & -0.0230 & 0.0100 & 0.0388 & 0.37 & 0.1481 & 100 \\
			& Param & 0.2775 & 0.2515 & 0.1161 & 0.4593 & 0.26 & -1.6167 & 1 \\
			\bottomrule
		\end{tabular}
	\end{center}
\end{table}

\begin{table}[H]
	\begin{center}
		\caption{Monte Carlo Experiment: Forecast, Larger $\sigma^2$}
		\label{tab:MC_fcst_large_sigma}
		\begin{tabular}{llrrr|rr|rr}
			\toprule
			& & \multicolumn{3}{c}{Point Forecast}  & \multicolumn{2}{c}{Set Forecast} & \multicolumn{2}{c}{Density Forecast} \\
			\cmidrule(lr){3-5} \cmidrule(lr){6-7} \cmidrule(lr){8-9} \noalign{\smallskip}
			& & \multicolumn{1}{c}{RMSFE } & \multicolumn{1}{c}{Error } & \multicolumn{1}{c}{Std } & \multicolumn{1}{c}{AvgL } & \multicolumn{1}{c}{Cov } & \multicolumn{1}{c}{LPS } & \multicolumn{1}{c}{CRPS }\\
			\midrule
			\multirow{4}{*}{\shortstack{DGP 1 \\ (Grp Ti Ho.)}}
			& Ti-Homo & 1.2281 & 0.0012 & 1.2210 & 4.8430 & 0.95 & -1.6271 & 0.6939 \\
			& Ti-Hetero & 1.2297 & 0.0041 & 1.2224 & 4.8338 & 0.95 & -1.6305 & 0.6952 \\
			& Tv-Homo & 1.2863 & 0.0180 & 1.2724 & 5.1561 & 0.95 & -1.6746 & 0.7266 \\
			& Tv-Hetero & 1.2902 & 0.0193 & 1.2760 & 5.1626 & 0.95 & -1.6776 & 0.7285 \\
			& Pooled & 1.3385 & 0.3476 & 1.2832 & 5.4955 & 0.96 & -1.7161 & 0.7567 \\
			& Flat & 1.2637 & -0.1433 & 1.2494 & 4.8660 & 0.95 & -1.6564 & 0.7147 \\
			& Param & 1.3415 & 0.3499 & 1.2856 & 8.2549 & 1 & -1.8465 & 0.7971 \\

			\midrule
			\multirow{4}{*}{\shortstack{DGP 2 \\ (Grp Ti He.)}}
			& Ti-Homo & 0.6989 & 0.0055 & 0.6949 & 2.7235 & 0.93 & -1.0583 & 0.3819 \\
			& Ti-Hetero & 0.6926 & 0.0024 & 0.6884 & 2.5818 & 0.96 & -0.8640 & 0.3600 \\
			& Tv-Homo & 0.8502 & 0.0101 & 0.8425 & 2.5678 & 0.89 & -1.2358 & 0.4496 \\
			& Tv-Hetero & 0.7313 & 0.0080& 0.7230 & 2.7180 & 0.95 & -0.9323 & 0.3812 \\
			& Pooled & 0.8699 & 0.4283 & 0.7507 & 3.7531 & 0.95 & -1.2926 & 0.4832 \\
			& Flat & 0.7172 & -0.0414 & 0.7120 & 2.8231 & 0.93 & -1.0899 & 0.3935 \\
			& Param & 0.8751 & 0.4284 & 0.7567 & 7.1947 & 1 & -1.5801 & 0.5658 \\

			\midrule
			\multirow{4}{*}{\shortstack{DGP 3 \\ (Grp Tv Ho.)}}
			& Ti-Homo & 1.2684 & -0.0325 & 1.2614 & 5.0355 & 0.95 & -1.6600 & 0.7162 \\
			& Ti-Hetero & 1.2686 & -0.0326 & 1.2616 & 5.0343 & 0.95 & -1.6603 & 0.7164 \\
			& Tv-Homo & 1.2750 & 0.0022 & 1.2616 & 5.0568 & 0.95 & -1.6653 & 0.7200 \\
			& Tv-Hetero & 1.2749 & 0.0026 & 1.2614 & 5.0580 & 0.95 & -1.6653 & 0.7197 \\
			& Pooled & 1.2678 & -0.0328 & 1.2608 & 5.0383 & 0.95 & -1.6596 & 0.7158 \\
			& Flat & 1.2781 & -0.0377 & 1.2711 & 4.7935 & 0.94 & -1.6690 & 0.7235 \\
			& Param & 1.2835 & -0.0214 & 1.2608 & 9.2910 & 1 & -1.8663 & 0.7878 \\

			\midrule
			\multirow{4}{*}{\shortstack{DGP 4 \\ (Std Ti Ho.)}}
			& Ti-Homo & 0.6435 & -0.0112 & 0.6400 & 2.5472 & 0.95 & -0.9807 & 0.3636 \\
			& Ti-Hetero & 0.6436 & -0.0103 & 0.6401 & 2.6020 & 0.95 & -0.9837 & 0.3639 \\
			& Tv-Homo & 0.7069 & 0.0247 & 0.6988 & 2.8327 & 0.95 & -1.0760 & 0.3993 \\
			& Tv-Hetero & 0.7072 & 0.0247 & 0.6991 & 2.8513 & 0.95 & -1.0771 & 0.3994 \\
			& Pooled & 0.8200 & 0.4198 & 0.7011 & 3.5977 & 0.97 & -1.2356 & 0.4660 \\
			& Flat & 0.6720 & -0.0580 & 0.6663 & 2.6339 & 0.95 & -1.0246 & 0.3800 \\
			& Param & 0.8243 & 0.4203 & 0.7059 & 7.0829 & 1 & -1.5529 & 0.5467 \\
			\bottomrule
		\end{tabular}
	\end{center}
\end{table}

\subsection{Main MC Simulation: Shorter Time Periods}

Here, we show the additional simulation results of DGP1, DGP2 and DGP4 with small period, i.e, $T = 5$. The rest settings remain the same: $N = 100$, $\sigma^2 =0.8^2$ and the true number of groups is $K^0 = 4$.

\begin{table}[H]
	\begin{center}
		\caption{Monte Carlo Experiment: Point Estimates, Smaller $T$}
		\label{tab:MC_est_small_T}
		\begin{tabular}{llrrrrrrc}
			\toprule
			& & \multicolumn{5}{c}{$\hat{\rho}$} & \multicolumn{1}{c}{$\hat{\alpha_i}$} & Cluster \\
			\cmidrule(lr){3-7} \cmidrule(lr){8-8} \cmidrule(lr){ 9-9} \noalign{\smallskip}
			& & \multicolumn{1}{c}{RMSE } & \multicolumn{1}{c}{Bias } & \multicolumn{1}{c}{Std } & \multicolumn{1}{c}{AvgL } & \multicolumn{1}{c}{Cov } & \multicolumn{1}{c}{PBias } & \multicolumn{1}{c}{Avg K} \\
			\midrule
			\multirow{7}{*}{\shortstack{DGP 1 \\ (Grp Ti Ho.)}}
			& Ti-Homo & 0.0379 & 0.0276 & 0.0198 & 0.0766 & 0.67 & -0.1414 & 3.16 \\
			& Ti-Hetero & 0.0387 & 0.029 & 0.0198 & 0.0772 & 0.66 & -0.1488 & 3.09 \\
			& Tv-Homo & 0.3577 & 0.3566 & 0.0199 & 0.0777 & 0.02 & -1.8223 & 1.37 \\
			& Tv-Hetero & 0.3654 & 0.3646 & 0.0195 & 0.0760 & 0.01 & -1.8584 & 1.63 \\
			& Pooled & 0.2789 & 0.2785 & 0.0120 & 0.0467 & 0.01 & -1.4050 & 1 \\
			& Flat & 0.0591 & -0.0554 & 0.0190 & 0.0738 & 0.16 & 0.2782 & 100 \\
			& Param & 0.3146 & 0.2783 & 0.1385 & 0.5503 & 0.43 & -1.4006 & 1 \\

			\midrule
			\multirow{7}{*}{\shortstack{DGP 2 \\ (Grp Ti He.)}}
			& Ti-Homo & 0.0523 & 0.0380 & 0.0245 & 0.0952 & 0.66 & -0.189 & 2.91 \\
			& Ti-Hetero & 0.0230 & 0.0126 & 0.0147 & 0.0574 & 0.86 & -0.0639 & 3.40 \\
			& Tv-Homo & 0.3099 & 0.3052 & 0.0288 & 0.1115 & 0.05 & -1.5558 & 5.65 \\
			& Tv-Hetero & 0.2099 & 0.2018 & 0.0377 & 0.1455 & 0.17 & -1.0403 & 2.87 \\
			& Pooled & 0.2613 & 0.2609 & 0.0138 & 0.0537 & 0 & -1.3253 & 1 \\
			& Flat & 0.0833 & -0.0798 & 0.0232 & 0.0902 & 0.04 & 0.4030 & 100 \\
			& Param & 0.2965 & 0.2602 & 0.1365 & 0.5417 & 0.51 & -1.3210 & 1 \\

			\midrule
			\multirow{7}{*}{\shortstack{DGP 4 \\ (Std Ti Ho.)}}
			& Ti-Homo & 0.2345 & 0.2329 & 0.0269 & 0.1051 & 0 & 0.0018 & 1 \\
			& Ti-Hetero & 0.2344 & 0.2328 & 0.0269 & 0.105 & 0 & 0.0017 & 1.01 \\
			& Tv-Homo & 0.2357 & 0.2342 & 0.0270 & 0.1051 & 0 & 0.0016 & 1 \\
			& Tv-Hetero & 0.2363 & 0.2347 & 0.0272 & 0.1062 & 0 & 0.0018 & 1.24 \\
			& Pooled & 0.2344 & 0.2328 & 0.0270 & 0.1051 & 0 & 0.0017 & 1 \\
			& Flat & 0.3601 & -0.3571 & 0.0459 & 0.1790 & 0 & -0.0029 & 100 \\
			& Param & 0.2516 & 0.2323 & 0.0941 & 0.3756 & 0.19 & 0.0066 & 1 \\
			\bottomrule
		\end{tabular}
	\end{center}
\end{table}

\begin{table}[H]
	\begin{center}
		\caption{Monte Carlo Experiment: Forecast, Smaller $T$}
		\label{tab:MC_fcst_small_T}
		\begin{tabular}{llrrr|rr|rr}
			\toprule
			& & \multicolumn{3}{c}{Point Forecast}  & \multicolumn{2}{c}{Set Forecast} & \multicolumn{2}{c}{Density Forecast} \\
			\cmidrule(lr){3-5} \cmidrule(lr){6-7} \cmidrule(lr){8-9} \noalign{\smallskip}
			& & \multicolumn{1}{c}{RMSFE } & \multicolumn{1}{c}{Error } & \multicolumn{1}{c}{Std } & \multicolumn{1}{c}{AvgL } & \multicolumn{1}{c}{Cov } & \multicolumn{1}{c}{LPS } & \multicolumn{1}{c}{CRPS }\\
			\midrule
			\multirow{4}{*}{\shortstack{DGP 1 \\ (Grp Ti Ho.)}}
			& Ti-Homo & 0.8491 & 0.0553 & 0.8407 & 3.3362 & 0.95 & -1.2561 & 0.4798 \\
			& Ti-Hetero & 0.8505 & 0.0585 & 0.8418 & 3.3743 & 0.95 & -1.2605 & 0.4810 \\
			& Tv-Homo & 0.9310 & 0.1221 & 0.9114 & 3.6884 & 0.95 & -1.3514 & 0.5275 \\
			& Tv-Hetero & 0.9352 & 0.1248 & 0.9156 & 3.6947 & 0.95 & -1.3549 & 0.5296 \\
			& Pooled & 1.0817 & 0.6143 & 0.8796 & 4.2897 & 0.96 & -1.4992 & 0.6152 \\
			& Flat & 0.8790 & -0.1242 & 0.8653 & 3.4044 & 0.95 & -1.2946 & 0.4980 \\
			& Param & 1.0852 & 0.6144 & 0.8836 & 7.5513 & 1 & -1.6938 & 0.6666 \\

			\midrule
			\multirow{4}{*}{\shortstack{DGP 2 \\ (Grp Ti He.)}}
			& Ti-Homo & 1.1264 & 0.0956 & 1.1131 & 4.2430 & 0.93 & -1.5308 & 0.6138 \\
			& Ti-Hetero & 1.0782 & 0.0397 & 1.0698 & 3.9245 & 0.95 & -1.3183 & 0.5621 \\
			& Tv-Homo & 1.2929 & 0.1517 & 1.2713 & 4.0156 & 0.89 & -1.6690 & 0.6880 \\
			& Tv-Hetero & 1.1323 & 0.1113 & 1.1129 & 4.1041 & 0.94 & -1.3952 & 0.5948 \\
			& Pooled & 1.2862 & 0.6034 & 1.1257 & 5.0186 & 0.93 & -1.6744 & 0.7084 \\
			& Flat & 1.1320 & -0.166 & 1.1129 & 4.3232 & 0.93 & -1.5477 & 0.6210 \\
			& Param & 1.2889 & 0.6022 & 1.1294 & 7.9859 & 0.99 & -1.8040 & 0.7577 \\

			\midrule
			\multirow{4}{*}{\shortstack{DGP 4 \\ (Std Ti Ho.)}}
			& Ti-Homo & 0.8550 & -0.0081 & 0.8500 & 3.4368 & 0.96 & -1.2665 & 0.4841 \\
			& Ti-Hetero & 0.8550 & -0.0082 & 0.8500 & 3.4362 & 0.96 & -1.2668 & 0.4841 \\
			& Tv-Homo & 0.8607 & -0.026 & 0.8501 & 3.4509 & 0.95 & -1.2733 & 0.4875 \\
			& Tv-Hetero & 0.8620 & -0.0259 & 0.8514 & 3.4513 & 0.95 & -1.2745 & 0.4883 \\
			& Pooled & 0.8550 & -0.0082 & 0.8500 & 3.4362 & 0.95 & -1.2665 & 0.4841 \\
			& Flat & 0.8907 & 0.0036 & 0.886 & 3.2123 & 0.92 & -1.3147 & 0.5052 \\
			& Param & 0.8671 & -0.0032 & 0.8498 & 8.5456 & 1 & -1.6545 & 0.5970 \\
			\bottomrule
		\end{tabular}
	\end{center}
\end{table}

\newpage

\subsection{Main MC Simulation: Different $K^0$}

In this section, we present the simulation results of DGP1, DGP2 and DGP3 with different number of groups. In particular, we consider $K^0 \in \{2,4,6,8\}$. The rest settings remain the same: $N = 100$, $T = 10$, and $\sigma^2 =0.8^2$.

\begin{figure}
	\caption{Monte Carlo Experiment: BGRE Estimator, Different $K^0$}

	\begin{subfigure}{.5\textwidth}
		\centering
		\caption{DGP1, Estimates}
		\includegraphics[width=.8\linewidth]{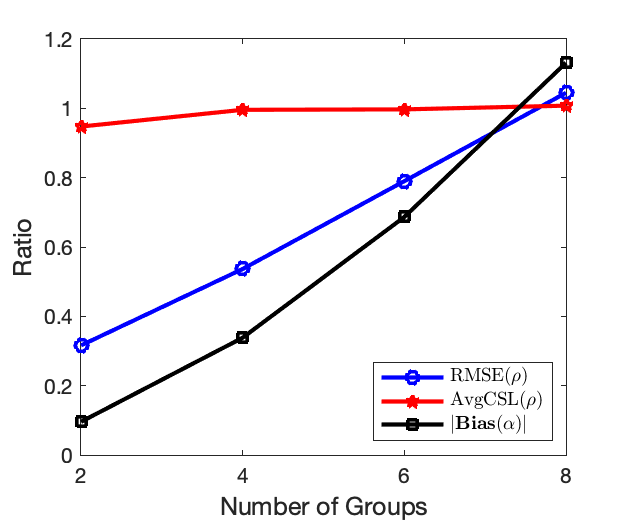}
		\label{fig:sub-first}
	\end{subfigure}
	\begin{subfigure}{.5\textwidth}
		\centering
		\caption{DGP1, Forecasts}
		\includegraphics[width=.8\linewidth]{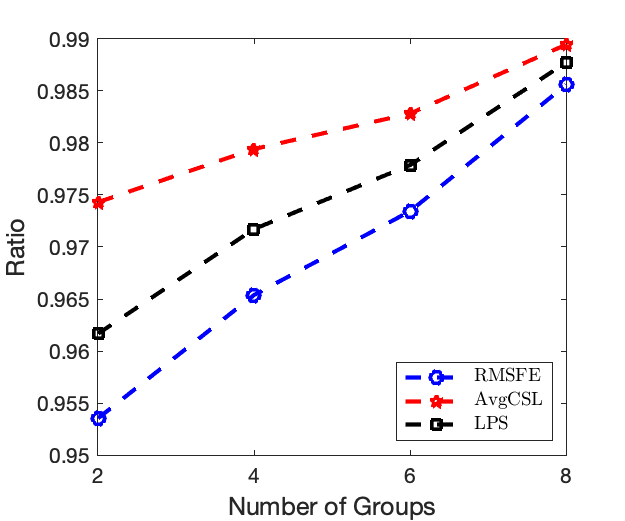}
		\label{fig:sub-second}
	\end{subfigure}

	\vspace{1cm}

	\begin{subfigure}{.5\textwidth}
		\centering
		\caption{DGP2, Estimates}
		\includegraphics[width=.8\linewidth]{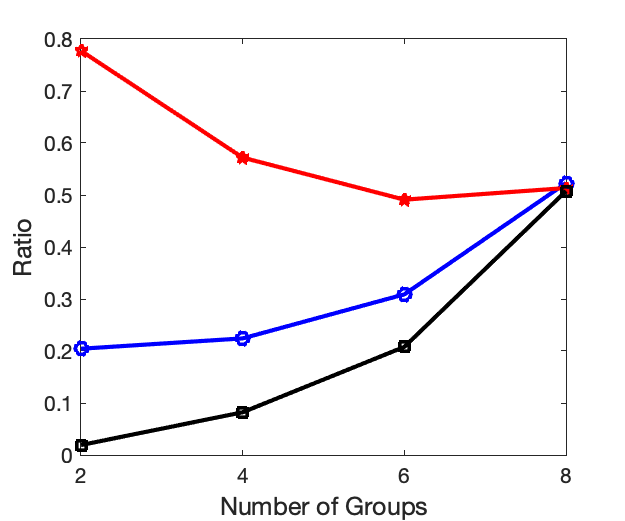}
		\label{fig:sub-third}
	\end{subfigure}
	\begin{subfigure}{.5\textwidth}
		\centering
		\caption{DGP2, Forecasts}
		\includegraphics[width=.8\linewidth]{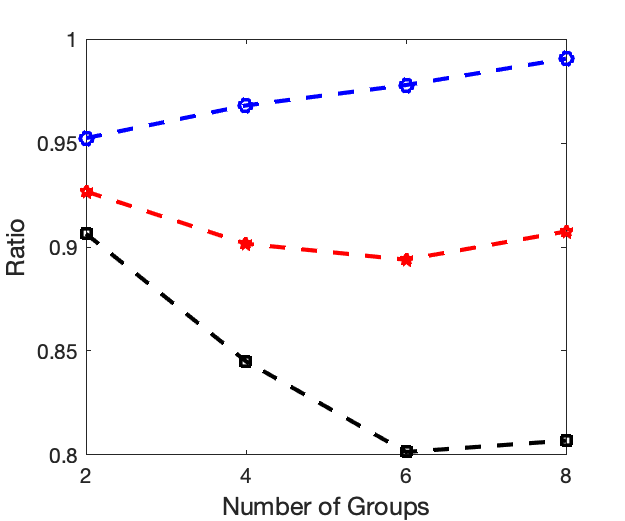}
		\label{fig:sub-fourth}
	\end{subfigure}

	\vspace{1cm}

	\begin{subfigure}{.5\textwidth}
		\centering
		\caption{DGP3, Estimates}
		\includegraphics[width=.8\linewidth]{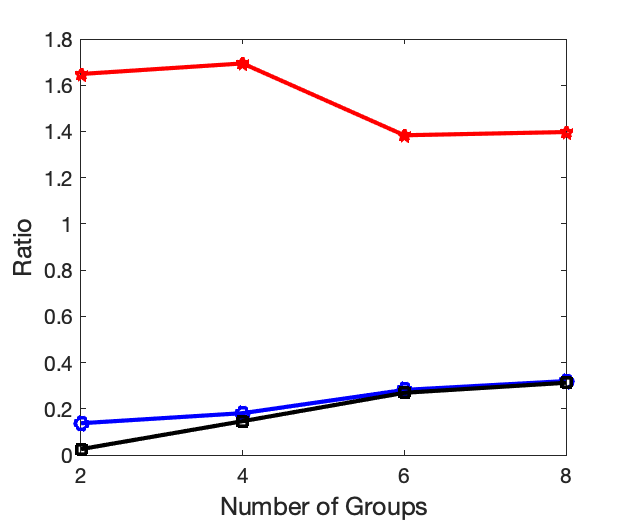}
		\label{fig:sub-fifth}
	\end{subfigure}
	\begin{subfigure}{.5\textwidth}
		\centering
		\caption{DGP3, Forecasts}
		\includegraphics[width=.8\linewidth]{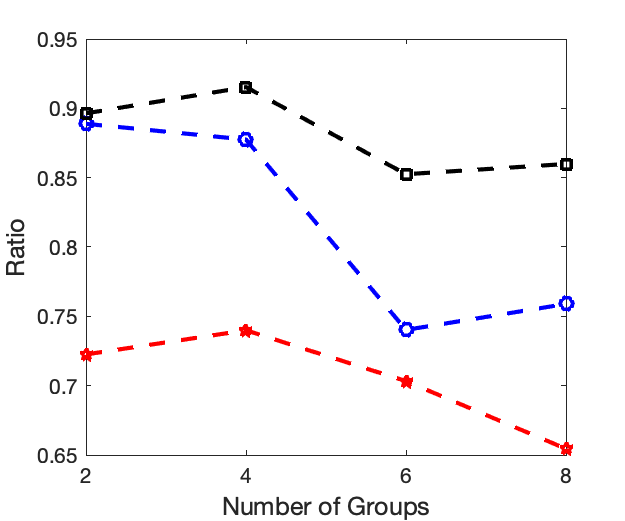}
		\label{fig:sub-sixth}
	\end{subfigure}

	\label{fig:MC_diffK}
\end{figure}

Figure \ref{fig:MC_diffK} presents the relative performance of the BGRE estimators against the flat estimator under different $K^0$. In particular, we show the results of the correctly specified estimators for each DGP, i.e., Ti-Homo estimator for DGP 1, Ti-Hetero estimator for DGP 2, and Tv-Homo estimator for DGP 3. For the DGP 1, the accuracy of the estimates and the predictive power of the BGRE estimator gradually vanish as $K^0$ increases. At $K^0 = 8$, the BGRE estimator still marginally dominates the flat estimator in all aspects besides the bias of $\alpha$. Moving to the DGP 2, the BGRE estimator offers better performance than the flat estimator for all $K^0$. This suggests that the BGRE estimator successfully captures heterogeneity in variance and sophisticated group patterns, even with eight different clusters. Regarding the DPG 3, where we introduce time variation in $\alpha$, the BGRE estimator outperforms the benchmark model in terms of forecasting. Moreover, we see RMSFE, the average length of the credible set, and LPS are all trending down. This suggests the more remarkable improvement in the predictive power of the BGRE estimator, as the true model becomes more sophisticated. It is also noteworthy that, while the average length of the credible set for $\rho$ is relatively large, the BGRE estimator generates a much lower RMSE of $\rho$ and absolute bias of $\alpha$ than the flat estimator.

\begin{table}[H]
	\begin{center}
		\caption{Monte Carlo Experiment: Point Estimates, Different $K^0$, DGP1 (Grp Ti Ho.)}
		\label{tab:MC_est_diffK_DGP1}
		\begin{tabular}{llrrrrrrc}
			\toprule
			& & \multicolumn{5}{c}{$\hat{\rho}$} & \multicolumn{1}{c}{$\hat{\alpha_i}$} & Cluster \\
			\cmidrule(lr){3-7} \cmidrule(lr){8-8} \cmidrule(lr){ 9-9} \noalign{\smallskip}
			& & \multicolumn{1}{c}{RMSE } & \multicolumn{1}{c}{Bias } & \multicolumn{1}{c}{Std } & \multicolumn{1}{c}{AvgL } & \multicolumn{1}{c}{Cov } & \multicolumn{1}{c}{PBias } & \multicolumn{1}{c}{Avg K} \\
			\midrule
			\multirow{7}{*}{$K^0 = 2$}

			& Ti-Homo & 0.0257 & 0.0073 & 0.0172 & 0.0667 & 0.89 & -0.0275 & 1.78 \\
			& Ti-Hetero & 0.0265 & 0.0085 & 0.017 & 0.0658 & 0.89 & -0.0296 & 1.77 \\
			& Tv-Homo & 0.1254 & 0.1160 & 0.0200 & 0.0776 & 0.39 & -0.4375 & 1.30 \\
			& Tv-Hetero & 0.1337 & 0.1240 & 0.0215 & 0.0815 & 0.39 & -0.4593 & 1.33 \\
			& Pooled & 0.1512 & 0.1481 & 0.0104 & 0.0404 & 0.18 & -0.5410 & 1 \\
			& Flat & 0.0813 & -0.0791 & 0.0182 & 0.0704 & 0.01 & 0.2873 & 100 \\
			& Param & 0.2616 & 0.1482 & 0.1886 & 0.7409 & 0.93 & -0.5365 & 1 \\

			\midrule
			\multirow{7}{*}{$K^0 = 4$}

			& Ti-Homo & 0.0198 & 0.0111 & 0.0120 & 0.0465 & 0.77 & -0.0735 & 3.63 \\
			& Ti-Hetero & 0.0200 & 0.0115 & 0.0120 & 0.0466 & 0.81 & -0.0758 & 3.58 \\
			& Tv-Homo & 0.2388 & 0.2371 & 0.0194 & 0.0736 & 0.06 & -1.5154 & 1.87 \\
			& Tv-Hetero & 0.2401 & 0.2385 & 0.0177 & 0.0680 & 0.06 & -1.5275 & 1.93 \\
			& Pooled & 0.2449 & 0.2448 & 0.0069 & 0.0267 & 0 & -1.5591 & 1 \\
			& Flat & 0.0369 & -0.0345 & 0.0121 & 0.0467 & 0.20 & 0.2173 & 100 \\
			& Param & 0.2695 & 0.2438 & 0.1100 & 0.4322 & 0.30 & -1.5545 & 1 \\

			\midrule
			\multirow{7}{*}{$K^0 = 6$}

			& Ti-Homo & 0.0158 & 0.0114 & 0.0086 & 0.0331 & 0.70 & -0.1094 & 5.21 \\
			& Ti-Hetero & 0.0163 & 0.0120 & 0.0086 & 0.0334 & 0.73 & -0.1152 & 5.12 \\
			& Tv-Homo & 0.3086 & 0.3083 & 0.0129 & 0.0497 & 0 & -2.8593 & 2.18 \\
			& Tv-Hetero & 0.3078 & 0.3075 & 0.013 & 0.0502 & 0 & -2.8522 & 2.22 \\
			& Pooled & 0.2708 & 0.2708 & 0.0055 & 0.0214 & 0 & -2.5079 & 1 \\
			& Flat & 0.0200 & -0.0172 & 0.0086 & 0.0332 & 0.48 & 0.1593 & 100 \\
			& Param & 0.2809 & 0.2703 & 0.0749 & 0.2944 & 0.02 & -2.5033 & 1 \\

			\midrule
			\multirow{7}{*}{$K^0 = 8$}

			& Ti-Homo & 0.0140 & 0.0114 & 0.0067 & 0.0261 & 0.63 & -0.1421 & 6.37 \\
			& Ti-Hetero & 0.0154 & 0.0128 & 0.0070 & 0.0269 & 0.60 & -0.1592 & 6.19 \\
			& Tv-Homo & 0.3236 & 0.3234 & 0.0105 & 0.0406 & 0 & -3.8671 & 2.38 \\
			& Tv-Hetero & 0.3247 & 0.3246 & 0.0103 & 0.0396 & 0 & -3.8807 & 2.41 \\
			& Pooled & 0.2797 & 0.2796 & 0.0050 & 0.0192 & 0 & -3.3408 & 1 \\
			& Flat & 0.0134 & -0.0105 & 0.0067 & 0.0259 & 0.64 & 0.1256 & 100 \\
			& Param & 0.2854 & 0.2794 & 0.0576 & 0.2262 & 0.02 & -3.3362 & 1 \\
			\bottomrule
		\end{tabular}
	\end{center}
\end{table}

\begin{table}[H]
	\begin{center}
		\caption{Monte Carlo Experiment: Forecast, Different $K^0$, DGP1 (Grp Ti Ho.)}
		\label{tab:MC_fcst_diffK_DGP1}
		\begin{tabular}{llrrr|rr|rr}
			\toprule
			& & \multicolumn{3}{c}{Point Forecast}  & \multicolumn{2}{c}{Set Forecast} & \multicolumn{2}{c}{Density Forecast} \\
			\cmidrule(lr){3-5} \cmidrule(lr){6-7} \cmidrule(lr){8-9} \noalign{\smallskip}
			& & \multicolumn{1}{c}{RMSFE } & \multicolumn{1}{c}{Error } & \multicolumn{1}{c}{Std } & \multicolumn{1}{c}{AvgL } & \multicolumn{1}{c}{Cov } & \multicolumn{1}{c}{LPS } & \multicolumn{1}{c}{CRPS }\\
			\midrule
			\multirow{7}{*}{$K^0 = 2$}

			& Ti-Homo & 0.8043 & -0.0166 & 0.7997 & 3.1591 & 0.95 & -1.2040 & 0.4546 \\
			& Ti-Hetero & 0.8045 & -0.0158 & 0.8000 & 3.1677 & 0.95 & -1.2049 & 0.4549 \\
			& Tv-Homo & 0.8281 & 0.002 & 0.8194 & 3.2726 & 0.95 & -1.2334 & 0.4681 \\
			& Tv-Hetero & 0.8305 & 0.0025 & 0.8219 & 3.2871 & 0.95 & -1.2364 & 0.4694 \\
			& Pooled & 0.8565 & 0.1246 & 0.8364 & 3.4496 & 0.95 & -1.2673 & 0.4844 \\
			& Flat & 0.8435 & -0.1001 & 0.8335 & 3.2424 & 0.95 & -1.2520 & 0.4771 \\
			& Param & 0.8660 & 0.1286 & 0.8426 & 7.0083 & 1 & -1.5565 & 0.5588 \\

			\midrule
			\multirow{7}{*}{$K^0 = 4$}

			& Ti-Homo & 0.8114 & -0.0075 & 0.8072 & 3.1933 & 0.95 & -1.2131 & 0.4586 \\
			& Ti-Hetero & 0.8124 & -0.0069 & 0.8081 & 3.2155 & 0.95 & -1.2159 & 0.4594 \\
			& Tv-Homo & 0.8638 & 0.0174 & 0.8545 & 3.4290 & 0.95 & -1.2755 & 0.4877 \\
			& Tv-Hetero & 0.8648 & 0.0177 & 0.8555 & 3.4423 & 0.95 & -1.2774 & 0.4884 \\
			& Pooled & 0.9619 & 0.3975 & 0.8685 & 4.0958 & 0.97 & -1.3905 & 0.5456 \\
			& Flat & 0.8405 & -0.0811 & 0.8325 & 3.2604 & 0.95 & -1.2484 & 0.4753 \\
			& Param & 0.9708 & 0.3987 & 0.8770 & 7.3543 & 1 & -1.6365 & 0.6132 \\

			\midrule
			\multirow{7}{*}{$K^0 = 6$}

			& Ti-Homo & 0.8172 & -0.0009 & 0.8130 & 3.2113 & 0.95 & -1.2196 & 0.4619 \\
			& Ti-Hetero & 0.8181 & 0.0006 & 0.8138 & 3.2487 & 0.95 & -1.2222 & 0.4624 \\
			& Tv-Homo & 0.8882 & 0.0349 & 0.8779 & 3.5329 & 0.95 & -1.3042 & 0.5016 \\
			& Tv-Hetero & 0.8885 & 0.0347 & 0.8782 & 3.5424 & 0.95 & -1.3055 & 0.502 \\
			& Pooled & 1.0947 & 0.6514 & 0.8741 & 4.9284 & 0.98 & -1.5306 & 0.6253 \\
			& Flat & 0.8395 & -0.0657 & 0.8329 & 3.2674 & 0.95 & -1.2472 & 0.4747 \\
			& Param & 1.1030 & 0.6533 & 0.8829 & 7.8565 & 1 & -1.7339 & 0.6855 \\

			\midrule
			\multirow{7}{*}{$K^0 = 8$}

			& Ti-Homo & 0.8271 & 0.0064 & 0.8227 & 3.2357 & 0.95 & -1.2315 & 0.4676 \\
			& Ti-Hetero & 0.8314 & 0.0110 & 0.8271 & 3.2960 & 0.95 & -1.2386 & 0.4703 \\
			& Tv-Homo & 0.8983 & 0.0460 & 0.8874 & 3.5889 & 0.95 & -1.3155 & 0.5076 \\
			& Tv-Hetero & 0.8990 & 0.0459 & 0.8881 & 3.6023 & 0.95 & -1.3171 & 0.5080 \\
			& Pooled & 1.2428 & 0.8744 & 0.8783 & 5.8013 & 0.99 & -1.6675 & 0.7153 \\
			& Flat & 0.8392 & -0.0567 & 0.8332 & 3.2702 & 0.95 & -1.2468 & 0.4745 \\
			& Param & 1.2505 & 0.8765 & 0.8872 & 8.4358 & 1 & -1.8330 & 0.7674 \\
			\bottomrule
		\end{tabular}
	\end{center}
\end{table}

\begin{table}[H]
	\begin{center}
		\caption{Monte Carlo Experiment: Point Estimates, Different $K^0$, DGP2 (Grp Ti He.)}
		\label{tab:MC_est_diffK_DGP2}
		\begin{tabular}{llrrrrrrc}
			\toprule
			& & \multicolumn{5}{c}{$\hat{\rho}$} & \multicolumn{1}{c}{$\hat{\alpha_i}$} & Cluster \\
			\cmidrule(lr){3-7} \cmidrule(lr){8-8} \cmidrule(lr){ 9-9} \noalign{\smallskip}
			& & \multicolumn{1}{c}{RMSE } & \multicolumn{1}{c}{Bias } & \multicolumn{1}{c}{Std } & \multicolumn{1}{c}{AvgL } & \multicolumn{1}{c}{Cov } & \multicolumn{1}{c}{PBias } & \multicolumn{1}{c}{Avg K} \\
			\midrule
			\multirow{7}{*}{$K^0 = 2$}

			& Ti-Homo & 0.0313 & 0.0021 & 0.0209 & 0.0814 & 0.88 & -0.0132 & 1.78 \\
			& Ti-Hetero & 0.0235 & 0.0014 & 0.0171 & 0.0667 & 0.92 & -0.0080 & 2.01 \\
			& Tv-Homo & 0.0501 & 0.0177 & 0.0300 & 0.1160 & 0.84 & -0.0776 & 8.44 \\
			& Tv-Hetero & 0.0351 & 0.0052 & 0.0249 & 0.0961 & 0.88 & -0.0242 & 2.34 \\
			& Pooled & 0.1214 & 0.1152 & 0.0128 & 0.0498 & 0.21 & -0.4354 & 1 \\
			& Flat & 0.1150 & -0.1127 & 0.0221 & 0.0858 & 0 & 0.4181 & 100 \\
			& Param & 0.2265 & 0.1143 & 0.1716 & 0.6747 & 0.94 & -0.4306 & 1 \\

			\midrule
			\multirow{7}{*}{$K^0 = 4$}

			& Ti-Homo & 0.0221 & 0.0091 & 0.0151 & 0.0586 & 0.87 & -0.0639 & 3.81 \\
			& Ti-Hetero & 0.0111 & 0.0034 & 0.0083 & 0.0322 & 0.93 & -0.0246 & 4.01 \\
			& Tv-Homo & 0.1927 & 0.1896 & 0.0215 & 0.0822 & 0.11 & -1.2371 & 11.51 \\
			& Tv-Hetero & 0.0950 & 0.0878 & 0.0244 & 0.0926 & 0.33 & -0.5819 & 3.42 \\
			& Pooled & 0.2319 & 0.2317 & 0.0079 & 0.0308 & 0 & -1.4910 & 1 \\
			& Flat & 0.0495 & -0.0471 & 0.0145 & 0.0563 & 0.07 & 0.2993 & 100 \\
			& Param & 0.2566 & 0.2310 & 0.1073 & 0.4215 & 0.3 & -1.4863 & 1 \\

			\midrule
			\multirow{7}{*}{$K^0 = 6$}

			& Ti-Homo & 0.0159 & 0.0098 & 0.0102 & 0.0398 & 0.85 & -0.0984 & 5.57 \\
			& Ti-Hetero & 0.0076 & 0.0041 & 0.0049 & 0.0191 & 0.89 & -0.0423 & 5.20 \\
			& Tv-Homo & 0.2855 & 0.2849 & 0.0154 & 0.0587 & 0 & -2.6521 & 14.68 \\
			& Tv-Hetero & 0.2374 & 0.2365 & 0.0178 & 0.0660 & 0 & -2.2017 & 3.51 \\
			& Pooled & 0.2644 & 0.2643 & 0.0060 & 0.0233 & 0 & -2.4582 & 1 \\
			& Flat & 0.0246 & -0.0222 & 0.0100 & 0.0389 & 0.38 & 0.2034 & 100 \\
			& Param & 0.2749 & 0.2641 & 0.0747 & 0.2937 & 0.03 & -2.4535 & 1 \\

			\midrule
			\multirow{7}{*}{$K^0 = 8$}

			& Ti-Homo & 0.0155 & 0.0122 & 0.0078 & 0.0302 & 0.68 & -0.1545 & 6.77 \\
			& Ti-Hetero & 0.0080 & 0.006 & 0.004 & 0.0153 & 0.78 & -0.0774 & 6.24 \\
			& Tv-Homo & 0.3215 & 0.3213 & 0.0119 & 0.0452 & 0 & -3.8562 & 15.33 \\
			& Tv-Hetero & 0.2990 & 0.2986 & 0.0128 & 0.0471 & 0 & -3.5879 & 3.60 \\
			& Pooled & 0.2772 & 0.2771 & 0.0052 & 0.0202 & 0 & -3.3266 & 1 \\
			& Flat & 0.0153 & -0.0129 & 0.0077 & 0.0298 & 0.63 & 0.1525 & 100 \\
			& Param & 0.2830 & 0.2769 & 0.0574 & 0.2254 & 0.02 & -3.3220 & 1 \\
			\bottomrule
		\end{tabular}
	\end{center}
\end{table}

\begin{table}[H]
	\begin{center}
		\caption{Monte Carlo Experiment: Forecast, Different $K^0$, DGP2 (Grp Ti He.)}
		\label{tab:MC_fcst_diffK_DGP2}
		\begin{tabular}{llrrr|rr|rr}
			\toprule
			& & \multicolumn{3}{c}{Point Forecast}  & \multicolumn{2}{c}{Set Forecast} & \multicolumn{2}{c}{Density Forecast} \\
			\cmidrule(lr){3-5} \cmidrule(lr){6-7} \cmidrule(lr){8-9} \noalign{\smallskip}
			& & \multicolumn{1}{c}{RMSFE } & \multicolumn{1}{c}{Error } & \multicolumn{1}{c}{Std } & \multicolumn{1}{c}{AvgL } & \multicolumn{1}{c}{Cov } & \multicolumn{1}{c}{LPS } & \multicolumn{1}{c}{CRPS }\\
			\midrule
			\multirow{7}{*}{$K^0 = 2$}

			& Ti-Homo & 1.2014 & 0.0055 & 1.1945 & 4.6638 & 0.94 & -1.6015 & 0.665 \\
			& Ti-Hetero & 1.1853 & 0.0042 & 1.1784 & 4.4307 & 0.95 & -1.4876 & 0.6381 \\
			& Tv-Homo & 1.3162 & 0.0050 & 1.3040 & 4.3151 & 0.91 & -1.6805 & 0.7173 \\
			& Tv-Hetero & 1.2017 & 0.0050 & 1.1883 & 4.4625 & 0.95 & -1.4999 & 0.6464 \\
			& Pooled & 1.2402 & 0.1183 & 1.2255 & 4.8993 & 0.94 & -1.6374 & 0.6879 \\
			& Flat & 1.2446 & -0.1098 & 1.2331 & 4.7811 & 0.93 & -1.6413 & 0.6909 \\
			& Param & 1.2481 & 0.1223 & 1.2325 & 7.8794 & 0.99 & -1.7799 & 0.7384 \\

			\midrule
			\multirow{7}{*}{$K^0 = 4$}

			& Ti-Homo & 1.0599 & 0.0110 & 1.0537 & 4.0879 & 0.93 & -1.4743 & 0.5798 \\
			& Ti-Hetero & 1.0427 & 0.0020 & 1.0364 & 3.7691 & 0.95 & -1.2646 & 0.5416 \\
			& Tv-Homo & 1.2697 & 0.0075 & 1.2586 & 3.7160 & 0.88 & -1.6528 & 0.6720 \\
			& Tv-Hetero & 1.0809 & 0.0022 & 1.0691 & 3.8909 & 0.94 & -1.3088 & 0.5624 \\
			& Pooled & 1.1839 & 0.3976 & 1.1072 & 4.8767 & 0.94 & -1.5952 & 0.6518 \\
			& Flat & 1.0771 & -0.0816 & 1.068 & 4.1807 & 0.93 & -1.4965 & 0.5906 \\
			& Param & 1.1914 & 0.3991 & 1.1148 & 7.8182 & 0.99 & -1.7574 & 0.7111 \\

			\midrule
			\multirow{7}{*}{$K^0 = 6$}

			& Ti-Homo & 0.9914 & 0.0155 & 0.9860 & 3.8175 & 0.93 & -1.4066 & 0.5383 \\
			& Ti-Hetero & 0.9813 & 0.0038 & 0.9759 & 3.4921 & 0.96 & -1.1432 & 0.4979 \\
			& Tv-Homo & 1.2778 & 0.0185 & 1.2680 & 3.3939 & 0.86 & -1.7016 & 0.6664 \\
			& Tv-Hetero & 1.0441 & 0.0185 & 1.0324 & 3.7406 & 0.95 & -1.2183 & 0.5310 \\
			& Pooled & 1.2388 & 0.6548 & 1.0457 & 5.3733 & 0.95 & -1.6475 & 0.6875 \\
			& Flat & 1.0035 & -0.0574 & 0.9966 & 3.9066 & 0.93 & -1.4264 & 0.5462 \\
			& Param & 1.2472 & 0.6565 & 1.0545 & 8.1351 & 0.99 & -1.8042 & 0.7455 \\

			\midrule
			\multirow{7}{*}{$K^0 = 8$}

			& Ti-Homo & 0.9721 & 0.0279 & 0.9659 & 3.7423 & 0.93 & -1.3865 & 0.5286 \\
			& Ti-Hetero & 0.9608 & 0.0115 & 0.9552 & 3.4452 & 0.96 & -1.1238 & 0.4876 \\
			& Tv-Homo & 1.2746 & 0.0282 & 1.2644 & 3.3205 & 0.85 & -1.7257 & 0.6624 \\
			& Tv-Hetero & 1.0348 & 0.0308 & 1.0223 & 3.7104 & 0.95 & -1.1883 & 0.5218 \\
			& Pooled & 1.3526 & 0.8862 & 1.0174 & 6.1307 & 0.96 & -1.7452 & 0.7616 \\
			& Flat & 0.9700 & -0.0444 & 0.9636 & 3.7966 & 0.93 & -1.3927 & 0.5271 \\
			& Param & 1.3601 & 0.8870 & 1.0264 & 8.6643 & 0.99 & -1.8832 & 0.8133 \\
			\bottomrule
		\end{tabular}
	\end{center}
\end{table}

\begin{table}[H]
	\begin{center}
		\caption{Monte Carlo Experiment: Point Estimates, Different $K^0$, DGP3 (Grp Tv Ho.)}
		\label{tab:MC_est_diffK_DGP3}
		\begin{tabular}{llrrrrrrc}
			\toprule
			& & \multicolumn{5}{c}{$\hat{\rho}$} & \multicolumn{1}{c}{$\hat{\alpha_i}$} & Cluster \\
			\cmidrule(lr){3-7} \cmidrule(lr){8-8} \cmidrule(lr){ 9-9} \noalign{\smallskip}
			& & \multicolumn{1}{c}{RMSE } & \multicolumn{1}{c}{Bias } & \multicolumn{1}{c}{Std } & \multicolumn{1}{c}{AvgL } & \multicolumn{1}{c}{Cov } & \multicolumn{1}{c}{PBias } & \multicolumn{1}{c}{Avg K} \\
			\midrule
			\multirow{7}{*}{$K^0 = 2$}

			& Ti-Homo & 0.2173 & 0.2169 & 0.0131 & 0.0509 & 0 & -2.0461 & 1 \\
			& Ti-Hetero & 0.2112 & 0.2108 & 0.0137 & 0.0529 & 0 & -2.0054 & 1.69 \\
			& Tv-Homo & 0.0321 & 0.0082 & 0.0234 & 0.091 & 0.95 & -0.0548 & 2.01 \\
			& Tv-Hetero & 0.0325 & 0.0079 & 0.0233 & 0.0906 & 0.93 & -0.0532 & 2.07 \\
			& Pooled & 0.2171 & 0.2168 & 0.0097 & 0.0377 & 0 & -2.0455 & 1 \\
			& Flat & 0.2337 & 0.2333 & 0.0142 & 0.0552 & 0 & -2.1542 & 100 \\
			& Param & 0.2472 & 0.2163 & 0.1176 & 0.4618 & 0.48 & -2.0408 & 1 \\

			\midrule
			\multirow{7}{*}{$K^0 = 4$}

			& Ti-Homo & 0.2741 & 0.2739 & 0.0119 & 0.0461 & 0 & -2.2118 & 2.03 \\
			& Ti-Hetero & 0.2750 & 0.2747 & 0.0120 & 0.0467 & 0 & -2.2170 & 2.48 \\
			& Tv-Homo & 0.0587 & 0.0536 & 0.0217 & 0.0837 & 0.33 & -0.3749 & 3.94 \\
			& Tv-Hetero & 0.0605 & 0.0555 & 0.0219 & 0.0850 & 0.29 & -0.3882 & 3.88 \\
			& Pooled & 0.1929 & 0.1927 & 0.0081 & 0.0313 & 0 & -1.6492 & 1 \\
			& Flat & 0.3242 & 0.3239 & 0.0127 & 0.0494 & 0 & -2.5618 & 100 \\
			& Param & 0.2167 & 0.1922 & 0.0983 & 0.3865 & 0.39 & -1.6445 & 1 \\

			\midrule
			\multirow{7}{*}{$K^0 = 6$}

			& Ti-Homo & 0.2897 & 0.2892 & 0.0158 & 0.0604 & 0 & -2.1634 & 1.89 \\
			& Ti-Hetero & 0.2984 & 0.2981 & 0.0137 & 0.0531 & 0 & -2.2277 & 2.08 \\
			& Tv-Homo & 0.0995 & 0.0971 & 0.0212 & 0.0819 & 0.01 & -0.7081 & 4.19 \\
			& Tv-Hetero & 0.1003 & 0.0979 & 0.0211 & 0.0817 & 0 & -0.7140 & 4.19 \\
			& Pooled & 0.2012 & 0.2010 & 0.0078 & 0.0302 & 0 & -1.5241 & 1 \\
			& Flat & 0.3534 & 0.3531 & 0.0152 & 0.0592 & 0 & -2.6306 & 100 \\
			& Param & 0.2208 & 0.2006 & 0.0907 & 0.3564 & 0.2 & -1.5195 & 1 \\

			\midrule
			\multirow{7}{*}{$K^0 = 8$}

			& Ti-Homo & 0.2666 & 0.2660 & 0.0171 & 0.0664 & 0 & -2.0004 & 1.93 \\
			& Ti-Hetero & 0.2834 & 0.2829 & 0.0157 & 0.0609 & 0 & -2.1269 & 2.02 \\
			& Tv-Homo & 0.1049 & 0.1023 & 0.0224 & 0.0865 & 0 & -0.7698 & 5.96 \\
			& Tv-Hetero & 0.1091 & 0.1066 & 0.0225 & 0.0858 & 0.01 & -0.8016 & 5.78 \\
			& Pooled & 0.2021 & 0.2019 & 0.0088 & 0.0342 & 0 & -1.5209 & 1 \\
			& Flat & 0.3279 & 0.3276 & 0.0159 & 0.0619 & 0 & -2.4628 & 100 \\
			& Param & 0.2228 & 0.2015 & 0.0935 & 0.3673 & 0.28 & -1.5162 & 1 \\
			\bottomrule
		\end{tabular}
	\end{center}
\end{table}

\begin{table}[H]
	\begin{center}
		\caption{Monte Carlo Experiment: Forecast, Different $K^0$, DGP3 (Grp Tv Ho.)}
		\label{tab:MC_fcst_diffK_DGP3}
		\begin{tabular}{llrrr|rr|rr}
			\toprule
			& & \multicolumn{3}{c}{Point Forecast}  & \multicolumn{2}{c}{Set Forecast} & \multicolumn{2}{c}{Density Forecast} \\
			\cmidrule(lr){3-5} \cmidrule(lr){6-7} \cmidrule(lr){8-9} \noalign{\smallskip}
			& & \multicolumn{1}{c}{RMSFE } & \multicolumn{1}{c}{Error } & \multicolumn{1}{c}{Std } & \multicolumn{1}{c}{AvgL } & \multicolumn{1}{c}{Cov } & \multicolumn{1}{c}{LPS } & \multicolumn{1}{c}{CRPS }\\
			\midrule
			\multirow{7}{*}{$K^0 = 2$}

			& Ti-Homo & 1.0475 & 0.1611 & 1.0286 & 4.8735 & 0.98 & -1.4966 & 0.5988 \\
			& Ti-Hetero & 1.0294 & 0.1392 & 1.0138 & 4.8565 & 0.98 & -1.4885 & 0.5900 \\
			& Tv-Homo & 0.9507 & 0.0186 & 0.9414 & 3.7414 & 0.95 & -1.3715 & 0.5381 \\
			& Tv-Hetero & 0.9515 & 0.0187 & 0.9422 & 3.7428 & 0.95 & -1.3733 & 0.5387 \\
			& Pooled & 1.0476 & 0.1609 & 1.0287 & 4.8783 & 0.98 & -1.4967 & 0.5991 \\
			& Flat & 1.0696 & 0.2195 & 1.0402 & 5.1782 & 0.98 & -1.5300 & 0.614 \\
			& Param & 1.0531 & 0.1601 & 1.031 & 7.6736 & 1 & -1.7061 & 0.6607 \\

			\midrule
			\multirow{7}{*}{$K^0 = 4$}

			& Ti-Homo & 1.3729 & 0.3097 & 1.3328 & 5.1851 & 0.94 & -1.7468 & 0.7851 \\
			& Ti-Hetero & 1.3553 & 0.3124 & 1.3137 & 5.1330 & 0.94 & -1.7239 & 0.7724 \\
			& Tv-Homo & 1.0879 & 0.0254 & 1.0792 & 3.9212 & 0.93 & -1.5055 & 0.6150 \\
			& Tv-Hetero & 1.0936 & 0.0257 & 1.0850 & 3.9377 & 0.93 & -1.5109 & 0.6182 \\
			& Pooled & 1.8946 & 0.1243 & 1.8874 & 5.5632 & 0.84 & -2.1561 & 1.1083 \\
			& Flat & 1.2398 & 0.4209 & 1.1606 & 5.2996 & 0.97 & -1.6448 & 0.7048 \\
			& Param & 1.8974 & 0.1246 & 1.8902 & 8.2086 & 0.98 & -2.0727 & 1.0854 \\

			\midrule
			\multirow{7}{*}{$K^0 = 6$}

			& Ti-Homo & 1.9244 & 0.4059 & 1.8756 & 5.6825 & 0.85 & -2.1845 & 1.1369 \\
			& Ti-Hetero & 1.8699 & 0.4200 & 1.8176 & 5.6099 & 0.87 & -2.1166 & 1.0933 \\
			& Tv-Homo & 1.2273 & 0.0505 & 1.2193 & 4.1107 & 0.90 & -1.6523 & 0.7007 \\
			& Tv-Hetero & 1.2294 & 0.0514 & 1.2213 & 4.1175 & 0.90 & -1.6512 & 0.7014 \\
			& Pooled & 2.5171 & 0.2716 & 2.5003 & 5.9269 & 0.71 & -2.7056 & 1.5662 \\
			& Flat & 1.6580 & 0.4999 & 1.5765 & 5.8450 & 0.92 & -1.9382 & 0.9455 \\
			& Param & 2.5192 & 0.2731 & 2.502 & 8.4685 & 0.93 & -2.3940 & 1.4845 \\

			\midrule
			\multirow{7}{*}{$K^0 = 8$}

			& Ti-Homo & 1.7915 & 0.1073 & 1.7850 & 6.1019 & 0.90 & -2.0260 & 1.0236 \\
			& Ti-Hetero & 1.7466 & 0.1157 & 1.7395 & 6.0285 & 0.91 & -2.0106 & 0.9988 \\
			& Tv-Homo & 1.1879 & 0.0152 & 1.1804 & 4.0868 & 0.91 & -1.6074 & 0.6746 \\
			& Tv-Hetero & 1.2079 & 0.0166 & 1.2005 & 4.1259 & 0.91 & -1.6221 & 0.6861 \\
			& Pooled & 2.2360 & 0.0755 & 2.2323 & 6.3016 & 0.83 & -2.3547 & 1.3270 \\
			& Flat & 1.5653 & 0.1355 & 1.5556 & 6.2445 & 0.95 & -1.8694 & 0.8838 \\
			& Param & 2.2393 & 0.0772 & 2.2350 & 8.7506 & 0.96 & -2.2403 & 1.2913 \\
			\bottomrule
		\end{tabular}
	\end{center}
\end{table}

\subsection{Two-Step GRE estimator} \label{appendix:2step_full_table}

\begin{table}[H]
	\begin{center}
		\caption{Monte Carlo Experiment: Two-Step GRE with Kmean, Point Estimates}
		\begin{tabular}{llrrrrrrc}
			\toprule
			& & \multicolumn{5}{c}{$\hat{\rho}$} & \multicolumn{1}{c}{$\hat{\alpha_i}$} & Cluster \\
			\cmidrule(lr){3-7} \cmidrule(lr){8-8} \cmidrule(lr){ 9-9} \noalign{\smallskip}
			& & \multicolumn{1}{c}{RMSE } & \multicolumn{1}{c}{Bias } & \multicolumn{1}{c}{Std } & \multicolumn{1}{c}{AvgL } & \multicolumn{1}{c}{Cov } & \multicolumn{1}{c}{PBias } & \multicolumn{1}{c}{Avg K} \\
			\midrule
			\multirow{4}{*}{\shortstack{DGP 1 \\ (Grp Ti Ho.)}}
			& Ti-Homo & 0.0627 & 0.0599 & 0.0114 & 0.0444 & 0.25 & -0.3836 & 2.2 \\
			& Ti-Hetero & 0.0611 & 0.0583 & 0.0116 & 0.0452 & 0.27 & -0.3734 & 2.2 \\
			& Tv-Homo & 0.1567 & 0.1544 & 0.0186 & 0.0721 & 0.11 & -0.9927 & 2.2 \\
			& Tv-Hetero & 0.1560 & 0.1537 & 0.0189 & 0.0738 & 0.10 & -0.9887 & 2.2 \\

			\midrule
			\multirow{4}{*}{\shortstack{DGP 2 \\ (Grp Ti He.)}}
			& Ti-Homo & 0.0550 & 0.0513 & 0.0133 & 0.0517 & 0.27 & -0.3388 & 2.2 \\
			& Ti-Hetero & 0.0456 & 0.0429 & 0.0099 & 0.0386 & 0.27 & -0.2822 & 2.2 \\
			& Tv-Homo & 0.1196 & 0.1143 & 0.0203 & 0.0789 & 0.21 & -0.7572 & 2.2 \\
			& Tv-Hetero & 0.1458 & 0.1427 & 0.0196 & 0.0764 & 0.15 & -0.9400 & 2.2 \\

			\midrule
			\multirow{4}{*}{\shortstack{DGP 3 \\ (Grp Tv Ho.)}}
			& Ti-Homo & 0.2863 & 0.2861 & 0.0114 & 0.0446 & 0.00 & -2.2885 & 2.26 \\
			& Ti-Hetero & 0.2807 & 0.2804 & 0.0115 & 0.0448 & 0.00 & -2.2489 & 2.26 \\
			& Tv-Homo & 0.1196 & 0.1171 & 0.0166 & 0.0648 & 0.08 & -0.8168 & 2.26 \\
			& Tv-Hetero & 0.1172 & 0.1144 & 0.017 & 0.0661 & 0.08 & -0.7982 & 2.26 \\

			\midrule
			\multirow{4}{*}{\shortstack{DGP 4 \\ (Std Ti Ho.)}}
			& Ti-Homo & 0.0832 & 0.0796 & 0.0210 & 0.0819 & 0.10 & 0.0014 & 2.03 \\
			& Ti-Hetero & 0.0833 & 0.0797 & 0.0210 & 0.0819 & 0.08 & 0.0015 & 2.03 \\
			& Tv-Homo & 0.0942 & 0.0908 & 0.0218 & 0.0851 & 0.05 & 0.0018 & 2.03 \\
			& Tv-Hetero & 0.0943 & 0.0908 & 0.0218 & 0.0851 & 0.06 & 0.0019 & 2.03 \\
			\bottomrule
		\end{tabular}
	\end{center}
\end{table}

\begin{table}[H]
	\begin{center}
		\caption{Monte Carlo Experiment: Two-Step GRE with Kmean, Forecast}
		\begin{tabular}{llrrr|rr|rr}
			\toprule
			& & \multicolumn{3}{c}{Point Forecast}  & \multicolumn{2}{c}{Set Forecast} & \multicolumn{2}{c}{Density Forecast} \\
			\cmidrule(lr){3-5} \cmidrule(lr){6-7} \cmidrule(lr){8-9} \noalign{\smallskip}
			& & \multicolumn{1}{c}{RMSFE } & \multicolumn{1}{c}{Error } & \multicolumn{1}{c}{Std } & \multicolumn{1}{c}{AvgL } & \multicolumn{1}{c}{Cov } & \multicolumn{1}{c}{LPS } & \multicolumn{1}{c}{CRPS }\\
			\midrule
			\multirow{4}{*}{\shortstack{DGP 1 \\ (Grp Ti Ho.)}}
			& Ti-Homo & 0.8607 & 0.0777 & 0.8502 & 3.3899 & 0.95 & -1.2715 & 0.4866 \\
			& Ti-Hetero & 0.8610 & 0.0751 & 0.8507 & 3.3949 & 0.95 & -1.2714 & 0.4867 \\
			& Tv-Homo & 0.8457 & 0.0084 & 0.8369 & 3.3324 & 0.95 & -1.2545 & 0.4775 \\
			& Tv-Hetero & 0.8457 & 0.0086 & 0.8369 & 3.3393 & 0.95 & -1.2551 & 0.4774 \\

			\midrule
			\multirow{4}{*}{\shortstack{DGP 2 \\ (Grp Ti He.)}}
			& Ti-Homo & 1.0969 & 0.0856 & 1.0860 & 4.2250 & 0.93 & -1.5155 & 0.6031 \\
			& Ti-Hetero & 1.0993 & 0.0723 & 1.0896 & 4.0301 & 0.94 & -1.4115 & 0.5857 \\
			& Tv-Homo & 1.0886 & 0.0054 & 1.0764 & 4.2115 & 0.93 & -1.5073 & 0.5971 \\
			& Tv-Hetero & 1.0900 & 0.0079 & 1.0775 & 3.9970 & 0.94 & -1.3806 & 0.5758 \\

			\midrule
			\multirow{4}{*}{\shortstack{DGP 3 \\ (Grp Tv Ho.)}}
			& Ti-Homo & 1.3326 & 0.3341 & 1.2852 & 5.0617 & 0.95 & -1.7082 & 0.7589 \\
			& Ti-Hetero & 1.3393 & 0.3219 & 1.2952 & 5.0363 & 0.95 & -1.7071 & 0.7616 \\
			& Tv-Homo & 1.3031 & 0.0273 & 1.2961 & 4.2823 & 0.90 & -1.7175 & 0.7477 \\
			& Tv-Hetero & 1.3052 & 0.0263 & 1.2982 & 4.2703 & 0.90 & -1.7150 & 0.7483 \\

			\midrule
			\multirow{4}{*}{\shortstack{DGP 4 \\ (Std Ti Ho.)}}
			& Ti-Homo & 0.8532 & -0.0232 & 0.8488 & 3.2353 & 0.94 & -1.2638 & 0.4822 \\
			& Ti-Hetero & 0.8532 & -0.0231 & 0.8488 & 3.2420 & 0.94 & -1.2641 & 0.4823 \\
			& Tv-Homo & 0.8537 & -0.0018 & 0.8458 & 3.2581 & 0.94 & -1.2640 & 0.4826 \\
			& Tv-Hetero & 0.8537 & -0.0014 & 0.8458 & 3.2654 & 0.94 & -1.2646 & 0.4826 \\
			\bottomrule
		\end{tabular}
	\end{center}
\end{table}

\subsection{Subjective Priors With Knowledge on Groups} \label{appendix:subPrior}

\begin{table}[H]
	\begin{center}
		\caption{Monte Carlo Experiment: Estimates, SGP prior, DGP3}
		\label{tab:MC_subPrior_est}
		\begin{tabular}{lrrrrrcc}
			\toprule
			& \multicolumn{5}{c}{$\hat{\rho}$} & \multicolumn{1}{c}{$\hat{\alpha_i}$} & Cluster \\
			\cmidrule(lr){2-6} \cmidrule(lr){7-7} \noalign{\smallskip}
			& \multicolumn{1}{c}{RMSE } & \multicolumn{1}{c}{Bias } & \multicolumn{1}{c}{Std } & \multicolumn{1}{c}{AvgL } & \multicolumn{1}{c}{Cov } & \multicolumn{1}{c}{Bias } & \multicolumn{1}{c}{Avg K} \\
			\midrule
			SGP-RE1 & 0.0396 & 0.0294 & 0.0225 & 0.0871 & 0.75 & -0.2072 & 4.00 \\
			SGP-RE2 & 0.0463 & 0.0378 & 0.0228 & 0.0892 & 0.64 & -0.2650 & 3.98 \\
			SGP-RE3 & 0.0760 & 0.0716 & 0.0214 & 0.0834 & 0.26 & -0.4993 & 3.58 \\
			SGP-RE4 & 0.0821 & 0.0793 & 0.0210 & 0.0818 & 0.02 & -0.5533 & 4.00 \\
			SGP-RE5 & 0.0691 & 0.0654 & 0.0214 & 0.0834 & 0.09 & -0.4583 & 6.00 \\
			TvHetero & 0.0599 & 0.0549 & 0.0220 & 0.0849 & 0.31 & -0.3842 & 3.85 \\
			Flat & 0.3243 & 0.3240 & 0.0126 & 0.0493 & 0 & -2.5626 & 100 \\
			\bottomrule
		\end{tabular}
	\end{center}
\end{table}

\begin{table}[H]
	\begin{center}
		\caption{Monte Carlo Experiment: Forecast, SGP prior, DGP3}
		\label{tab:MC_subPrior_fcst}
		\begin{tabular}{lrrrrrrr}
			\toprule
			& \multicolumn{3}{c}{Point Forecast}  & \multicolumn{2}{c}{Set Forecast} & \multicolumn{2}{c}{Density Forecast} \\
			\cmidrule(lr){2-4} \cmidrule(lr){5-6} \cmidrule(lr){7-8} \noalign{\smallskip}
			& \multicolumn{1}{c}{RMSFE } & \multicolumn{1}{c}{Error } & \multicolumn{1}{c}{Std } & \multicolumn{1}{c}{AvgL } & \multicolumn{1}{c}{Cov } & \multicolumn{1}{c}{LPS } & \multicolumn{1}{c}{CRPS }\\
			\midrule
			SGP-RE1 & 1.0266 & 0.0198 & 1.0178 & 3.7766 & 0.93 & -1.4553 & 0.5823 \\
			SGP-RE2 & 1.0465 & 0.0214 & 1.0377 & 3.8382 & 0.93 & -1.4716 & 0.5926 \\
			SGP-RE3 & 1.1337 & 0.0287 & 1.1250 & 4.1172 & 0.93 & -1.5542 & 0.6455 \\
			SGP-RE4 & 1.0682 & 0.0310 & 1.0590 & 4.0305 & 0.94 & -1.4934 & 0.6063 \\
			SGP-RE5 & 1.0784 & 0.0293 & 1.0696 & 3.9689 & 0.93 & -1.5070 & 0.6114 \\
			Tv-Hetero & 1.0952 & 0.0255 & 1.0866 & 3.9531 & 0.93 & -1.6450 & 0.6192 \\
			Flat & 1.2400 & 0.4211 & 1.1608 & 5.3189 & 0.97 & -1.6450 & 0.7048 \\
			\bottomrule
		\end{tabular}
	\end{center}
\end{table}

\subsubsection{Fixed $K$ Estimator: Imposing the true number of groups}

As shown in the previous subsection, the Bayesian GRE estimator works reasonably well in finite samples to determine the number of groups. In this subsection, we assume that the number of groups is known and focus on clustering. We present a table of the accuracy of clustering, where each row shows the faction of units that are correctly assigned to the true group. As an orthodox clustering algorithm, the results for Kmeans are also included as the benchmark. To avoid cluttering the tables in the main text, we don't present the results for suboptimal estimators. To be more precise, for the DGP involving time-invariant random effects, we only document the result for Ti-Homo and Ti-Hetero since other estimators are arguably worse in clustering, based on the simulation presented in the previous section. The same rule applies for time-varying DGPs.

Table \ref{tab:MC_fixK_prob} shows the accuracy of clustering for each estimator. Overall, the accuracy is high for Kmeans and correctly specified estimators in each DGP, while our BGRE estimators are slightly dominated by the Kmeans algorithm. The reasons are straightforward. Our BGRE estimators simultaneously estimate parameters and group units while Kmeans merely performs clustering. The additional estimation steps in our block Gibbs sampler depend on priors and parametric assumptions that could affect the clustering. On the other hand, the Kmeans algorithm forms clusters through spatial relationships between units, free of any assumption. Such differences yield the discrepancies in accuracy between Kmeans and BGRE estimators. But they are acceptable as the discrepancies are within 10\% most of the time. Comparing the performance of Kmeans in the two-step GRE estimator (Table \ref{tab:MC_grp_kmean}), imposing the correct number of groups indeed improves the clustering ability of Kmeans. Nevertheless, it is uncommon to know the truth in practice.


\begin{table}[htp]
	\begin{center}
		\caption{Monte Carlo Experiment: Accuracy of Clustering, Fixed $K^0$}
		\label{tab:MC_fixK_prob}
		\begin{tabular}{ll|cccc}
			\toprule
			& &\textbf{Group 1}&\textbf{Group 2}&\textbf{Group 3}&\textbf{Group 4} \\\hline
			\multirow{3}{*}{\shortstack{DGP 5 \\ (Grp Ti Ho.)}}
			& Ti-Homo & 86.87\% & 83.62\% & 60.42\% & 91.64\% \\
			& Ti-Hetero & 73.59\% & 66.17\% & 56.47\% &95.82\% \\
			& Kmeans & 88.00\% & 96.00\% & 68.00\% & 100.00\% \\

			\midrule
			\multirow{3}{*}{\shortstack{DGP 6 \\ (Grp Ti He.)}}
			& Ti-Homo & 78.17\% & 78.66\% & 66.58\% & 99.14\% \\
			& Ti-Hetero & 89.63\% & 84.22\% & 85.18\% & 99.98\% \\
			& Kmeans & 84.00\% & 100.00\% & 88.00\% & 100.00\% \\

			\midrule
			\multirow{3}{*}{\shortstack{DGP 7 \\ (Grp Tv Ho.)}}
			& Tv-Homo & 99.33\% & 68.99\% & 93.40\% & 84.61\% \\
			& Tv-Hetero & 98.92\% & 71.53\% & 93.41\% & 75.87\% \\
			& Kmeans & 96.00\% & 88.00\% & 100.00\% & 88.00\% \\
			\bottomrule
		\end{tabular}
	\end{center}
\end{table}

Next, we visualize the clusters to provide a clear view of the performance of clustering. We construct a posterior similarity matrix, a matrix containing the posterior probabilities of observations $i$ and $j$ being in the same cluster (estimated empirically from the MCMC draws). This design avoids the problem of reassigning group members to give posterior draw and show a clear group structure.

\begin{figure}[tbp]
	\caption{Heatmap for Similarity Matrix, DGP5, fix $K^0$}
	\label{fig:heatmap_dgp5}
	\begin{center}
		\includegraphics[scale= 0.5]{{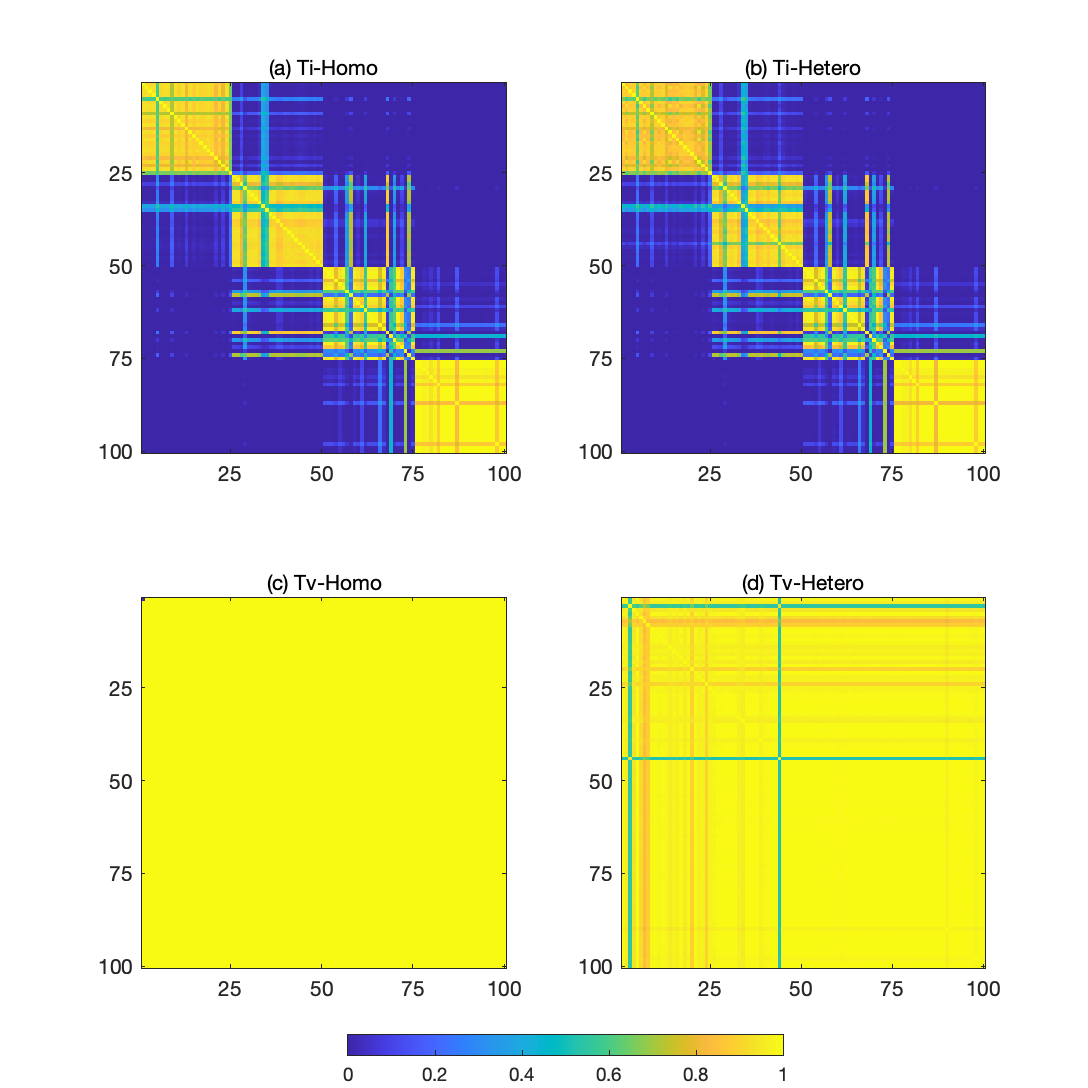}}
	\end{center}
\end{figure}

\begin{figure}[tbp]
	\caption{Heatmap for Similarity Matrix, DGP6, fix $K^0$}
	\label{fig:heatmap_dgp6}
	\begin{center}
		\includegraphics[scale= 0.5]{{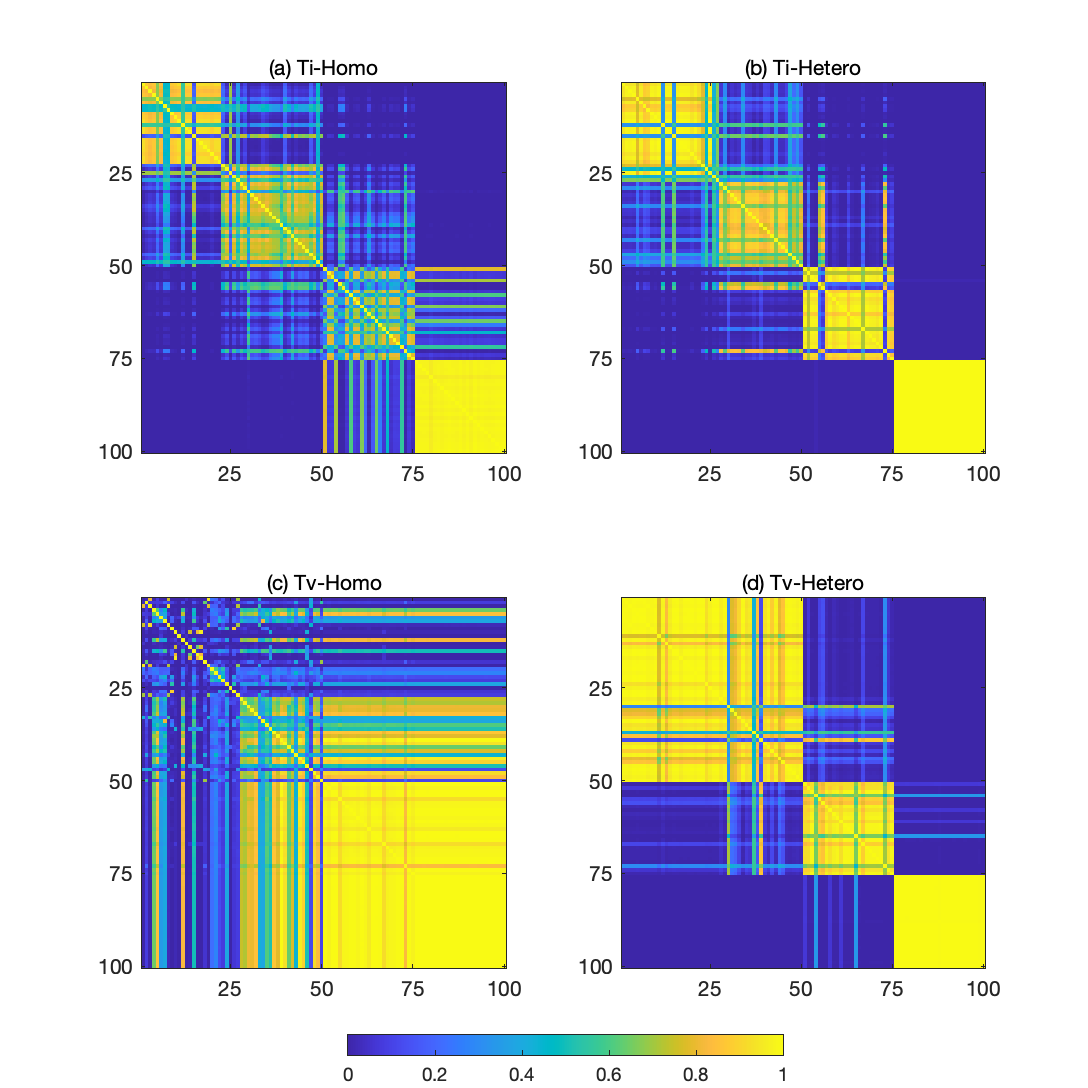}}
	\end{center}
\end{figure}

\begin{figure}[tbp]
	\caption{Heatmap for Similarity Matrix, DGP7, fix $K^0$}
	\label{fig:heatmap_dgp7}
	\begin{center}
		\includegraphics[scale= 0.5]{{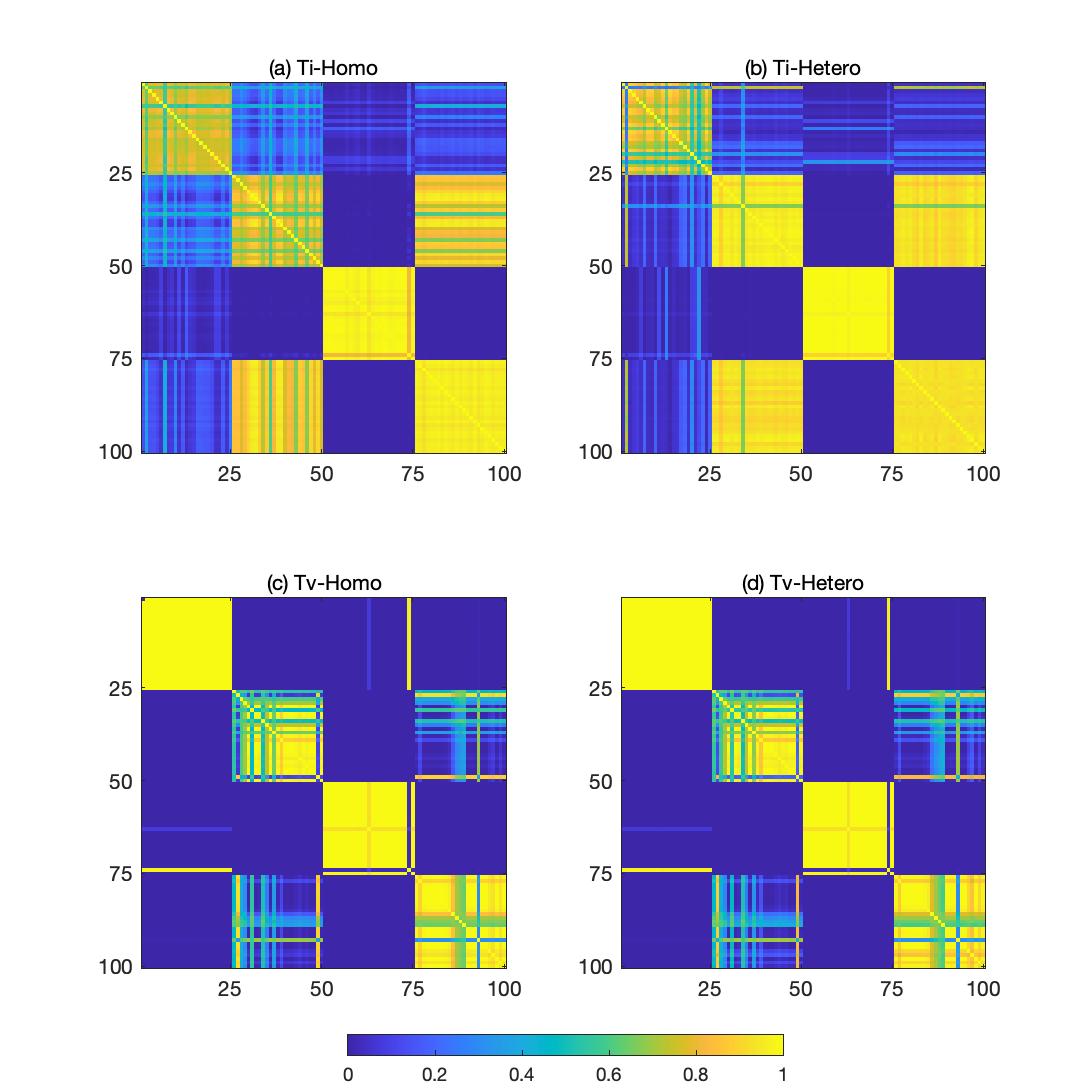}}
	\end{center}
\end{figure}

Figure \ref{fig:heatmap_dgp5}, \ref{fig:heatmap_dgp6} and \ref{fig:heatmap_dgp7} present the similarity matrices for the simulation using DGP5, DGP6 and DGP7, respectively. The colors depict the degree of similarity. Ideally, a perfect estimator should reveal four light yellow squares in the heatmap, leaving the remaining area in dark blue. As DGP5 implements fixed-effects and assumes homoskedasticity, Ti-Homo and Ti-Hetero estimator reveal a clear partition that matches the design of DGP5. Though a few units are incorrectly clustered, four yellow squares on the diagonal indicate that the posterior partition is reliable. However, Tv-Homo and Tv-Hetero estimators deliver inferior estimates and present one major group instead of four.

Turning to DGP6, the best partition is generated by the Ti-Hetero estimator, which is correctly specified under this DGP. Even though the data density of group 2 heavily overlaps with the one of group 1 and group 3, due to the relatively small mean and large variance in $\alpha_{i}$, Ti-Hetero estimator succeeds in delivering a clear group pattern that clearly distinguishes these three groups. The Ti-Homo estimator also has an excellent performance with ignorance of heteroskedasticity, but it generates much more vague boundaries between groups 1, 2, and 3. The Tv-Homo and Tv-Hetero results are incredibly messed, none of which depicts the correct partition.

As for DGP7, Tv-Homo and Tv-Hetero are the best, which is expected in this DGP. We see a clear four-group pattern from the similarity matrix in panel (c) and (d). A few yellow and light blue stripes in the off-diagonal block suggest Tv-Homo and Tv-Hetero estimators wrongly allocate a few units in posterior draws, especially for the units in group 2 and 4. Indeed, the paths of random effects in these two groups share great similarities. As depicted in Figure \ref{plot:dgp3_mean}, the red line (group 2) can be roughly viewed as the step function approximation of the green line (group 4). Ti-Home and Ti-Hetero struggle as they ignore the time effect in $\alpha_{i}$ by construction.

\newpage
\section{Data Set} \label{appendix:data}

The individual company raw annual data are obtained from the COMPUSTAT database. We constructed the sample using the data from the year 1999 to 2019. The reason to not use the data back to the 1970s is to avoid potential structure breaks in the variable of interest and to reflect the advanced speed of capital accumulation in recent decades. The primary variables of interest are:
\begin{itemize}
	\item K = Capital stock: net property, plant, and equipment. [PPENT]
	\item I = Investment: capital expenditures in property, plant, and equipment. [CAPX]
	\item Y = Sales: net sales revenues. [SALE]
	\item CF = Cash Flow: income after taxes and interest plus depreciation minus dividends. [EBITDA - TXT - XINT - DVT]
\end{itemize}
Additional variables used in the alternative model specification:
\begin{itemize}
	\item Q = Tobin's Q: define as (E+B-INV) / K - 1.
	\item E = Market value of equity: the sum of common equity and preferred equity. [$\text{PRCC}_f$*CSHO + PSTK]
	\item B = Book value of debt: the sum of short-term and long-term debt. [DLC + DLTT]
	\item INV = Market value of inventories. [INVT].
\end{itemize}
The variable names and formula in the bracket are corresponding items in COMPUSTAT. We process the raw data according to the following guidance:
\begin{enumerate}
	\item Observations where capital stock and sales are either zero or negative are eliminated.
	\item Firms that have missing values in the primary variables of interest during 1999-2019 are excluded.
	\item We eliminate any firm-year observation if the firm involved in merger and acquisition.
	\item Each firm must have valid annual observations from the year 1999 to 2019.
\end{enumerate}

The final sample comprises 337 firms and the observations on each firm is 20. The summary statistics are reported in Table \ref{tab:summary_stat}.
\begin{table}[htp]
	\begin{center}
		\begin{threeparttable}
			\caption{Descriptive Statistics for the Variables of Interest}
			\label{tab:summary_stat}
			\begin{tabular}{c|rrrrrrrrr}
				\toprule
				& Min & 25\% & Median & Mean & 75\% & Max & StD & Skew. & Kurt. \\
				\midrule
				I/K & 0.03 & 0.11 & 0.16 & 0.17 & 0.21 & 0.53 & 0.09 & 1.41 & 2.53 \\
				CF/K & -1.13 & 0.12 & 0.26 & 0.38 & 0.51 & 2.55 & 0.48 & 1.55 & 5.94 \\
				Y/K & -1.53 & 0.54 & 1.35 & 1.19 & 1.95 & 4.19 & 1.17 & -0.23 & -0.09 \\
				N/K & -8.63 & -5.36 & -4.19 & -4.56 & -3.46 & -1.77 & 1.52 & -0.74 & -0.12 \\
				log(K) & -0.37 & 5.16 & 6.77 & 6.60 & 8.32 & 9.82 & 2.26 & -0.63 & 0.21 \\
				q & -0.55 & 0.83 & 2.96 & 7.37 & 8.32 & 90.06 & 12.92 & 4.00 & 19.63 \\
				\bottomrule
			\end{tabular}
			\begin{tablenotes}
				\small
				\item Notes: The descriptive statistics are computed across $N$ and $T$ dimension of the panel.
			\end{tablenotes}
		\end{threeparttable}
	\end{center}
\end{table}

\section{Additional Empirical Results} \label{appendix:extra_empical}
In this section, we present the full result of empirical analysis in which detailed yearly estimate results are listed here.

\begin{table}[H]
	\begin{center}
		\caption{Empirical Application: Predict Investment Rate, RMSFE}
		\begin{tabular}{lllllll}
			\toprule
			& & 2015 & 2016 & 2017 & 2018 & 2019 \\
			\midrule
			\multirow{7}{*}{\shortstack{Homogenous \\ Coef.}}
			& Ti-Homo & 0.0917 & 0.1395 & 0.2625 & 0.1166 & 0.1108 \\
			& Ti-Hetero & 0.0750 & 0.1159 & 0.3550 & 0.0674 & 0.0822 \\
			& Tv-Homo & 0.0927 & 0.1382 & 0.2590 & 0.1165 & 0.1177 \\
			& Tv-Hetero & 0.0783 & 0.1156 & 0.3686 & 0.0692 & 0.0812 \\
			& Pooled & 0.0926 & 0.1386 & 0.2625 & 0.1160 & 0.1150 \\
			& Flat & 0.1034 & 0.1491 & 0.2703 & 0.1328 & 0.1100 \\
			& Param & 0.1958 & 0.2295 & 0.2466 & 0.2492 & 0.2043 \\
			\midrule
			\multirow{5}{*}{\shortstack{Heterogenous \\ Coef.}}
			& Ti-Homo & 0.1103 & 0.1006 & 1.8575 & 0.1041 & 0.1144 \\
			& Ti-Hetero & 0.1104 & 0.0999 & 1.8802 & 0.1028 & 0.1152 \\
			& Tv-Homo & 0.1582 & 0.1729 & 0.2863 & 0.1782 & 0.1070 \\
			& Tv-Hetero & 0.1097 & 0.1009 & 1.8644 & 0.1062 & 0.1101 \\
			& Flat & 0.1649 & 0.1906 & 1.7937 & 0.1833 & 0.1164 \\
			\bottomrule
		\end{tabular}
	\end{center}
\end{table}

\begin{table}[H]
	\begin{center}
		\caption{Empirical Application: Predict Investment Rate, Average Number of Groups}
			\begin{tabular}{lllllll}
				\toprule
				& & 2015 & 2016 & 2017 & 2018 & 2019 \\
				\midrule
				\multirow{7}{*}{\shortstack{Homogenous \\ Coef.}}
				& Ti-Homo & 2 & 2 & 1 & 2 & 2 \\
				& Ti-Hetero & 6.62 & 7.8 & 6.66 & 7.79 & 7.86 \\
				& Tv-Homo & 1 & 1 & 1 & 1 & 1 \\
				& Tv-Hetero & 6.75 & 6.79 & 6.8 & 7.64 & 6.75 \\
				& Pooled & 1 & 1 & 1 & 1 & 1 \\
				& Flat & 337 & 337 & 337 & 337 & 337 \\
				& Param & 1 & 1 & 1 & 1 & 1 \\
				\midrule
				\multirow{5}{*}{\shortstack{Heterogenous \\ Coef.}}
				& Ti-Homo & 7.75 & 6.66 & 6.78 & 8.05 & 7.48 \\
				& Ti-Hetero & 7.09 & 6.65 & 7.23 & 6.67 & 6.68 \\
				& Tv-Homo & 1 & 1 & 1 & 1 & 1 \\
				& Tv-Hetero & 6.07 & 6.11 & 6.79 & 6.57 & 7.46 \\
				& Flat & 337 & 337 & 337 & 337 & 337 \\
				\bottomrule
			\end{tabular}
	\end{center}
\end{table}

\begin{table}[H]
	\begin{center}
	\caption{Empirical Application: Predict Investment Rate, Frequentist Coverage Rate}
		\begin{tabular}{lllllll}
			\toprule
			& & 2015 & 2016 & 2017 & 2018 & 2019 \\
			\midrule
			\multirow{7}{*}{\shortstack{Homogenous \\ Coef.}}
			& Ti-Homo & 0.9822 & 0.9703 & 0.9733 & 0.9733 & 0.9614 \\
			& Ti-Hetero & 0.9525 & 0.9466 & 0.9674 & 0.9585 & 0.9021 \\
			& Tv-Homo & 0.9822 & 0.9644 & 0.9792 & 0.9733 & 0.9525 \\
			& Tv-Hetero & 0.9466 & 0.9347 & 0.9496 & 0.9466 & 0.8813 \\
			& Pooled & 0.9852 & 0.9644 & 0.9792 & 0.9763 & 0.9555 \\
			& Flat & 0.9792 & 0.9792 & 0.9703 & 0.9733 & 0.9703 \\
			& Param & 1 & 1 & 1 & 1 & 1 \\
			\midrule
			\multirow{5}{*}{\shortstack{Heterogenous \\ Coef.}}
			& Ti-Homo & 0.9407 & 0.9585 & 0.9555 & 0.9525 & 0.8724 \\
			& Ti-Hetero & 0.9436 & 0.9466 & 0.9674 & 0.9496 & 0.8724 \\
			& Tv-Homo & 0.9852 & 0.9792 & 0.9792 & 0.9792 & 0.9703 \\
			& Tv-Hetero & 0.9466 & 0.9466 & 0.9407 & 0.9258 & 0.8338 \\
			& Flat & 0.9733 & 0.9822 & 0.9733 & 0.9763 & 0.9733 \\
			\bottomrule
		\end{tabular}
	\end{center}
\end{table}

\begin{table}[H]
	\begin{center}
		\caption{Empirical Application: Predict Investment Rate, Length of 95\% Credible Set}
		\begin{tabular}{lllllll}
			\toprule
			& & 2015 & 2016 & 2017 & 2018 & 2019 \\
			\midrule
			\multirow{7}{*}{\shortstack{Homogenous \\ Coef.}}
			& Ti-Homo & 0.5737 & 0.5692 & 0.5710 & 0.5912 & 0.5908 \\
			& Ti-Hetero & 0.4149 & 0.3129 & 0.3223 & 0.3030 & 0.4012 \\
			& Tv-Homo & 0.5665 & 0.5620 & 0.5628 & 0.5851 & 0.5875 \\
			& Tv-Hetero & 0.2886 & 0.2801 & 0.2809 & 0.3810 & 0.3867 \\
			& Pooled & 0.5759 & 0.5716 & 0.5716 & 0.5946 & 0.5966 \\
			& Flat & 0.5709 & 0.5664 & 0.5729 & 0.5976 & 0.6041 \\
			& Param & 6.7334 & 6.7548 & 6.9387 & 6.8507 & 6.8211 \\
			\midrule
			\multirow{5}{*}{\shortstack{Heterogenous \\ Coef.}}
			& Ti-Homo & 0.2881 & 0.3008 & 0.403 & 0.2925 & 0.2837 \\
			& Ti-Hetero & 0.2889 & 0.3024 & 0.4033 & 0.2921 & 0.2841 \\
			& Tv-Homo & 0.6368 & 0.6605 & 0.7106 & 0.6414 & 0.6344 \\
			& Tv-Hetero & 0.2827 & 0.2961 & 0.3978 & 0.2840 & 0.2741 \\
			& Flat & 0.6660 & 0.7119 & 0.7948 & 0.6826 & 0.6753 \\
			\bottomrule
		\end{tabular}
	\end{center}
\end{table}

\begin{table}[H]
	\begin{center}
		\caption{Empirical Application: Predict Investment Rate, LPS}
		\begin{tabular}{lllllll}
			\toprule
			& & 2015 & 2016 & 2017 & 2018 & 2019 \\
			\midrule
			\multirow{7}{*}{\shortstack{Homogenous \\ Coef.}}
			& Ti-Homo & 0.8021 & 0.5577 & -0.3237 & 0.6752 & 0.7039 \\
			& Ti-Hetero & 1.5788 & 1.5833 & 1.6552 & 1.7086 & 1.3724 \\
			& Tv-Homo & 0.8044 & 0.5601 & -0.3055 & 0.6754 & 0.6671 \\
			& Tv-Hetero & 1.5618 & 1.5680 & 1.6063 & 1.6896 & 1.2981 \\
			& Pooled & 0.7952 & 0.5556 & -0.3197 & 0.6716 & 0.6746 \\
			& Flat & 0.7532 & 0.4900 & -0.5030 & 0.5868 & 0.6935 \\
			& Param & -1.2611 & -1.2671 & -1.2779 & -1.2760 & -1.2722 \\
			\midrule
			\multirow{5}{*}{\shortstack{Heterogenous \\ Coef.}}
			& Ti-Homo & 1.3901 & 1.5670 & 1.2802 & 1.6146 & 1.1904 \\
			& Ti-Hetero & 1.3059 & 1.5676 & 1.5278 & 1.6140 & 1.1883 \\
			& Tv-Homo & 0.6598 & 0.5030 & 0.4711 & 0.5313 & 0.6879 \\
			& Tv-Hetero & 1.5067 & 1.5490 & 1.5286 & 1.5520 & 1.1013 \\
			& Flat & 0.5675 & 0.4889 & 0.4966 & 0.4385 & 0.6205 \\
			\bottomrule
		\end{tabular}
	\end{center}
\end{table}

\begin{table}[H]
	\begin{center}
		\caption{Empirical Application: Predict Investment Rate, CRPS}
		\begin{tabular}{lllllll}
			\toprule
			& & 2015 & 2016 & 2017 & 2018 & 2019 \\
			\midrule
			\multirow{7}{*}{\shortstack{Homogenous \\ Coef.}}
			& Ti-Homo & 0.0529 & 0.0633 & 0.0712 & 0.0599 & 0.0605 \\
			& Ti-Hetero & 0.0398 & 0.0399 & 0.0511 & 0.0315 & 0.0464 \\
			& Tv-Homo & 0.0532 & 0.0635 & 0.0721 & 0.0599 & 0.0634 \\
			& Tv-Hetero & 0.0354 & 0.0394 & 0.0517 & 0.0332 & 0.0454 \\
			& Pooled & 0.0535 & 0.0637 & 0.0712 & 0.0603 & 0.0627 \\
			& Flat & 0.0537 & 0.0640 & 0.0723 & 0.0620 & 0.0604 \\
			& Param & 0.3510 & 0.3550 & 0.3649 & 0.3601 & 0.3554 \\
			\midrule
			\multirow{5}{*}{\shortstack{Heterogenous \\ Coef.}}
			& Ti-Homo & 0.0389 & 0.0382 & 0.1257 & 0.0343 & 0.0485 \\
			& Ti-Hetero & 0.0391 & 0.0381 & 0.1271 & 0.0346 & 0.0485 \\
			& Tv-Homo & 0.0639 & 0.072 & 0.0790 & 0.0696 & 0.0615 \\
			& Tv-Hetero & 0.0384 & 0.0379 & 0.1248 & 0.0353 & 0.0486 \\
			& Flat & 0.0702 & 0.0751 & 0.1587 & 0.0721 & 0.065 \\
			\bottomrule
		\end{tabular}
	\end{center}
\end{table}

\begin{figure}[H]
	\caption{Posterior Predictive Density for each industries, Year = 2019}
	\begin{center}
		\includegraphics[scale=0.7]{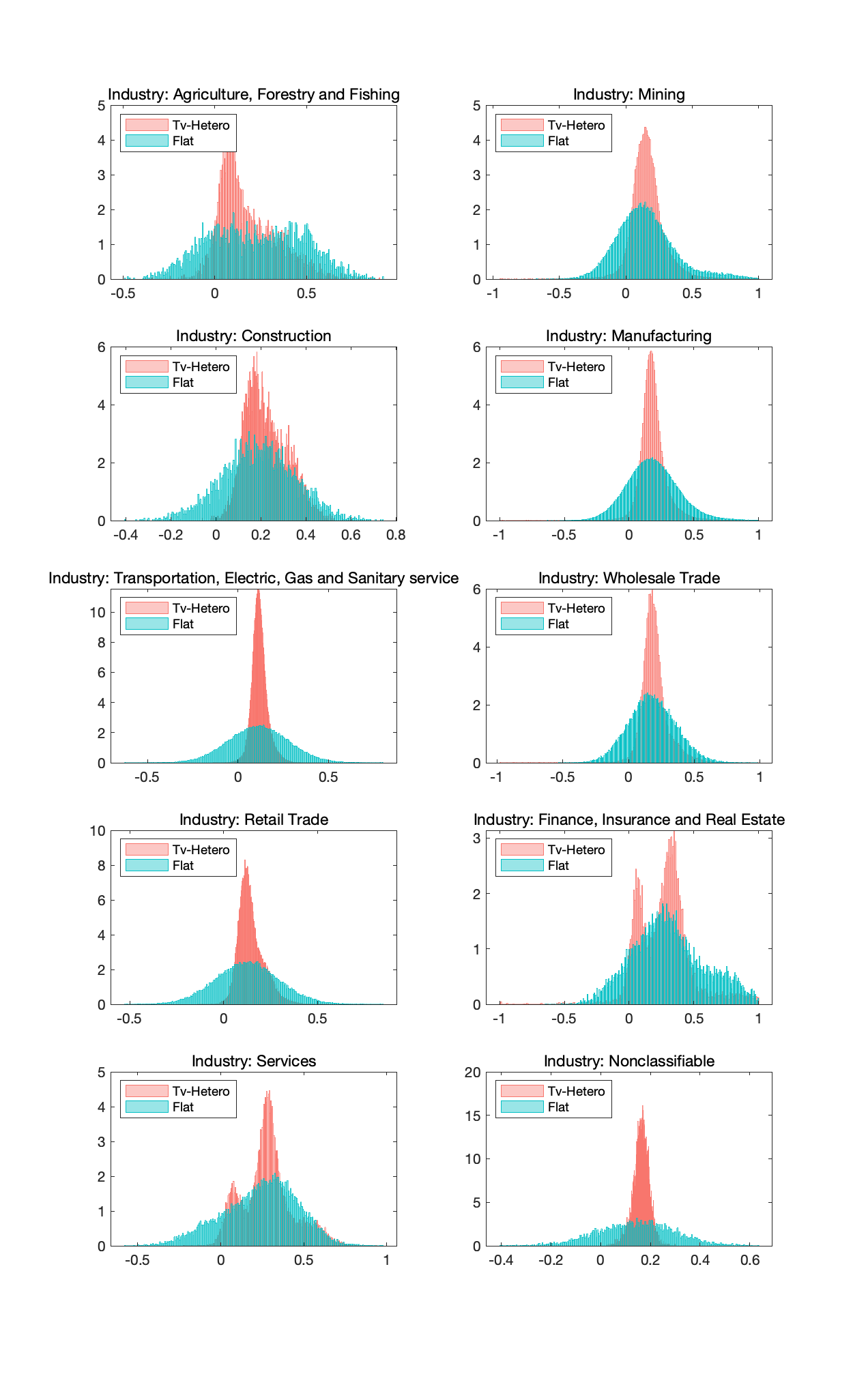}
	\end{center}
	\label{fig:yfcst_industry_panel}
\end{figure}

\end{document}